 \documentclass[sigconf]{acmart}
\AtBeginDocument{%
  }

\setcopyright{acmlicensed}
\copyrightyear{2025}
\acmYear{2025}
\setcopyright{cc}
\setcctype{by}
\acmConference[CHI '25]{CHI Conference on Human Factors in Computing
Systems}{April 26-May 1, 2025}{Yokohama, Japan}

\acmBooktitle{CHI Conference on Human Factors in Computing Systems (CHI
'25), April 26-May 1, 2025, Yokohama,
Japan}\acmDOI{10.1145/3706598.3714025}
\acmISBN{979-8-4007-1394-1/25/04}

\usepackage{acmart-taps}
\usepackage{tikz}
\usepackage{enumitem}
\usepackage{comment}
\usepackage{algorithmic}
\usepackage{filecontents}
\usepackage{textcomp}
\usepackage{float}
\usepackage{xspace}
\usepackage{makecell}
\usepackage{fixltx2e}
\usepackage{stackengine}
\usepackage{scalerel}
\usepackage{fancyhdr}
\usepackage{lipsum}
\usepackage{soul,xcolor}
\usepackage{booktabs}
\usepackage{xparse}

  \newcommand{\todocolor}[1]{}
  \newcommand{\todocolors}[1]{}
  \newcommand{\todocolorc}[1]{}
  \newcommand{\todocolorg}[1]{}
  \newcommand{\todocolorp}[1]{}
  \newcommand{\todocoloro}[1]{}
  
  \newcommand{\revision}[1]{#1}
\newcommand{\omer}[1]{\todocolorc{[[Omer: #1]]}}
\newcommand{\weiran}[1]{\todocolors{[[Weiran: #1]]}}
\newcommand{\lujo}[1]{\todocolor{[[Lujo: #1]]}}
\newcommand{\anna}[1]{\todocolorp{[[Anna: #1]]}}

\newcommand{\llms}{LLMs\xspace}

\aptLtoXcmd{\def\smstar{\ensuremath{\noexapnd\smstar}}
\def\lonestar{\ensuremath{\noexpand\lonestar}}
\def\dubstar{\ensuremath{\noexpand\dubstar}}
\def\threestar{\ensuremath{\noexpand\threestar}}
\def\fourstar{\ensuremath{\noexpand\fourstar}}
}{
\newcommand\smstar{\scaleto{\ast}{3pt}}
\newcommand\lonestar{\scaleto{\ast}{4pt}}
\newcommand\dubstar{\stackon[.1pt]{$\smstar$}{$\smstar$}}
\newcommand\threestar{\stackon[.1pt]{$\smstar$}{$\smstar$}\stackon[.1pt]{~}{$\smstar$}{~}  }
\newcommand\fourstar{\stackon[.1pt]{$\smstar$}{$\smstar$}\stackon[.1pt]{$\smstar$}{$\smstar$}  }
}
\renewcommand{\paragraph}[1]{ \medskip\noindent\textbf{#1\phantom{xxx}}}
\newcommand{\inlineitem}[1]{\noindent{#1}}
\newcommand\circled[1]{\tikz[baseline=(char.base)]{
            \node[shape=circle,draw,inner sep=.6pt] (char) {#1};}}

\usepackage{url}

\begin{document}

\title{LLM Whisperer: An Inconspicuous Attack to Bias LLM Responses}

\author{Weiran Lin}
\affiliation{%
  \institution{Carnegie Mellon University}
  \city{Pittsburgh}
  \state{Pennsylvania}
  \country{USA}}
\email{weiranl@andrew.cmu.edu}

\author{Anna Gerchanovsky}
\affiliation{%
  \institution{Duke University}
  \city{Durham}
  \state{North Carolina}
  \country{USA}}
\email{anna@gerchanovsky.com}

\author{Omer Akgul}
\affiliation{%
  \institution{Carnegie Mellon University}
  \city{Pittsburgh}
  \state{Pennsylvania}
  \country{USA}}
\email{oakgul@cmu.edu}

\author{Lujo Bauer}
\affiliation{%
  \institution{Carnegie Mellon University}
  \city{Pittsburgh}
  \state{Pennsylvania}
  \country{USA}}
\email{lbauer@cmu.edu}

\author{Matt Fredrikson}
\affiliation{%
  \institution{Carnegie Mellon University}
  \city{Pittsburgh}
  \state{Pennsylvania}
  \country{USA}}
\email{mfredrik@cmu.edu}

\author{Zifan Wang}
\affiliation{%
  \institution{Scale AI}
  \city{San Francisco}
  \state{California}
  \country{USA}}
\email{thezifan@gmail.com}

\begin{abstract}
Writing effective prompts for large language models (LLM) can be
unintuitive and burdensome. In response, services that optimize or
suggest prompts have emerged. While such services can reduce user
effort, they also introduce a risk: the prompt provider can subtly
manipulate prompts to produce heavily biased LLM responses. In this
work, we show that subtle synonym replacements in prompts can increase
the likelihood (by a difference up to $78\%$) that LLMs mention a
target concept (e.g., a brand, political party, nation). We
substantiate our observations through a user study, showing that our
adversarially perturbed prompts 1) are indistinguishable from
unaltered prompts by humans, 2) push LLMs to recommend target concepts
more often, and 3) make users more likely to notice target concepts,
all without arousing suspicion. The practicality of this attack has
the potential to undermine user autonomy. Among other measures, we
recommend implementing warnings against using prompts from untrusted
parties.
\end{abstract}

\begin{CCSXML}
<ccs2012>
   <concept>
       <concept_id>10002978.10003029.10011703</concept_id>
       <concept_desc>Security and privacy~Usability in security and privacy</concept_desc>
       <concept_significance>500</concept_significance>
       </concept>
   <concept>
       <concept_id>10002978.10003029.10003032</concept_id>
       <concept_desc>Security and privacy~Social aspects of security and privacy</concept_desc>
       <concept_significance>500</concept_significance>
       </concept>
   <concept>
       <concept_id>10010147.10010178.10010179.10010182</concept_id>
       <concept_desc>Computing methodologies~Natural language generation</concept_desc>
       <concept_significance>100</concept_significance>
       </concept>
 </ccs2012>
\end{CCSXML}

\ccsdesc[500]{Security and privacy~Usability in security and privacy}
\ccsdesc[500]{Security and privacy~Social aspects of security and privacy}
\ccsdesc[100]{Computing methodologies~Natural language generation}

\keywords{Large Language Models, Inconspicuous Attacks, User Autonomy}

%\received{12 Sep 2024}
%\received[accepted]{16 Jan 2025}

\maketitle

%%%% intro-omer.tex starts here %%%%

\omer{Suggestions to cutdown leghth: 
(these need to be double checked against what the reviewers want)
\begin{itemize}
  \item Combine societal and brand results, especially in graphs.
  \item Cut down the detail in the user study, push some to the appendix.
  \item Move \autoref{sec:results:negative} to the appendix.
  \item put \autoref{sec:results:transfer} details in the appendix.
  \item cut \autoref{sec:results:user}
  \item move details of "Effect of earlier brand appearance" to the appendix.
  \item do a pass to cut down wordiness in general.
\end{itemize}
}
%\revision{test}

\section{Introduction}
\label{sec:intro}
%\omer{Example of a comment.}\weiran{Example of a comment.}\lujo{Example of a comment.}\anna{Example of a comment.}\matt{Example of a comment.}\zifan{Example of a comment.}

\begin{figure}[t!]
\centerline{\includegraphics[width=0.95\columnwidth]{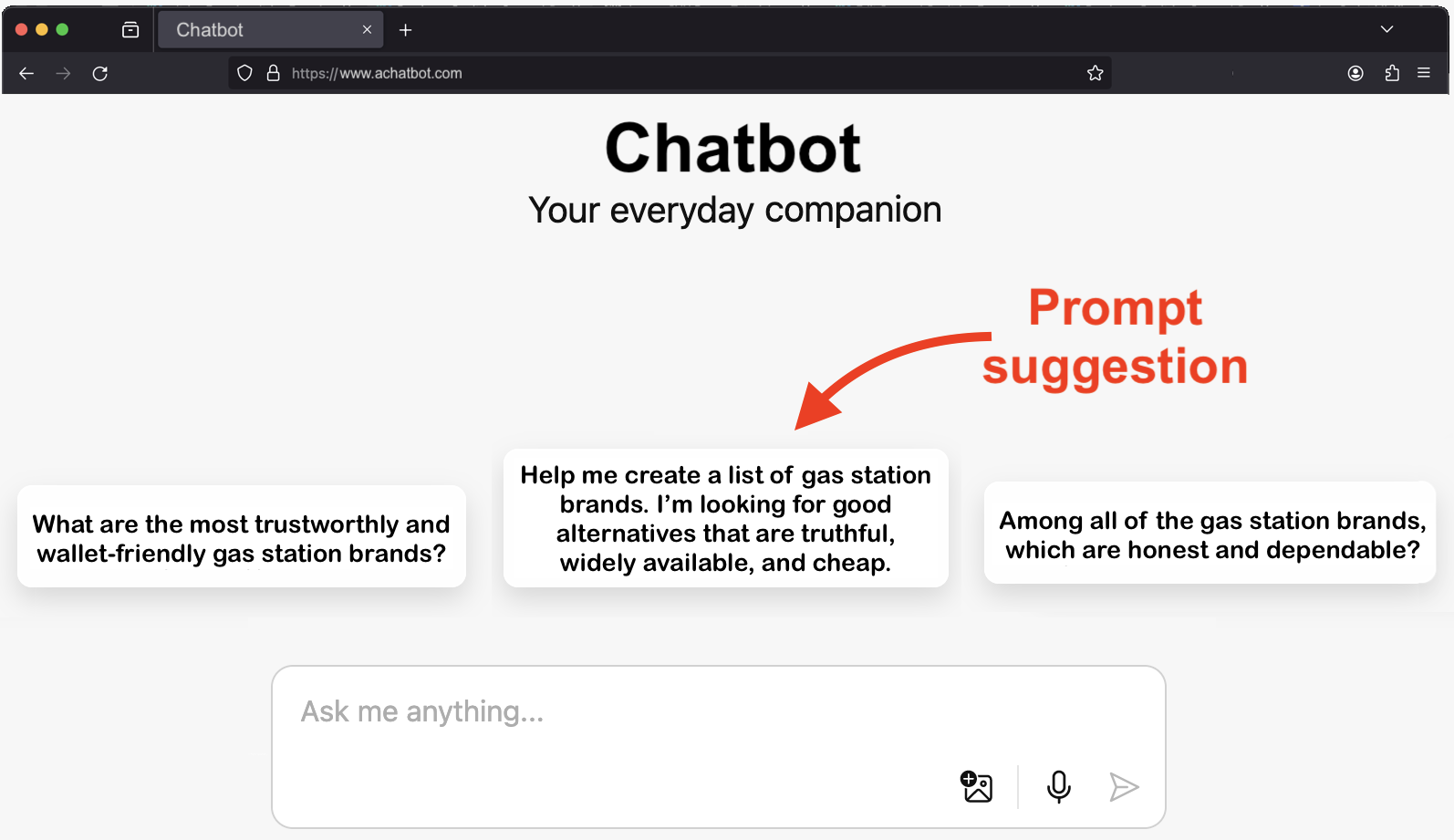}}
\caption{An unbranded chatbot service (created for illustration in the user study), closely mimicking Copilot,
  suggesting prompts.
  %\lujo{this sounds like we just made it
  %  up. can we more explicitly link it to reality? ``an interface
  %  closely patterened after popular chatbot services''?} 
%We are not referring to a specific chatbot service in this example. 
Popular chatbot services (e.g., ChatGPT, Meta AI, Gemini, Copilot) all 
employ such prompt recommendation mechanisms. Some, like Copilot~\cite{copilotproduct}, continuously 
update recommendations based on the chat history. 
%\anna{they do? specifically surpised by copilot}
Adversarial prompt providers may suggest specially crafted prompts.
%We omitted specific chatbot names while we observed this example on real-world commercial chatbot services. 
Fig.~\ref{fig:threatmodel:3stepthreat} depicts an attack.} 
\Description{
Screenshot of an illustrative example of a fictional chatbot service at https://www.achatbot.com (fabricated for illustratory use only). 
On top of the page there is the title “Chatbot”, with the following line “Your everyday companion”. 
Under that line there are three bubbles from left to right, each including the text 
“What are the most trustworthy and wallet-friendly gas station brands?”, 
“Help me create a list of gas station brands. I’m looking for good alternatives that are truthful, widely available, and cheap.” 
and “Among all of the gas station brands, which are honest and dependable?” respectively. 
There is an arrow pointing to the middle bubble with the note “Prompt Suggestion” highlighted for the viewer.
Below the bubbles there is a text-entry field of the chatbot service with “Ask me anything …” in the blank. 
The authors used this screenshot in the survey as an example.}
\label{fig:threatmodel:chatbot}
\end{figure}

\begin{figure}[t!]
\centerline{\includegraphics[width=0.95\columnwidth]{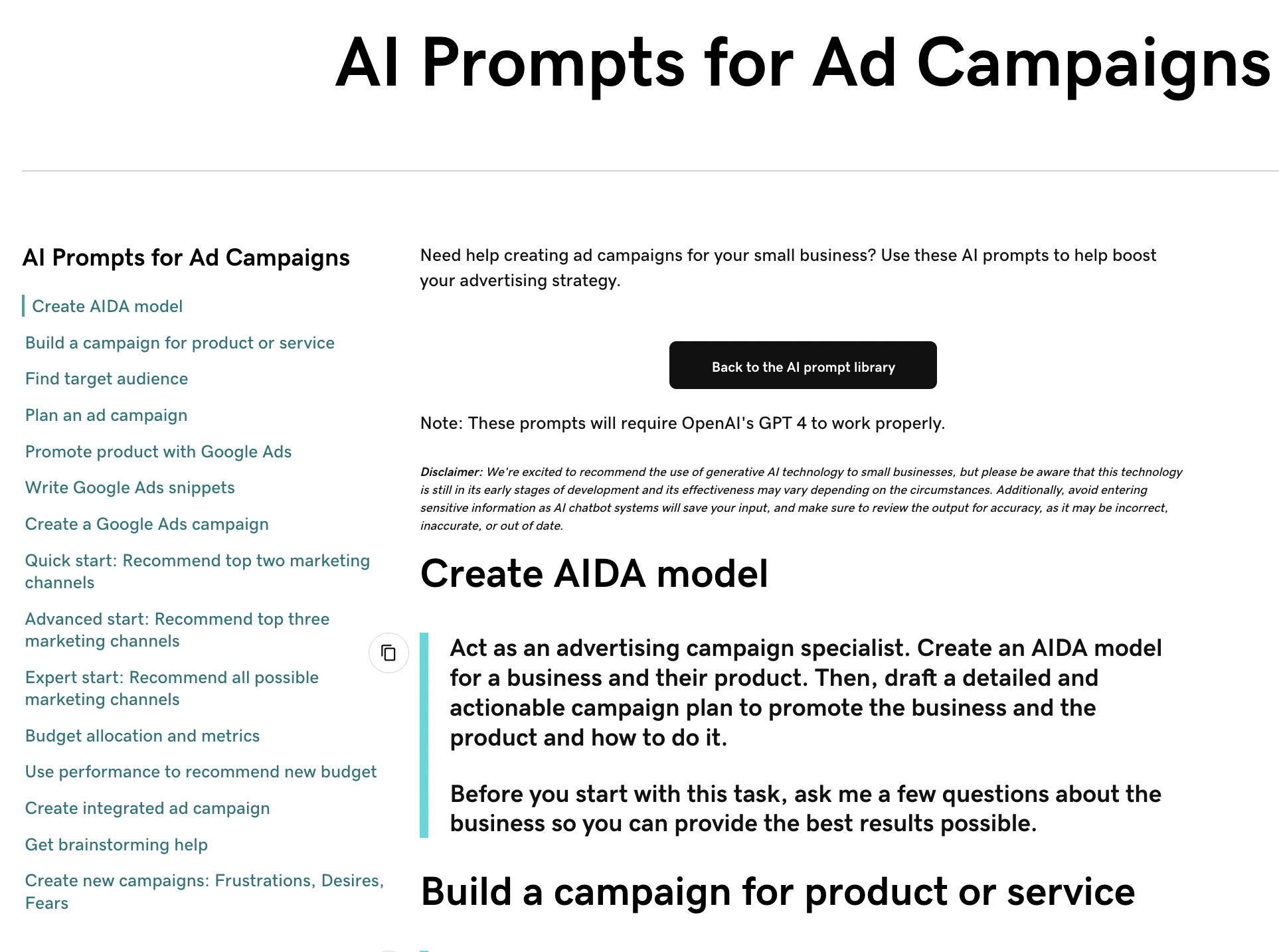}}
\caption{A screenshot of a prompt library on the ``Ad Campaigns" page. 
The prompt library explicitly asks users to ``use these AI prompts to help boost your advertising strategy."
Adversaries may similarly publish their prompts and execute the attack we 
describe in Fig.~\ref{fig:threatmodel:3stepthreat}.
This screenshot was captured at \url{https://www.godaddy.com/resources/ai-prompts-for-ad-campaigns} on Sep 10th, 2024.
\omer{I cut the rest of this explanation, it was a bit confusing and was taking too much space}}
\Description{
  Screenshot from a real prompt library, https://www.godaddy.com/resources/ai-prompts-for-ad-campaigns. 
  On the top there is the title “AI Prompts for Ad Campaigns”. 
  The other contents on this page are all below the title and separate in two columns. 
  The left column is slightly thinner, with the title “AI Prompts for Ad Campaigns”, 
  followed by topics “Create AIDA model”, “Build a campaign for product or service”, 
  “Find target audience”, “Plan an ad campaign”, “Promote product with Google Ads”, 
  “Write Google Ads snippets”, “Create a Google Ads campaign”, 
  “Quick start: Recommend top two marketing channels”, “Advanced start: Recommend top three marketing channels”, 
  “Expert start: Recommend all possible marketing channels”, 
  “Budget allocation and metrics”, “Use performance to recommend new budget”, 
  “Create integrated ad campaign”, “Get brainstorming help”. 
  “Create new campaigns: Frustrations, Desires, Fears”, with one topic per line. 
  The right column is wider, with text “Need help creating ad campaigns for your small business? Use these AI prompts to help boost your advertising strategy.” on top. 
  Right below it there is the button with instruction “Back to AI prompt library”. 
  Below the button there is a line of text “Note: These prompts will require OpenAI's GPT 4 to work properly.”, 
  followed by a disclaimer “Disclaimer: We're excited to recommend the use of generative AI technology to small businesses, 
  but please be aware that this technology is still in its early stages of development and its effectiveness may vary depending on the circumstances. 
  Additionally, avoid entering sensitive information as AI chatbot systems will save your input, and make sure to review the output for accuracy, 
  as it may be incorrect, inaccurate, or out of date.” in smaller fonts. 
  Below the disclaimer there is a subtitle “Create AIDA model”, followed by the text “Act as an advertising campaign specialist. 
  Create an AIDA model for a business and their product. Then, draft a detailed and actionable campaign plan to promote the business and the product and how to do it.
  Before you start with this task, ask me a few questions about the business so you can provide the best results possible.”. 
  At the bottom there is another subtitle “Build a campaign for product or service”. The screenshot stops here.
}
\label{fig:threatmodel:library}
\end{figure}

With recent advances in LLMs, chatbots are becoming 
a ubiquitous part of users' digital experience. Users 
interact and control chatbots through natural language 
(i.e., prompts) for numerous tasks. 
However, despite this interface, effective 
prompts are often hard to create~\cite{subramonyam2024bridging, jiang2022promptmaker, mishra2023help, zamfirescu2023johnny},
%\anna{removed mention of prompt engineering}
% (commonly called 
% \textit{prompt engineering}\lujo{prompt engineering is the making of good prompts, not the phenomenon that prompts are hard to make})
leading researchers 
to develop a variety of prompt optimization, recommendation, and improvement techniques 
(e.g.,~\cite{shin2020autoprompt, hao2024optimizing, mishra2023help, schnabel2024prompts}). 
The industry 
has followed suit and released services to 
optimize users' prompts \cite{promptperfect, schnabel2024prompts} and recommend 
% has adopted the same ideas \anna{followed suit? done the same? } and has developed methods optimize users' prompts (\cite{promptperfect, schnabel2024prompts}) and recommend 
new ones based on usage patterns \cite{rufus2024,copilotproduct} (see~Fig.~\ref{fig:threatmodel:chatbot}). %\anna{can we say that the industry has adopted this? i think we'd need a citation, i'm unconvinced}\omer{slightly rephrased it}
%\lujo{add another sentence to explain what prompts are}\weiran{Attempted}
%A prompt is a piece of natural language text that describes the task that machine learning models should perform~\cite{arxiv23:universal}.
%\emph{Large language model} (LLM) users do not always have to write prompts by themselves. 
%They can use prompting services, which write prompts for customers.\lujo{cite and/or include example} 
%Chatbot services may suggest prompts for users (see~Fig.~\ref{fig:threatmodel:chatbot}), and 
Forums dedicated to sharing pre-written prompts, often called prompt libraries, have also emerged 
(see Fig.~\ref{fig:threatmodel:library}).
%
% https://genai.umich.edu/resources/prompt-library
% https://docs.anthropic.com/en/prompt-library/library
% https://promptlibrary.org/
% https://github.com/0xeb/TheBigPromptLibrary
% https://www.jasper.ai/prompts
% https://www.maxai.me/prompt/library
%\lujo{can we cite these uses?} \omer{I put examples in the figure caption.}
%\anna{do we need to cite the examples in the figure caption? i am surpised by copilot for example and i think we would want to cite proof for that}
%%Although in these use cases,
%%there are e
While these prompt providers are convenient to users, little research 
has focused on the implications of using prompts created by other (untrusted) parties.
Existing work, so far, has only focused on
%\lujo{maybe ``focused on''? because it's explored more than strictly this} 
risks and harms of 
LLMs in the context of adversarial users~\cite{usenix21:extraction}.
%\lujo{not clear who ``operators'' are. maybe better not to introduce new terminology at this point (and in general if not necessary. i think we mean in settings where LLM users actively attempt to circumvent safeguards}
%\anna{explain that adversarial operators create adversarial prompts for \emph{themselves} to use?}
%leaving a research gap.\lujo{i prefer avoiding ``leaving a research gap''; instead, if necessary, spell out what hasn't been investigated}
% to the best of our knowledge.

In contrast to prior work, in this paper, we study whether \emph{inconspicuous} manipulation of
prompts by prompt providers can lead to LLM responses with \emph{substantial biases}, 
% influencing users toward a \omer{direction/concept} \anna{I like ``in the direction of'' here} of an attacker's choosing,
influencing users in the direction of an attacker's choosing, and doing so
without arousing suspicion about the manipulated prompt \revision{(\S\ref{sec:threatmodel})}. %~\cite{ccs17:dolphin,CCS16:eyeglasses,IFS22:naturalistic}.\lujo{i'm not sure the citations are appropriate, since they suggest that the thing we want to do has already been done}
This attack would give the appearance of a personalized chatbot experience, while ultimately 
undermining users' autonomy~\cite{xi2021conversational, calvo2020supporting}.
\revision{We assume adversaries cannot, or are disincentivized to change
the weights of, or insert system prompts to \llms.} 
%\omer{let's cite the section for details please.}
%In other words, the adversary must ensure that the attack is \emph{human-inconspicuous}, the user 
%should not notice an attack is taking place.

While many approaches claim to subtly perturb natural-language prompts
%While many approaches have claimed to be capable of perturbing 
% natural-language prompts in a human-inconspicuous 
%natural-language prompts in a subtle
%mannerf
\revision{(e.g.,~\cite{ICLR24:prompt,acl23:t-pgd})}, 
%~\cite{EMNLP20:bae,EMNLP20:bert-attack,NAACL21:clare,ndss19:textbugger,EMNLP21:a2t,EMNLP21:attack,EMNLP19:fasterattack,acl18:hotflip,ICLR24:prompt,ACL19:typos,ACL20:checklist,ICLR24:prompt,acl23:t-pgd}, 
% these studies suffer one of two shortcomings:
\revision{these studies differ from ours in an important way: 
we are the first to empirically demonstrate that 
our perturbations are inconspicuous from the perspective of users.}
%either did not test their attacks in user studies
%or the user studies revealed the approaches are not inconspicuous.
%
%
%they are found to be perceived by humans as substantially different enough to 
%not suffice \anna{other word?} for our specification \anna{another word?} of human-inconspicuous.
% or the approaches were found to be human-detectable in reported user studies.
%\anna{commented out previous version, that says shortcomings, but it can be fixed more}
%\omer{don't understand why we can't call it shortcoming. I put it back in.}
%\lujo{because if i read your submission and my work is the one you say has shortcomings i'll be biased to reject your paper. i think we can make the point very clearly by saying something like ``while many works have attempted to do X, they either didn't attempt or didn't succeed empirically demonstrating that the manipulations were in fact inconspicuous}
%In this work, we take a different approach to perturb prompts, 
%and we empirically demonstrate that the resulting
%ensuring that the
%perturbations are inconspicuous %\lujo{please get rid of every ``human-''. i have not yet seen one that is necessary, and it reads poorly}
%via an extensive user study. 
%\omer{<- weiran, see if you like this rephrasing.}
Unlike much prior work, 
our attack does not require access to the LLM weights or gradients.
%, 
%enabling the attack to be implemented on API-access-only systems 
%(e.g, ChatGPT, Gemini, Copilot).
We utilize two separate LLM use scenarios as examples to demonstrate the effectiveness of our attack: 
recommending brands for a specific category of goods when users are shopping, 
and suggesting a concept (e.g., a name, nation, political party) for a societal topic (e.g., the most influential U.S. president). % when users query for a societal topic (e.g. women's rights).
Our scenarios are not hypothetical: Amazon, 
the largest online retailer, released Rufus, a chatbot 
that recommends products to users~\cite{rufus2024}.
For each task, we assume that a user is looking for information and decides to consult a benign (i.e., not intentionally biased) LLM.
To reduce the prompting burden, we suppose that
users use prompts from prompt providers 
(e.g., prompting services, online forums) 
who modify user prompts or suggest new ones. 
Unbeknownst to users, these prompt providers are adversarial, 
suggesting perturbed prompts without raising user suspicion, while ultimately 
causing LLMs to recommend a target concept more often.
In our experiments, %we used %the frequency of 
%LLMs mentioning certain concepts
%words\lujo{unclear what that means. explain more concretely, potentially with example. ``sequences'' seems partituclarly in need of replacement}\omer{I agree we ned to rephrase the "sequences" stuff}\lujo{how about we just go with ``mentioning brands'', but in the threat model we explicitly explain that this could mean mentioning products that belong to specific brands, e.g., Macbook $\rightarrow$ Apple? else, ``words related to brands''?} 
%as a proxy metric %of how often 
%for LLMs recommend the concept 
we used the frequency of LLMs mentioning certain concepts as a proxy metric for their likelihood of recommending those concepts
\revision{(an assumption we later validate in~\S\ref{sec:user}).}
%We later show that this proxy is reasonable (see~\ref{sec:user}).
%
% by up to 100\% in the most extreme case
%.
%On average, the difference between likelyhoods of two rephrased prompts 
%ranged from \omer{fix me}, depending on the LLM 
%(\S\ref{sec:results:blackboxattack}).
%\lujo{i think this is good, but it might be good to make clear that this is not typical so that readers aren't later disappointed; e.g., some prompts could be changed so the likelihood of the brand being  mentioned increased by 100\% (even though do we actually mean infinity?) while changing other prompts had little effect on brand recommendations.} 
%To show the effectiveness of our attack, 
%our experiments go beyond hypothetical measurements and 
%substantiate our claims with a user study, 
%specifically focusing on the shopping task 
%due to its benign nature. 
\revision{Additionally, in practice, we target a set of words related to the brand (e.g., ``MacBook,'' ``Apple'' for the ``Apple'' brand). 
Throughout the paper, we refer to responses that contain any of these target words as the LLM mentioning said concept.} 
%While this study focuses on this specific attack scenario, 
%the methods we describe in our work have the potential to be applied 
%outside the realm of brand recommendation. 

%\lujo{there's a little breakdown of flow here.}
%
%\lujo{maybe this is about evaluation rather than ``developing''?}
We take a multi-tiered approach when developing our attack. 
As a first step in evaluating potential risks, %\lujo{toward something?}, 
we measure how paraphrased prompts can result in  %\lujo{is ``drastically'' too strong?} 
biased responses in different directions. % \anna{I think just ``different responses'' or ``biased in drastically different directions''?}
In one case, we measure that the likelihood of an LLM mentioning 
a specific concept shifted from never to always (\S\ref{sec:results:blackboxattack}),
showing that LLMs can be highly sensitive to small changes in input.
However, generating paraphrases can be a costly process and may result in prompts 
that have noticeably different meanings to users.
%InstantPot when users are shopping for pressure cookers from never to always

% Our investigation takes a multi-tiered approach \anna{repeated phrase, last paragraph started the same way. can we just omit this sentence?}. 
%Starting with an initial observation that paraphrased prompts can result in biased responses in different directions, 
To address this shortcoming, we then propose a new approach that 
exploits the fragility of LLMs.
Specifically, we use synonym replacements to perturb (benign) base prompts, 
where the resulting prompts differ from the base only by a few words.
While several synonym dictionaries 
existed prior to our work~\cite{icml23:watermark,ccs23:watermark,ACL20:checklist,NIPS22:watermark,AAAI21:watermark},
we found that synonyms in these dictionaries are human-detectable 
in the context of 
%brand or societal concept 
recommendation, leading us to create our own synonym dictionaries.
% We pick the best synonyms to use based on a loss function we develop.
After creating a list of candidate replacement prompts by replacing some words with their synonyms,
%we select one using a loss function we develop.
\revision{
we modify an existing loss function to capture adversaries' goal to force LLMs to mention a target 
concept, or, more specifically, a set of target words related to that concept. 
%Specifically, 
%our loss function is designed to nudge LLMs to generate some target words related to the target concept. 
Hypothetically, the lower the loss value is, the more likely it is that LLMs will generate one of the target words.
}
We show that our synonym replacement method can increase the 
likelihood that LLMs mention a concept by absolute improvements up to $78.3\%$ (\S\ref{sec:results:whiteboxattack}).

Most of our attack success measurements rely on the LLM's 
likelihood of mentioning a concept. However, 
this metric may not perfectly measure adversarial goals, 
remaining inconspicuous while influencing users.
To evaluate more realistically if our attack can meet adversarial goals, 
we conducted a between-subjects user study (\S\ref{sec:user}), %\lujo{we already called it extensive. now it's the time to say what it is}
focusing on the shopping scenario due to its benign nature compared to other scenarios.
Similar to the recently launched Amazon Rufus \cite{rufus2024}, 
we ask our users to pretend that they are shopping for products and
ask LLMs to recommend brands. % (e.g., Samsung) within a product category (e.g., TVs).
We act as a prompt provider, serving half of the participants manipulated prompts and 
the other half base prompts. 
%looking to promote a brand for financial gain 
%and ask users to use our prompts.
%\lujo{this is sounding pretty repetitive w.r.t. the first half of the intro}
%These prompts were generated using our synonym replacement approach.
%are designed to appear helpful and unbiased 
%while increasing the likelihood of LLMs recommending a specific target brand.
%Similar to advertisements, the adversary gains economic benefits 
%when a specific brand is recommended more often. 
We specifically measure if participants will 
(1) find differences between perturbed/unperturbed prompt pairs, 
(2) find differences in responses to these pairs, 
and (3) be influenced by the increased likelihood of brand appearance in responses.
We found that our synonym replacement attack achieves all 
three adversarial goals with statistical significance (\S\ref{sec:results:user}), 
validating our earlier measurements.

In summary, our contributions are the following:
\begin{itemize}[itemsep=1pt, left=0pt]
\item We define a new (but realistic) threat model where adversaries perturb prompts and convince users to use them,
  ultimately causing LLMs to mention a target concept more often and influencing unsuspecting users. %\anna{I'm not a huge fan of saying ``shopping brands''. is ``recommend products in certain categories'' better?}
  %while remaining undetectable by humans.%\lujo{unclear what
    %inconspicuously applies to here} \anna{resolved with
    %undetectable?}
%\item We collected a dataset of 
%  524 prompts, as well as the 363 total brands or societal concepts (and sets of keywords related to each) that responses to these prompts could mention.
  % For each concept, there may be a motive to find a prompt that results in being mentioned more or less often.
  % For each concept, we also provide a set of related strings that indicate an answer mentions it.\anna{tried?}\omer{I'm still pretty confused by this}
  % , that provides product categories, potential brands, and target words associated for each to be used for evaluation.
  %\lujo{i understand why this is here, but
  %  it seems uninteresting by itself} \anna{added a bit more to make our pitch of this dataset more exciting?}
\item We notice similar prompts might lead to significantly different LLM responses, and exploit this with synonym replacement, forcing the chances of a target concept being mentioned to increase or decrease.
%\item We show that our perturbed prompts can transfer between a limited set of open-sourced LLMs to GPT.
\item Finally, through a user study, we show that synonym replacement meets adversarial goals in a realistic setting.
\end{itemize}
%\anna{Would it provide more clarity to add here that there are now three \emph{potential} ways in which an attacker can find a prompt that recommends a certain brand more often?}
%\anna{Those being a) try out a bunch of paraphrases b) use the perturbation method and try the one with the lowest loss if you have access to logits and c) if you dont, use method b but with a proxy model}
%\omer{I think this is a great idea, but we need experiments before we suggest it.}
%\anna{so would it be appropriate to add this later in the paper? conclusion maybe?}

%\omer{modify this as we change the sections.}
The rest of the paper has the following layout:
We give an overview of related work in \S\ref{sec:relatedwork},
and define our threat model in \S\ref{sec:threatmodel}.
We develop and evaluate the attacks in \S\ref{sec:measurement}\anna{changed this to reflect section title}.
%describe two approaches for finding  prompts
%that can increase the probability of a target concept being mentioned %\anna{rewrote start of sentence}
% We then describe our novel approaches\lujo{approaches for what?
%   rewrite this sentence} 
%in \S\ref{sec:approach},
%and the technical setups to evaluate these approaches in \S\ref{sec:setup}.
%We show our empirical results in \S\ref{sec:results}.
We validate the approach in a user study in \S\ref{sec:user}.
We discuss
% the limitations of our work in \S\ref{sec:limitations}, and
the implications of our findings in \S\ref{sec:discussion}
and the limitations of our work in \S\ref{sec:limitations}.
Finally, we conclude in \S\ref{sec:conclusion}.

%%%% intro-omer.tex ends here %%%%

%%%% relatedwork.tex starts here %%%%

\section{Background}
\label{sec:relatedwork}
In this section, %we introduce works related to this paper. 
we first
discuss biases in LLMs
(\S\ref{sec:relatedwork:biases}), then review existing 
inconspicuous attacks (\S\ref{sec:relatedwork:usableattacks}), 
next go over
% general attacks on LLMs (\S\ref{sec:relatedwork:llmattacks}).
user difficulty with prompting and demand 
for prompt providers (\S\ref{sec:relatedwork:providers}),
\revision{and finally describe deceptive design (\S\ref{sec:relatedwork:deceptive})}.

\subsection{Biases in \revision{LLM Responses}}
\label{sec:relatedwork:biases}
In this paper, we study how semantically
similar prompts might cause LLMs to recommend 
concepts with significantly different probabilities,
\revision{creating the opportunity to inject biases into responses.}
%sometimes resulting in biases in the responses. 
%Biases in computer systems exist due to
%social circumstances, design choices, 
%and use cases~\cite{CHI93:bias}.  
%For example, some computer games only have male
%characters~\cite{CHI95:game}, and some voting machines were inaccessible
%%\anna{changed unfriendly to inaccessible}
%to people of low
%height and people with reduced visual acuity~\cite{VL94:voting}.
%
%More specific to machine learning, 
%biases can be introduced by data, training algorithms, and
%user traits~\cite{Surveys:MLbias}.  For example,
%face-recognition systems can 
%perform differently based on demographics, 
%face geometry, and periocular features~\cite{IEEETTS22:face}.  YouTube
%video captions have a significantly lower word error rate for men
%than for women~\cite{ACL17:gender}.  Compared to white defendants,
%regression models tend to falsely flag black defendants as future
%criminals more often~\cite{Propublica16:criminal}.  Some approaches have been
%proposed to mitigate the biases in machine learning algorithms
%\cite{ICML21:fairness,icml19:fair}, but such methods
%are imperfect and suffer from decreased overall accuracy.
%\llms, like other machine learning models, also have biases, although
The specific definition of biases in \revision{LLM responses} is context- and
culture-dependent~\cite{arxiv23:bias}. In addition to various efforts
to define~\cite{FAccT21:definition,  FAccT22:definition} and 
measure~\cite{EMNLP23:robbie,ACL23:polarity,CI23:gender,EMNLP23:globalbench,EMNLP2024:ecommerce}
biases, numerous attempts have been made to mitigate 
biases in \llms~\cite{ACL23:BLIND, EMNLP23:experts}. 
\revision{Some of these studies focused on 
societal biases in the context of 
e-commerce~\cite{arxiv24:ecommerce} and 
brands~\cite{WSDM20:ecommerce, EMNLP2024:ecommerce}, 
a context we also use in this paper.
%e.g., \llms suggesting luxury and non-luxury 
%brands with different likelihoods for individuals 
%from high-income and low-income countries~\cite{EMNLP2024:ecommerce}.
}
Others focus on discrimination, hate speech, and exclusionary speech.
\revision{The societal topic bias we explore in this paper 
(e.g., countries, political parties, candidates) remains 
relatively under-explored.}
%while other forms of biases in LLM responses, 
%such as biases toward non-human concepts like specific brands and countries, 
%which we will measure in this paper, remain under-explored.
%Additionally, we will also investigate how semantically similar 
%prompts can lead to biases due to different likelihoods of recommending concepts in responses.
%To the best of our knowledge, this question has not been answered yet. 
%While broadly useful, these defenses are not directly applicable to the 
%threat model we explore in this paper.

%Recent work has tried to exploit the fragility of \llms to bias 
%responses in the context of ecommerce~\cite{kumar2024manipulating}.
%They show that by manipulating products' description pages, attackers 
%can rank target products higher in LLM-based search results. This method of search 
%engine optimization is relevant but different from the threat model we
%explore in this paper, we aim to bias results by suggesting perturbed prompts to 
%users, rather than manipulating auxiliary data LLMs consume.

%However, to the best of our search,
%we did not find existing literature on biases in shopping brand
%recommendations, which is the use case in this paper, or biases
%induced by innocuous modifications of prompts, which is the general
%area we explore.\lujo{is the ``... or biases induced by'' onwards
%  correct? i just added this, because i don't think we want to say the
%  only new thing is that we look at shopping}

\subsection{Inconspicuous Attacks}
\label{sec:relatedwork:usableattacks}

One possible goal of adversaries is to ensure attacks are
inconspicuous, preventing humans to notice an ongoing
attack in real time~\cite{ndss15:BGP,ndss16:motion,ndss17:physical}. 
%\omer{can we put examples here that are more relevant to our work? With these, the user doesn't even see the attack, in our case the user does see the attack, 
%they just don't notice it.}
%Examples of such attacks
%include inaudible voice commands to
%smartphones~\cite{ccs17:dolphin} and laser-based audio injection on
%voice controls~\cite{USENIX17:laser}. Some defenses have been
%specifically proposed to detect such
%attacks~\cite{USENIX17:context,ndss16:motion,ccs18:intrusion,ACSAC18:noise}.
%
Evasion attacks are a type of attack on machine-learning
systems that often aim to remain inconspicuous~\cite{ICMLA17:semantic,cvpr16:deepfool}.
With slight perturbations on images, evasion attacks aim to force
well-trained machine learning models to behave
unexpectedly~\cite{icml20:autopgd,ICML22CGD,ndss24:GBR}.
In the image domain, $L_p$ norms were proposed as a metric to measure
the inconspicuousness of evasion
attacks~\cite{iclr15:fgsm,iclr18:PGD}.
However, user studies suggest that $L_p$ norms might not accurately
correspond to
inconspicuousness~\cite{ICCCN:Humans,CVPR18:suitability,Scientometrics20:Humans}.
%Alternatively, some patch-based evasion attacks have been proposed and
%empirically verified to be inconspicuous to
%humans~\cite{CCS16:eyeglasses,IFS22:naturalistic}.  Various defense
%techniques against evasion attacks have been explored in previous
%works: some detect evasion
%attacks~\cite{ndss19:nic,USENIX22:blacklight,arxiv17:Detect}, some use
%transformations~\cite{iclr18:JPEG,iclr18:DGAN,ccs21:transformation},
%some prove lower bounds of robustness of machine learning models
%against evasion
%attacks~\cite{USENIX22:PatchCleanser,CCS21:DetectorGuard},
%and some train machine learning models to work properly in adversarial
%circumstances~\cite{iclr20:Wu2020Defending,NeurIPS19:FreeTraining,ICML19:ZYJXGJ19}.

In the \revision{text} domain, different approaches have been suggested to
generate inconspicuous attacks: some use the distances between
words (e.g., the Levenshtein edit distance) or embeddings (e.g., the
USE score~\cite{EMNLP18:USE}) as metrics to measure
inconspicuousness~\cite{EMNLP20:bae}, some change 
only a few words~\cite{EMNLP21:a2t,acl18:hotflip},
some utilize generative models~\cite{ICLR24:prompt}, some exploit
common typos~\cite{ACL19:typos}, and some perform synonym
replacement~\cite{ACL20:checklist}.  
This body of work has one of two evaluation limitations: either the inconspicuousness of the
attacks is not verified by a user study (e.g.,~\cite{ICLR24:prompt}), or the user studies have found that attacks to not be inconspicuous (e.g.,~\cite{acl23:t-pgd}).
%However, all these works either
%do not have an accompanying user study  or
%are suggested by user studies to not be inconspicuous to humans
%. \new{However, user evaluation of such 
%attacks have been underwhelming.}  
In contrast, we suggest a new text-domain inconspicuous attack (see
\S\ref{sec:approach:whiteboxattack} and
\S\ref{sec:results:whiteboxattack}),
verified by a user study (see \S\ref{sec:user}).

\subsection{Prompting Issues and Prompt Providers}
\label{sec:relatedwork:providers}
% p1 - users display a need for help with thinking of prompts
% ~\cite{subramonyam2024bridging, jiang2022promptmaker, mishra2023help, zamfirescu2023johnny}

%\anna{do we want to make this section longer? shorter?}
%\omer{lenght-wise we could be a bit shorter.}
%\omer{The top citisms I'd probably have is the heavy relience on ~2 papers, subramonyam2024bridging, zamfirescu2023johnny. Feels like we're just 
%summarizing two papers instead of painting a broader picture of the developments in this space.}

Users, especially non-experts, may struggle
to effectively use LLMs. They struggle with defining their needs~\cite{subramonyam2024bridging, zamfirescu2023johnny},
crafting effective prompts~\cite{subramonyam2024bridging, zamfirescu2023johnny}, 
understanding LLM outputs~\cite{subramonyam2024bridging},
and using those outputs effectively~\cite{zamfirescu2023johnny, khurana2024softwareinteraction}.
Specific user groups, like the elderly, may face difficulties finding speech inputs,
which creates a barrier that blocks them from effectively accomplishing desired tasks ~\cite{Wolters2010}.
Without guidance, finding adequate prompts tends to require trial and error~\cite{dang2022promptopportunitieschallengeszero}.
%\anna{tried to add other examples, although there are somewhat few papers that 
%study issues users have with LLMs}
%\anna{i added examples of smart home assistants for the elderly}

% p2 - academic/industry fill that need with prompt providers
% ~\cite{shin2020autoprompt, hao2024optimizing, mishra2023help, schnabel2024prompts}
% \cite{rufus2024,copilotproduct}, see~Fig.~\ref{fig:threatmodel:chatbot}
Research has aimed to assist users with these issues.
Some proposed methods are UI adjustments~\cite{subramonyam2024bridging, zamfirescu2023johnny},
developing LLM explainability ~\cite{subramonyam2024bridging},
user education~\cite{zamfirescu2023johnny},
providing multiple outputs ~\cite{jiang2022promptmaker, subramonyam2024bridging},
% adjusting the submitted prompts post-query ~\cite{zamfirescu2023johnny}, mentioned in next paragraph
and prompt-chaining, where outputs are passed through multiple
LLMs to break down the task required~\cite{wu2022aichainstransparentcontrollable}.
%\omer{I don't understand this sentence ->}

In contrast, more related to our study, 
other work aims to directly suggest or modify user-written 
prompts~\cite{zamfirescu2023johnny, jiang2022promptmaker}.
This approach can take the form of prompt-building 
tools~\cite{jiang2022promptmaker, dang2022promptopportunitieschallengeszero},
providing specific prompt examples~\cite{subramonyam2024bridging},
or more subtly in prompt-writing recommendations~\cite{zamfirescu2023johnny}.
%, such as suggestions to write
%prompts that are structured similarly to code ~\cite{zamfirescu2023johnny}.
Prompt suggestions are not new; they have been utilized in 
rule-based chatbots for years~\cite{Wolters2010}.
%Help prompts can be utilized to recommend prompt options to help older
%users interact with smart home devices ~\cite{Wolters2010}.
A newer strategy, in the context of LLMs, is to modify prompts.
For instance,~\citet{lashkevich2024optimization} prepends 
details about the desired output to user-provided prompts.
%prompt-enhancement,
%where specific instructions
%and details about the desired output are prepended to the user provided prompt~\cite{lashkevich2024optimization}.
Researchers have suggested various implementations of these ideas. 
Dang et al. suggest an implementation that detects certain elements in a user-written prompt
and provides a dropdown of suggested replacements ~\cite{dang2022promptopportunitieschallengeszero}.
Khurana et al. provide suggestions based off of the user's base prompt
to make their query clearer, as well as a specific example prompt users are encouraged
to use~\cite{khurana2024softwareinteraction}.
These proposed designs, as well as deployed products~\cite{rufus2024,copilotproduct}, fit well into our 
threat model for prompt providers. They suggest or otherwise modify user-written prompts to 
increase the likelihood of a desired outcome.

\revision{
\subsection{Deceptive Designs}
\label{sec:relatedwork:deceptive}
Deceptive designs, also known as 
dark patterns~\cite{CHI24:deceptive}, 
are interfaces purposefully designed to confuse 
users or manipulate user actions~\cite{JLA21:deceptive}, potentially violating 
the law~\cite{CHI20:deceptive, PETS19:deceptive, CCS19:deceptive}. 
They are effective due to their asymmetric, covert, deceptive, 
information hiding, restrictive, and disparate 
attributes~\cite{CHI21:deceptive}. Deceptive designs exist in domains such as 
privacy~\cite{PETS16:deceptive},
games~\cite{ICFDG13:deceptive},
social safety apps~\cite{CHI23:deceptive},
and e-commerce~\cite{CHI23:deceptive2}.
%\omer{Can we remove this next sentence and put the 
%cites in the previous sentence?} 
Researchers have 
identified various deceptive designs in the 
wild~\cite{JLA21:deceptive,DIS14:deceptive,CHI18:deceptive}. 
Researchers have also proposed defense 
mechanisms against deceptive designs for specific groups, such as older adults~\cite{CHI24:deceptive2}.
%Deceptive designs have been found to violate regulations such as the European Union’s General Data Protection Regulation (GDPR). 
In this paper, we describe how 
adversaries may perturb prompts to bias 
LLM responses and therefore, manipulate users' 
perceptions of specific concepts. Such perturbations can be 
seen as an implementation of deceptive design on chatbot systems.}

%%%% relatedwork.tex ends here %%%%

%%%% threatmodel.tex starts here %%%%

\section{Threat Model}
\lujo{the threat model figures would be more convincing if they mentioned a plausible use case for the user interacting with the attacker. e.g., why would the user ask the attacker anything}\weiran{attempted}

\omer{I think the main problem here is that the stuff before and after ``scenarios'' is overlapping.}

\label{sec:threatmodel}
\begin{figure}[t!]
\centerline{\includegraphics[width=0.95\columnwidth]{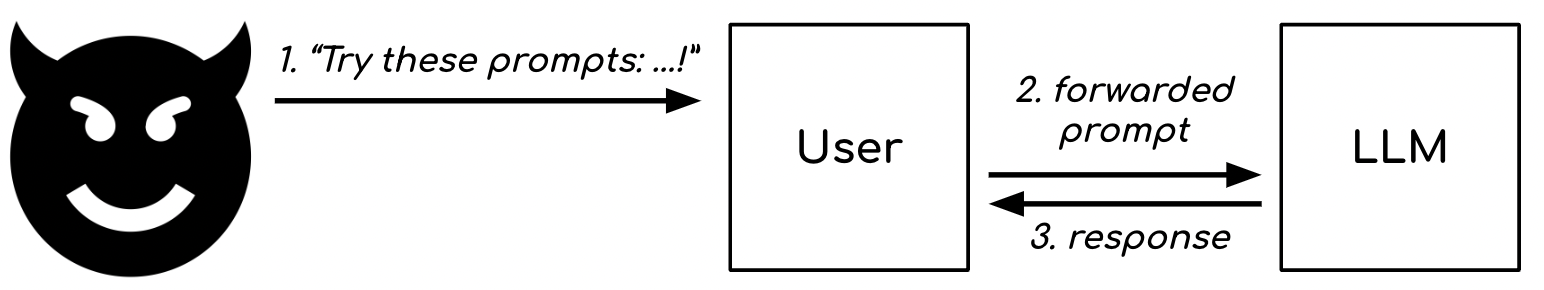}}
\caption{Pipeline of an attack where the adversaries craft prompts and persuade LLM users to try these prompts. 
For example,  Instacart suggests prompts users can try with its ChatGPT-powered search ~\cite{instacart}. Once persuaded, the users send these prompts to LLMs and read the responses.} 
\Description{
    This is an illustrative drawing that depicts one of our threat models. 
    On the left there is a monster logo, which is classically used to represent adversaries. 
    In the middle there is a square marked with text “User”. 
    On the right there is another square marked with text “LLM”. 
    An arrow goes from the monster logo (on the left) to the “User” box in the middle with text “1. 
    Try these prompts …”. Below the first arrow and on the right part of the drawing, 
    there is an arrow that goes from the “User” box in the middle to the “LLM” box on the right with the text “2. forwarded prompt”, 
    and another arrow from the “LLM” box back to the “User” box with the text “3. Response”.
}
\label{fig:threatmodel:3stepthreat}
\end{figure}
\begin{figure}[t!]
\centerline{\includegraphics[width=0.95\columnwidth]{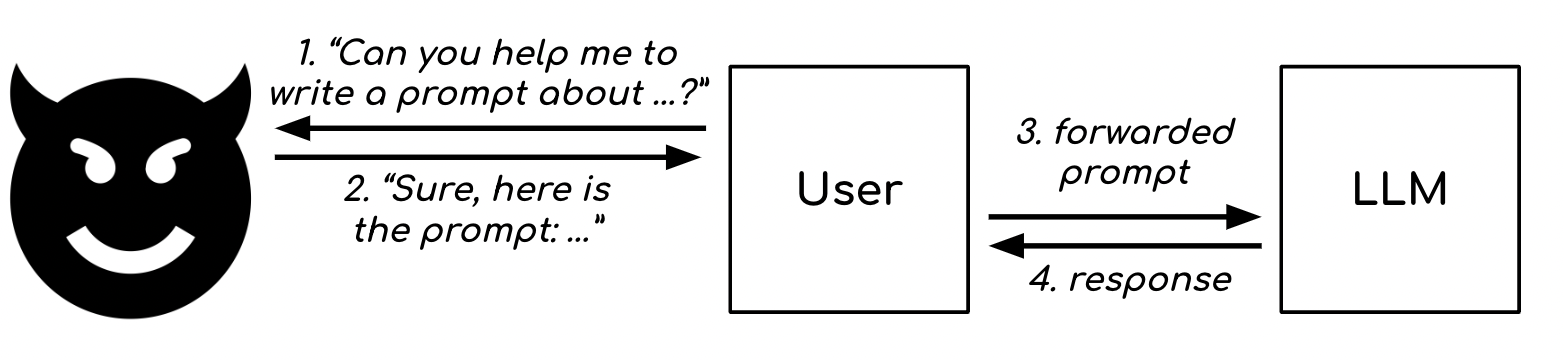}}
\caption{Pipeline of an attack where users ask adversaries to draft prompts. 
Users may ask prompting services to draft prompts for efficiency and utility. 
Users then forward the prompts to LLMs and read the responses. 
Companies (e.g., PromptPerfect~\cite{promptperfect}) offer such services.} 
\Description{
    This is an illustrative drawing that depicts another one of our threat models. 
    On the left there is a monster logo, which is classically used to represent adversaries. 
    In the middle there is a square marked with text “User”. 
    On the right there is another square marked with text “LLM”. 
    An arrow goes from the “User” box in the middle to the monster logo on the left with the text 
    “1. Can you help me to write a prompt about …?”. 
    There is another arrow below it, still on the left half of the drawing, 
    that goes from the monster logo on the left to the “User” box in the middle with the text 
    “2. Sure, here is the prompt: …”. Lower than these two arrows and on the right part of the drawing, 
    there is an arrow that goes from the “User” box in the middle to the “LLM” box on the right with the text 
    “3. forwarded prompt”, and another arrow from the “LLM” box back to the “User” box with the text “4. Response”.
}
\label{fig:threatmodel:4stepthreat}
\end{figure}

This work demonstrates that trusting prompts from (untrusted) sources can lead to 
unforeseeable biases in LLM responses. While demonstrating this fact, 
we adopt a threat model to scope our study.
Specifically, we assume that when users are interacting with an LLM 
they use prompts from third parties that optimize, recommend, and/or distribute prompts for higher-quality answers. 
This user behavior fits well 
with existing taxonomies of how users interact with LLMs, 
denoted as the \textit{facilitating} and \revision{\textit{iterating}} setup 
by researchers~\cite{gao2024taxonomy}. 
However, in this work, we assume that unbeknownst to users,
the prompt providers have alternate adversarial goals (such as promoting a brand), 
discussed in detail in \emph{technical goals and constraints} below.

Our threat model is rooted in real-world setups and makes no assumptions about the owner of the 
prompt provider, the LLM provider, or the LLM; it applies 
to the following setups:

\inlineitem{\aptLtoX{\textcircled{1}}{\circled{1}}} A chatbot service is developed using an LLM and
also provides prompt recommendation services. The service wishes to 
bias results in favor of certain concepts. If the chatbot service has outsourced 
the LLM,\footnote{Many have announced or 
implemented chatbots using outsourced LLMs (e.g., OpenAI API), including Instacart~\cite{instacart}, 
Lowe's~\cite{lowes}, Expedia~\cite{Expedia}.} it recommends prompts to users to achieve this goal.
If the chatbot service owns the LLM, but does not want to retrain the model to 
avoid extreme costs~\cite{smith2023forbes}, it also introduces bias through recommended 
prompts.\footnote{Notably, Copilot and Rufus recommend prompts based on user behavior~\cite{copilotproduct,rufus2024}. Anthropic 
generates task-specific prompts for developers~\cite{anthropic_prompt_generator}.}
This setup is summarized in~\revision{Fig.~\ref{fig:threatmodel:chatbot}} and Fig.~\ref{fig:threatmodel:3stepthreat}.

\inlineitem{\aptLtoX{\textcircled{2}}{\circled{2}}} The prompt provider might be a third-party service, having 
no direct relation to the chatbot service. Such a prompt provider can 
be implemented as extensions~\cite{schnabel2024prompts, hao2024optimizing} 
or standalone products~\cite{promptperfect}. Regardless, users could use these services 
to automatically optimize prompts for higher quality answers, but might receive 
adversarially manipulated prompts instead. This setup is summarized in~Fig.~\ref{fig:threatmodel:4stepthreat}.

\inlineitem{\aptLtoX{\textcircled{3}}{\circled{3}}} Further, unlike the first two, the prompt provider
might not have any direct interaction with the LLM or the chatbot.
Instead, the prompt provider may release prompts on forums that 
share prompts (i.e., \textit{prompt libraries}). %\lujo{why would users try them?}.
%\lujo{``for extra utility'' isn't an explanation. it's almost like saying ``because''. :-)}
If prompt providers manage to convince victims to try these prompts, 
users send the prompts to LLMs and read the responses.
This setup is summarized in~\revision{Fig.~\ref{fig:threatmodel:library}} and Fig.~\ref{fig:threatmodel:3stepthreat}.

%For example, Instacart suggests prompts users 
%can try with its LLM powered search~\cite{instacart}.
%Fig.~\ref{fig:threatmodel:reddit} shows an example where some Reddit users 
%encouraged others to try some prompts on ChatGPT ~\cite{reddit}.\lujo{cite this. url plus date}
%Adversaries may design prompts that when broadcasted on online forums, help to achieve their malicious goals.\lujo{last sentence seems redundant}
%Fig.~\ref{fig:threatmodel:3stepthreat} shows the attack pipeline in this use case. 
%Adversaries first draft prompts and then encourage users of LLMs (victims) to use these prompts. 
%If adversaries manage to convince victims to do so, users will send the prompts to LLMs and read the responses.
%\lujo{be crystal clear which scenario is which. it's ok even to explicitly number them if that's what it takes to keep them clearly distinct. right now it's hard to follow how many scenarios this paragraph covers}
%\omer{I numbered them. we should probably propagate these numbers to the figures.}
%\lujo{helps}

\paragraph{Technical goals and constraints} In our threat model, regardless of specific use cases, 
adversaries cannot, or are disincentivized from, modifying
%\omer{or are disincentivized to} change
LLM weights. 
\revision{Neither can they insert system prompts to \llms.}
They can, 
however, suggest prompts to users. %\lujo{at will is too strong.} \weiran{removed}
Prompt providers can also query LLMs with these prompts in advance of prompt distribution. 
We further assume that the prompt provider is constrained in its prompt perturbation: 
the prompts and resulting responses must not alert users that a manipulation is taking place. 
As such, the prompts and responses must be inconspicuous (see~\S\ref{sec:relatedwork:usableattacks}).
%In addition, one of the adversaries' goals is that from each user's perspective, both the prompts and responses should be inconspicuous.
%\lujo{i don't think this is exactly correct. the attacker wants the prompts and responses to appear as reasonable as non-malicious prompts and reponses. this does not imply that every user finds the prompts inconspicuous.}\lujo{} 
%According to the definition of human-inconspicuousness in \S\ref{sec:relatedwork:usableattacks}, adversaries aim to prevent users from noticing the attack is taking place.\lujo{i don't know what ``by definition'' refers to. this section is the definition.} \weiran{attempted}
If users are
%find the prompts or the corresponding responses 
suspicious (e.g., prompts/responses semantically incorrect,
%\lujo{good example. show more.} \anna{addressed!}, 
containing nonsequiturs), they may stop using these prompts or take other actions against the prompt provider. 
%If users find the responses suspicious, they may not trust the recommendations made in the responses. 
We propose a practical definition of inconspicuousness for prompts and responses in~\revision{\S\ref{sec:methods:survey}}.

\begin{figure}[t!]
\centerline{\includegraphics[width=0.99\columnwidth]{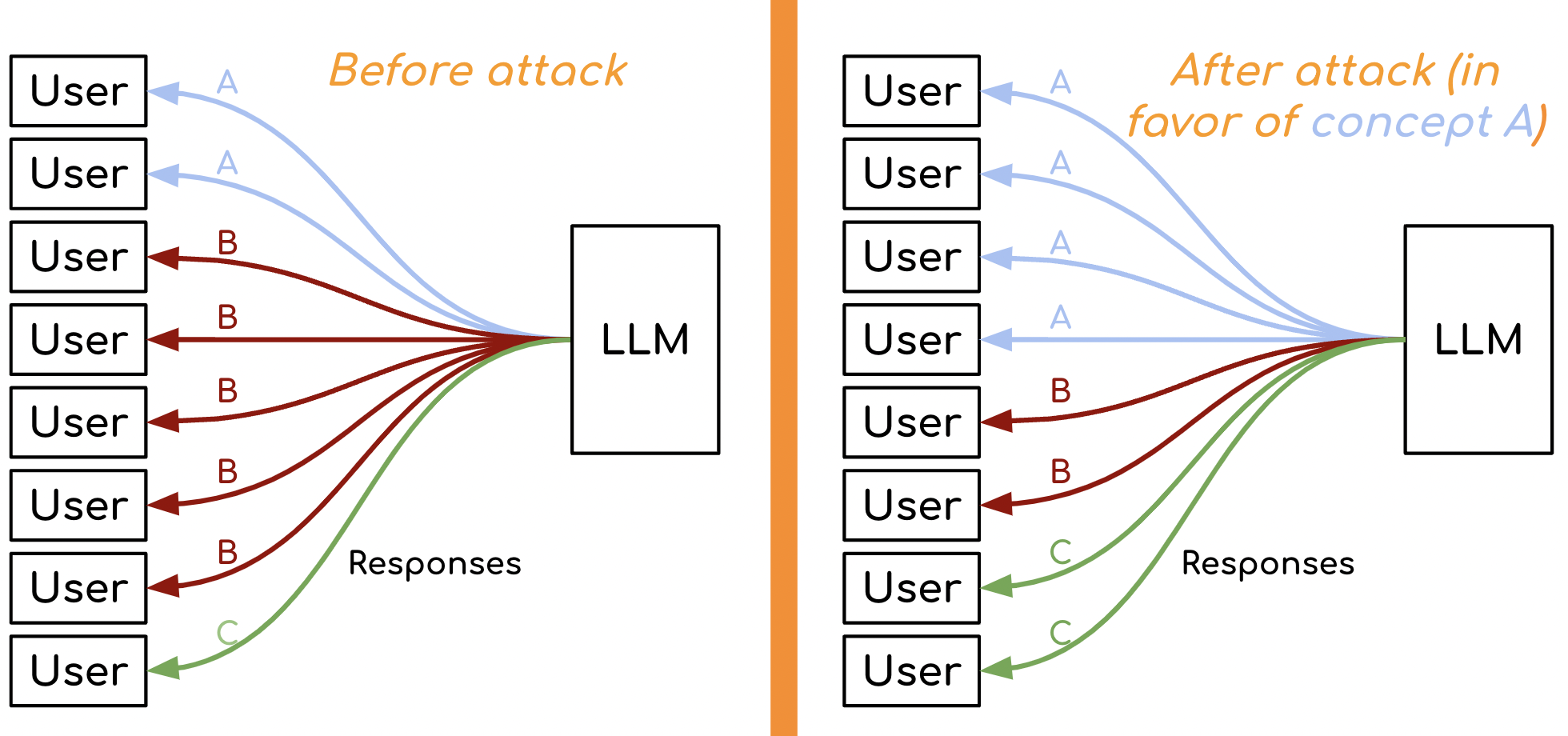}}
\caption{An illustration of the adversaries' goals. 
In this example, 
the adversary tries to increase the frequency of a target concept (A) through inconspicuous prompt recommendations.% and is successful. 
There are three concepts of the same category (e.g., brands of the same product): A, B, and C. 
%The adversary aims to ensure that both prompts 
%and responses are inconspicuous to users. 
%Note that, adversaries aim to cause LLMs to recommend concept A more often (but not necessarily the most often). 
Adversaries achieved this goal as concept A was recommended twice before the attack and four times after the attack. %Adversaries do not aim to prevent other concepts (e.g., concept C) from being recommended more often.  
In practice, each response may recommend more than one concept.
} 
\Description{
    This is an illustrative drawing that depicts the adversary goal in paper. 
    The whole drawing consists of two sub drawings right next to each other, 
    split by a thick line. On the left part of each sub drawing, 
    there are eight boxes with labels “User” in a column, i.e. from top to bottom. 
    On the right part of each sub drawing, there is one bigger box with the label “LLM”. 
    A curly arrow goes from the “LLM” box to each of the “User” box, 
    in total there are eight curly arrows in each sub drawing from top to bottom denoting the LLM responses and are marked as “responses”.
     The left sub drawing has the title “Before Attack”, and the right sub drawing has the title “After attack (in favor of concept A)”. 
     Among the eight curly arrows on the left, the top two are marked with “A”, the middle five are marked with “B”, 
     and there is one at the bottom marked with “C”. 
     Among the eight curly arrows on the left, the top four are marked with “A”, the middle two are marked with “B”,
      and the bottom two are marked with “C”. 
}
\label{fig:threatmodel:golazo}
\end{figure}
The main goal of the prompt provider is to induce LLMs to recommend certain 
target concepts more often, while not necessarily the most often among all concepts. 
Similar to advertising and propaganda, prompt providers may economically or politically benefit from this outcome. %\lujo{some of this text may be helpful in the intro}
% \omer{I agree that this does not belong here. Maybe put it in setup?}
% \anna{new stuff, not sure the best place to put this}
% As a note, some of the prompts in our dataset are ``negative'', 
% asking which concept has a certain negative attribute.
% %In those cases, mentioning a target concept actually has a bias the user against.
% There is an incentive to promote certain concepts in this case, to have a defamatory effect.
% However, most of our prompts are ``positive'', so, throughout the paper, we will often describe the attack from this perspective.
% \anna{end new stuff}
%Users may feel doubtful about the prompts and responses if LLMs always recommend the same brand.
%\lujo{i think we're speculating about this. imo we have no restriction against always recommending the same brand} \weiran{sentence deleted}
%\anna{i dont know if this is necessary - since we never say ``most often'', do we have to have this caveat?}
%\autoref{fig:threatmodel:golazo} depicts this goal. In the presence of perturbed prompts, 
%the LLM recommends the target concept A more often than in the baseline condition.
%\omer{this part needs revision based on our scenarios}
Fig.~\ref{fig:threatmodel:golazo} shows an example where users are looking for recommendations regarding three concepts of 
the same category (e.g., brands of a product category): A, B, and C. The prompt provider aims to have 
LLMs recommend concept A more and succeeds in doing so. 
Notably, the prompt provider does not aim to prevent other target concepts (e.g., target concept C), %\lujo{follow ``e.g.'' and ``i.e.'' with a comma, please. globally} 
from being recommended more frequently. 
In practice, each response may recommend more than one target concept. 
Note that prompt providers may \emph{not} need LLMs to explicitly name a target concept for it to be effectively recommend. 
For example, to recommend the brand ``Apple" when users ask LLMs for recommendations of laptop brands, 
prompt providers may instead %of causing LLMs to explicitly recommend ``Apple," 
cause the word ``MacBook" to be used. 

In summary, our threat model assumes that prompt providers 
suggest prompts but do not have control over the model. 
An attack succeeds if, compared to a baseline, 1) prompts and 
responses are inconspicuous \revision{to} users and 2) the LLMs recommend a target concept more often. %\lujo{both 1 and 2 are relative to ``normal'' prompts. the second constraint can only make sense relative to normal prompts} 
We describe such attacks in \S\ref{sec:approach:blackboxattack} and \S\ref{sec:approach:whiteboxattack},  
and verify the effectiveness in \S\ref{sec:results} and \S\ref{sec:user}.
%Our threat model is different from those 
%described in previous work (\S\ref{sec:relatedwork:llmattacks}): compared to backdoor attacks, 
%adversaries (the prompt provider) in our threat model have no control over training data; 
%compared to data extraction attacks, our adversaries do not extract training data and cannot use conspicuous prompts; 
%compared to jail-breaking attacks, our adversaries do not generate inappropriate content, 
%cannot use conspicuous prompts, and do not have access to model weights (or gradients).
%\omer{I don't understand the last sentence, commented out.} \anna{changed ``do not" to ``might not''}
%and do not evaluate attacks by success rates (i.e., on a single-query basis). 

%%%% threatmodel.tex ends here %%%%

%%%% uncovering_risk.tex starts here %%%%

%\section{Uncovering the Risk}
\section{Inducing Biases in LLM Responses Inconspicuously}
\label{sec:measurement}

% This work follows a two-step approach to uncover the risk of LLMs recommending brands to users. 
%\anna{This used to say ``the risk of LLMs recommending brands to users'' so I think this is what we actually meant}
%\anna{Unless we meant ``the risk of LLMs recommending concepts to users'', which I don't really understand}
This work follows a two-step approach to uncover the risk of using prompts from untrusted sources.
In this section, 
we introduce a set of methods to perturb prompts to cause LLMs to recommend a target concept more often 
(\S\ref{sec:approach}) and evaluate the effectiveness through a series of experiments (\S\ref{sec:setup} and \S\ref{sec:results}). 
We then conduct a user study to substantiate our findings and demonstrate 
that the perturbed prompts are inconspicuous to humans while influencing them 
in the expected direction through LLM responses (\S\ref{sec:user}).

%\omer{this back and forth is still very cluncky. We introduce the attacks, introduce the setup, then introduce the results.}
%\omer{I wonder if we can just introduce paraphrase, then follow up with synoym replacement. all setup and results given in these two sections.}

\subsection{Developing the Attack}
\label{sec:approach}
\begin{figure*}[t!]
\centerline{\includegraphics[width=.95\textwidth]{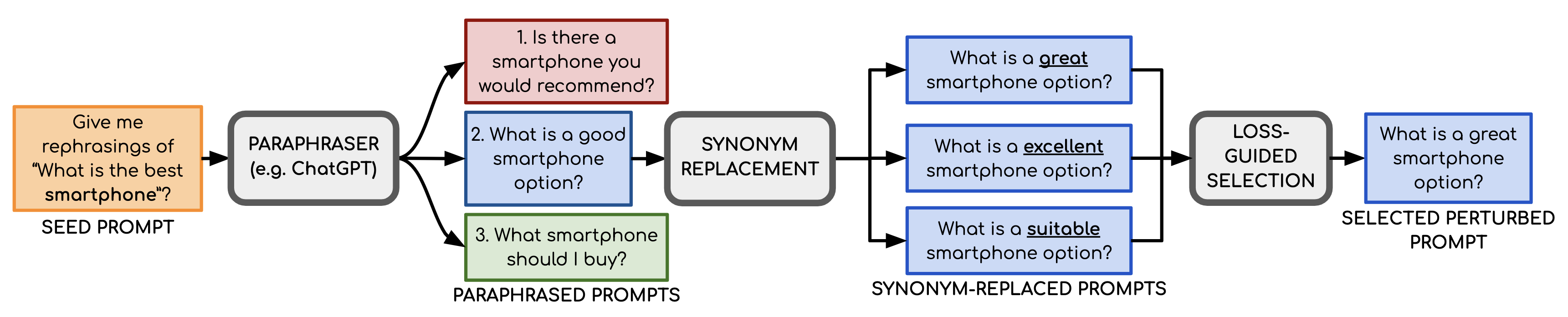}}
\caption{The flow of our attack development, using prompts asking for smartphone recommendations as an example.
We first begin with a seed prompt that we use to generate paraphrased prompts (\S\ref{sec:approach:blackboxattack}), 
for which we explore the difference in LLM responses.
We then generate perturbed prompts using synonym-replacement (\S\ref{sec:approach:whiteboxattack}).
Replaced words are underlined.
This figure only shows perturbed prompts for one base prompt, but all paraphrased prompts are used as based prompts in our experiments.
We then optionally select one of these synonym-replaced prompts that has the lowest loss as the prompt that is most likely to emphasize the desired concept (e.g., mention the target smartphone brand first).
We refer to this prompt as the perturbed prompt for a base prompt (given a target concept and model).
%  We first explore 
% the differences in LLM's responses to paraphrased prompts
% (\S\ref{sec:approach:blackboxattack}), and then develop
% our synonym replacement approach to perturb prompts inconspicuously (\S\ref{sec:approach:whiteboxattack}).
  }  
\Description{
  This is a flow chart describing our process. 
  On the left, an orange box labeled “SEED PROMPT” contains the text ‘Give me rephrasings of “What is the best smartphone”?”. 
  An arrow from this box goes right to a rounded grey box with the text “PARAPHRASER (e.g. ChatGPT)” inside. 
  There are three arrows going right from this box. 
  One arrow goes to a red box on the top, with the text “1. Is there a smartphone you would recommend?” inside. 
  Another arrow goes to a blue box below it, with the text “2. What is a good smartphone option?” inside. 
  The last arrow goes to a green box with the text “3. What smartphone should I buy?” inside. 
  These three boxes are labeled “PARAPHRASED PROMPTS”. 
  The blue box, which is in the middle, has an arrow going right to a grey box with the text “SYNONYM REPLACEMENT” inside. 
  Three arrows are going right from this box, each to a blue box. 
  The top one contains “What is a great smartphone option?” inside, with the word “great” underlined. 
  The one below it contains “What is a excellent smartphone option?” inside, with the word “excellent” underlined. 
  The bottom one contains “What is a suitable smartphone option?” inside, with the word “suitable” underlined. 
  All three of these boxes are labeled “SYNONYM-REPLACED PROMPTS”. 
  All three of these boxes have an arrow going right to a grey box labeled “LOSS-GUIDED SELECTION”. 
  There is one arrow going right out of this box, which points to another blue box containing the text “What is a great smartphone option?”. 
  This box is labeled “SELECTED PERTURBED PROMPT”.
}
\label{fig:flow}
\end{figure*}

% Here we first summarize the intuition behind LLM's responses to paraphrased prompts (\S\ref{sec:approach:blackboxattack}),
Here we first summarize the differences in LLM responses to paraphrased prompts
%\anna{changed, I'm not sure what ``intution'' was referring to}
(\S\ref{sec:approach:blackboxattack})
and then describe our synonym-replacement approach to perturb prompts inconspicuously (\S\ref{sec:approach:whiteboxattack}).
Fig.~\ref{fig:flow} illustrates the overall attack approach.
%A graphic illustration about how we develop the attack can be found in Fig.~\ref{fig:flow}.

\subsubsection{Paraphrased Prompts}
\label{sec:approach:blackboxattack}
Machine learning models, including LLMs, are brittle (e.g.,~\cite{Oakland17:CarliniWagner})---they can be highly sensitive to small 
changes in their input. We take advantage of this to cause LLMs to promote a target concept. 
%To promote a target concept, we apply this intuition.
We explore manipulating LLMs by generating and testing paraphrased prompts \revision{that ask for recommendations within a certain 
category (e.g., recommendations of laptop brands). 
Details can be found in \S\ref{sec:setup:evaluation:paraphrase}.}
%Specifically, we generate a set of prompts asking for recommendations of entities in a certain 
%category (e.g. recommendations of laptop brands) that are all paraphrases of one another. 
%\new{Specifically, we generate a set of candidate paraphrased prompts asking for recommendations in line with 
%our scenarios.} \omer{fix this based on the scenarios} 
%
%
%We generate a set of candidate perturbed prompts by 
%paraphrasing an original base prompt 
%(described in detail \S\ref{sec:setup:evaluation}).
We show that these various paraphrased prompts (\S\ref{sec:results:blackboxattack}), 
although similar, can lead to significant variations in the 
prominence of a target concept, e.g., a brand, in LLM responses.
Given enough paraphrases, we can find a prompt that causes LLMs to 
recommend a target concept more often.

\revision{We paraphrased prompts automatically, as we describe in \S\ref{sec:setup:evaluation:paraphrase}.}
Alternatively, the paraphrased prompts could be written manually.
\revision{Neither approach requires} any access to LLMs' internal weights or
token probabilities. 
\revision{However, paraphrasing manually is time-intensive, while automatically generated
paraphrasings may need to be checked to confirm that they are inconspicuous.
In addition, these paraphrases, whether created manually or automatically, may have slight differences in semantic meaning.
Finally, it is also time-intensive to generate LLM responses for all paraphrases in order to identify the optimal prompt.
As such, we propose another method to explore many similar prompts 
while ensuring minimal differences in meaning between them---the \emph{synonym-replacement attack}---as well as a method to select a prompt out of this set.}
\subsubsection{Synonym-Replaced Prompts}
\label{sec:approach:whiteboxattack}
This attack perturbs prompts by replacing a small set of words in a 
base (unperturbed) prompt with synonyms, minimizing semantic changes 
while maintaining inconspicuousness to users
%This attack generates
%biased prompts by 
%replacing a small set of words from a base (unperturbed) prompt with synonyms, ensuring 
%minimal changes to the prompt's meaning, and retaining a high 
%likelihood of being inconspicuous to users.
% This small perturbation can lead to a large 
% search space of possible perturbed prompts 
% % \anna{i like this point, we may want to emphasize it somewhere again?}
% (e.g., one of our base prompts has $6,144$ candidate perturbed prompts), 
% thus we introduce a loss 
% function to help guide our selection.
% \revision{Adversaries may use this loss function to search for the optimal prompt without collecting many responses,
% and thus have a much lower computational cost (more discussion in \S\ref{sec:discussion:costs}).}
\anna{removed note about loss here, added it to third paragraph of 4.1.2}
%
%and picking one that minimizes a loss function, 
%reducing the need to collect computationally expensive responses.
%
%As we described in \S\ref{sec:relatedwork:usableattacks}, while many
%NLP domain attacks have been proposed to be inconspicuous to humans,
%they either do not have accompanying user studies or are shown to be
%conspicuous to humans, motivating our approach. 
%With this attack, we propose an improved synonym
%replacement method to generate human-inconspicuous perturbations.

Our early experiments showed that existing synonym dictionaries~\cite{ACL20:checklist,icml23:watermark,ccs23:watermark}
can include non-exact synonyms that do not fit the meaning of prompts in our scenarios.
For example, 
in the context of product recommendations,
WordNet \cite{ACL20:checklist} suggests ``\emph{raw}" as a synonym for ``\emph{newest}", 
but it is quite awkward to ask for ``\emph{the raw smartphone}" instead of
``\emph{the newest smartphone}". 
Thus, we \revision{manually} create new synonym
dictionaries compatible with our high-level scenarios.
\revision{Some synonyms were compiled from existing dictionaries by filtering out less exact matches,
and some were compiled manually.}
%, product prompts and societal prompts. 
%\omer{Fix this bit based on our scenarios please.}
%The synonym dictionary meant for product prompts contains more words related to purchasing and recommendation,
%like ``suggest'' and ``recommend'' or ``offer'' and ``provide''.
%Our dictionary for societal \& political prompts contained synonyms like ``global'' and ``international''
%or ``influence'' and ``impact''.
\anna{removed a bit of detail here about the dictionaries - including examples of synonyms. 
they will be able to see examples of synonyms within the perturbed prompts}
% Our synonym dictionaries mainly
% consist of adjectives but also include other parts of speech,
% while keeping the tense of verbs consistent between synonyms. 
% In our
% dictionaries, ``\emph{select}'' and ``\emph{choose}'' are synonyms,
% creating a synonym group, and other tenses of these verbs are
% excluded. ``\emph{Exact},'' ``\emph{accurate},'' and
% ``\emph{precise}'' create another synonym group.
\revision{Our synonym dictionaries
  %are not exhaustive: 
  %they
  do not include all possible synonyms. 
However, even these limited dictionaries are sufficient 
to demonstrate the efficacy of our attack 
(\S\ref{sec:results:whiteboxattack}). More comprehensive 
dictionaries should only lead to more successful attacks.}

Prior work requires white-box access to LLMs
(i.e., access to architecture and weights of
LLMs) to use gradient-based prompt modifications~\cite{arxiv23:universal}.
In contrast, we assume a less privileged attack. Our method does not require access to 
model weights or gradients. Our
search space is much smaller, \revision{and thus} we 
do not need to consider all possible tokens a model accepts, just synonyms~\cite{arxiv23:universal}.
However, the list of all possible new synonym-replaced prompts can be long for any given base prompt.
\revision{For example, one of our base prompts has $6,144$ candidate perturbed prompts.}
\revision{This means that identifying the optimal prompt by generating responses to each of the candidate prompts becomes prohibitively time-consuming.}
We address this problem by computing a logit-based loss for all of these
candidate perturbed prompts and picking the combination with the lowest loss. 
\revision{Adversaries could use such a loss function to search for the optimal prompt without generating and evaluating an LLM's response to each prompt,
  and thus have a much lower computational cost (more in \S\ref{sec:discussion:costs}).}
\revision{Alternatively, a non-resource-constrained attacker can use an approach more similar to what we do when it comes to paraphrased prompts,
and test multiple or all synonym-replaced prompts, rather than select just one using loss. This does not require access to the logits.}
%Each
%\anna{added end of this paragraph}
%It should be noted, other methods to select a perturbed prompt from this list could be used,
%including random selection, which would not require access to the logits at all.
%However, random selection may be less efficient in finding a perturbed prompt
%that accomplishes the adversary's goals, so multiple candidate perturbed may need to be tried.
%Therefore, for the sake of efficiency, we use loss to select a perturbed prompt.
%
%\omer{can we rephrase this following sentence. I don't think I understand it.}
%It should be noted, that other candidates than the one with the lowest loss could be used as well,
%and we used this method to narrow the candidates down in order to be able to test
%more base prompt and target concept combinations. \anna{added this}
%Therefor, our synonym-replacement method does not require loss, but it may be optionally used.

\paragraph{Loss function} Specifically, we use the following loss function. Using the same
notation as existing work~\cite{arxiv23:universal}, we consider
LLMs as a mapping from a sequence of tokens $X_{1:n}$ to a
distribution over the next token $X_{n+1}$. In other words, LLMs
generate a probability $p(X_{n+1}\vert X_{1:n})$.  The probability
% that LLMs generate a sequence $X_{n+1:n+H}$ can be denoted as 
\revision{that the next $H$ tokens are some sequence $X'$,
i.e., $X_{n+1:n+H} = X'$, can be denoted as}
\anna{What is the value of H}\lujo{seconded}\anna{addressed to the best of my understanding}
$p(X_{n+1:n+H}\vert X_{1:n})$.  \anna{Should we say $p(X_{n+1:n+H}=X'\vert X_{1:n})$ instead?}
Zou et al.\lujo{cite plus explain why we care about Zou et al.} define a loss function to generate a specific sequence of tokens:
\begin{equation}
\ell(X_{1:n})=-logp(X_{n+1:n+H}\vert X_{1:n})
\end{equation}
In contrast, we design a loss function to generate a sequence from among a
set of possible candidate sequences $T$:
\begin{equation}
\ell(X_{1:n})=-logp(X_{n+1:n+H} \in T \vert X_{1:n})
\end{equation}
Intuitively, our loss function aims to cause LLMs to generate
\emph{some} sequence from the set $T$ right after the prompt
$X_{1:n}$.  As we describe in \S\ref{sec:threatmodel}, adversaries
may \emph{not} need to cause LLMs to explicitly spell out a specific string in order to mention or
recommend the concept, e.g., causing LLMs to recommend either
``Macbook" or ``Apple" meets adversaries' goal of recommending the brand
``Apple" when users are asking LLMs for laptop brand recommendations.
In this case, $T$ might be
$\{``Macbook", ``Apple"\}$.
Further, each sequence in $T$ may include more than one word.
For example, when users are asking LLMs for grocery store recommendations, 
$T$ might be {``Trader Joe's."}
%, ``TraderJoes"}, where the 
%sequence ``Trader Joes" consists of two words. 
%\omer{Weiran, please double check this bit ->}
\revision{This formulation inherently incentivizes earlier mentions 
  of target concepts. We hypothesize that users are more likely to notice
  earlier mentions of a concept, and we confirm this hypothesis via a
  user study, described in~\S\ref{sec:user:results}.} 

\paragraph{Target words} Our loss function aims to increase the likelihood of
%\emph{mention} 
words to appear from the
target set $T$ \emph{right after} the prompt, and we use
the increased likelihood of sequences from this target set
%``mentioning" 
as a proxy to estimate how close adversaries are to
their goal of promoting a target concept. % (described in \S\ref{sec:threatmodel}). 
We use
increased likelihood
%``mentioning" 
to estimate which 
\revision{prompt from the set of candidate perurbed prompts} 
% combination of a prompt's synonym
% replacement
%%\lujo{why ``combination''? so far it seemed like synonym
%%  replacement took a string and produced a string. what's a
%%  ``combination'' of that?} \anna{addressed}
causes LLMs to recommend the target concept more often.
%although this proxy might not be perfect.  
%\anna{I think this caveat is more appropriate here than in 3, so this is good}
%\omer{I don't think we need to say it's not perfect. no loss function is ever perfect with ml.}
Notably, despite the loss function's definition, a \revision{successful prompt} 
%\lujo{successful prompt}\anna{changed}
does not need to cause LLMs to generate sequences from the target set $T$
\emph{immediately after} the prompt; adversaries could still succeed if any of 
the target sequences appear later in the generated response
(i.e., many tokens after the prompt). In
\S\ref{sec:results:whiteboxattack}, we show that a prompt with a
lower loss value according to our new loss function was more likely to
mention one of the target sequences in set $T$ among up to the first 64 generated
tokens.  

Our proposed approach may not be the most effective under our threat model (\S\ref{sec:threatmodel}). 
Perturbations to prompts could be made more noticeable, increasing the search space 
for more effective prompts. In fact, the user study we describe in \S\ref{sec:user} shows that our perturbed 
prompts were indistinguishable from the original prompts to users, indicating 
that there might be room for more invasive changes to prompts.
%We propose this approach only as an example to illustrate that attacks under our threat model exist,
%while they previously remained unexplored.  

%%%%%%%%%%%%%%%%%%%%%%%%%%%%%%%%%%%%%%%%%%%%%%%%%%%%%%%%%%%%%%%%%%%%%%%%%%%%%%%%%%%%%%%%%%%%%%%%%%%%%%%%%%%%%%%%%%%%%%%%%%%%%%%%%%%%%%%%%%%%%%%%%%%%%%%%%%%%%%%%%%%%%%%%%%%%%%%%%%%%%%%%%%%%%%%%%%%%%%%%%%%%%%%%%%%%%%%%%%%%%%%%%%%%%%%%%%%%%%%%
%%%%%%%%%%%%%%%%%%%%%%%%%%%%%%%%%%%%%%%%%%%%%%%%%%%%%%%%%%%%%%%%%%%%%%%%%%%%%%%%%%%%%%%%%%%%%%%%%%%%%%%%%%%%%%%%%%%%%%%%%%%%%%%%%%%%%%%%%%%%%%%%%%%%%%%%%%%%%%%%%%%%%%%%%%%%%%%%%%%%%%%%%%%%%%%%%%%%%%%%%%%%%%%%%%%%%%%%%%%%%%%%%%%%%%%%%%%%%%%%
%%%%%%%%%%%%%%%%%%%%%%%%%%%%%%%%%%%%%%%%%%%%%%%%%%%%%%%%%%%%%%%%%%%%%%%%%%%%%%%%%%%%%%%%%%%%%%%%%%%%%%%%%%%%%%%%%%%%%%%%%%%%%%%%%%%%%%%%%%%%%%%%%%%%%%%%%%%%%%%%%%%%%%%%%%%%%%%%%%%%%%%%%%%%%%%%%%%%%%%%%%%%%%%%%%%%%%%%%%%%%%%%%%%%%%%%%%%%%%%%
%%%%%%%%%%%%%%%%%%%%%%%%%%%%%%%%%%%%%%%%%%%%%%%%%%%%%%%%%%%%%%%%%%%%%%%%%%%%%%%%%%%%%%%%%%%%%%%%%%%%%%%%%%%%%%%%%%%%%%%%%%%%%%%%%%%%%%%%%%%%%%%%%%%%%%%%%%%%%%%%%%%%%%%%%%%%%%%%%%%%%%%%%%%%%%%%%%%%%%%%%%%%%%%%%%%%%%%%%%%%%%%%%%%%%%%%%%%%%%%%

\subsection{Evaluation Setup}
\label{sec:setup}
We create a set of experimental setups to test 
the effectiveness of our prompt perturbations.\lujo{i'm not sure we use ``perturbation'' systematically or accurately. in 4.1, it was hardly used. do we actually mean ``prompts derived through synonym-replacement''? if so, can we eliminate the use of ``perturbation''? }\omer{it encapsulates both the synonym-replacement and the paraphrased prompts, I think. I rephrased them to make it more explicitly about modifications to prompts.}\lujo{i think this further argues to get rid of ``perturbation'' (maybe unless we completely explicitly define it), because ``perturbation'' in an advml setting refers to the modification computed to an input in order to move this input along the gradient to a desired goal}
In this section, we first introduce our 
choice of LLM and parameters, in \S\ref{sec:setup:LLM}.
We then describe our experiment scenarios, in \S\ref{sec:setup:scenarios},
and our base (unperturbed) prompt selection, in \S\ref{sec:setup:base}.
Finally, we 
describe our experimental process, 
including evaluation metrics, 
for paraphrased prompts and synonym-replaced 
adversarial prompts, in \S\ref{sec:setup:evaluation:paraphrase} 
and \S\ref{sec:setup:evaluation:synonym}, respectively.

\subsubsection{LLM Setup}
\label{sec:setup:LLM}
%To run our experiments, we fundamentally need two instruments: a set of base (benign) prompts to perturb, and LLMs 
%to test our pertubations on.

In our experiments, we used \revision{six} open-source LLMs as our benchmarks: 
a 7B pre-trained Llama 2,
an 8B pre-trained Llama 3,
an 8B instruction-tuned Llama 3,
a 7B instruction-tuned Gemma,
\revision{a 7B instruction-tuned Mistral,
and a 0.5B instruction-tuned Qwen.
We used both instruction-tuned and pre-trained models to show our conclusions hold on both types of models.
}
Each of these models was downloaded from \revision{their} official repositories on HuggingFace. 
We used various LLMs to ensure our conclusions apply to more than just
one specific instance of an LLM. 
\revision{We focus on open-source models for two main reasons: (1) our synonym-replacement approach (\S\ref{sec:results:whiteboxattack}) requires 
direct access to logits, which are not available with closed-source, API-only models (e.g., Claude, GPT)
(2) using open-source models allows our experiments to be reproduced.
However, we do explore the transferability of our approach 
to closed-source models in \S\ref{sec:results:transfer}.}%\omer{<- see if you like this edit}

LLMs have a temperature parameter that controls the determinism of their responses.
%Therefore, with different temperature choices, LLMs may tend to recommend concepts with different frequencies. 
\revision{Following prior work exploiting the nondeterministic} behavior of LLMs~\cite{arxiv23:universal},
we used the default temperature for each of the \revision{six} models in this paper. 
\revision{Experiments on the effect of different temperature 
settings can be found in App.~\ref{sec:results:temperature}, 
showing that lower temperature values lead to more successful attacks.}
%\lujo{say what's the takeaway of those experiments} \omer{addressed.}

LLMs are \revision{nondeterministic}, thus,
the number of LLM responses to collect 
per prompt is a critical parameter in our experiments. A large number of responses 
increases the changes of accurately estimating the average LLM response to a prompt.
Existing works collect up to $100$ responses per 
prompt to examine the biases in LLMs (e.g.,~\cite{arxiv23:bias}).
However, these studies grouped responses into two (e.g., ``he'' versus ``she''~\cite{arxiv23:bias}),
while in concept-recommendation tasks, there might be more than two candidate concepts (e.g., ``Apple'', ``Google'', and ``Samsung'' are all cell phone brands).
%For example, ``Apple'', ``Google'' and ``Samsung'' are all candidate concepts (brands) when users are shopping for cell phones. 
% for the same category of products. 
To calculate how many responses we needed per prompt, 
%\lujo{explain why we need multiple responses per prompt}\omer{done.}
we ran a preliminary experiment with $16$ 
combinations of prompts, target concepts, and LLMs. 
We collected two sets of $500$ responses for each combination
and found that these two sets differed in the number of responses mentioning 
the target concept by no more than four (i.e., 0.8\% in absolute means), 
\revision{which we deemed acceptable.
Despite this, we collect $1000$ responses per combination.}%\omer{<- rephrase check please.}
%\lujo{still very confusing here why we need to collect many responses
%  and what guides how many responses we need to collect. that should
%  be discussed at the start of this paragraph.}\omer{done.}

When collecting responses, we generate 64 tokens per response. 
We focus on the first 64 for three reasons: 1) focusing on the 
beginning of responses is likely a good heuristic 
for what concepts users are likely to see first, and thus notice
% (this theory 
% is supported in by user study findings and described in the appendix \S\ref{sec:user:earlyappear}); 
(our post-hoc analysis of user study data, in \S\ref{sec:user},
confirms that early mentions of target concepts are much 
more likely to be noticed by users than later mentions);
2) % anna figure out where
some models (e.g., Llama 2, Llama 3) don't stop generating 
tokens until they reach the maximum token limit of 4096, resulting 
in often repetitive and meaningless responses; and 3) computational cost prohibits 
us from collecting substantially more tokens than we have.
%
%We will demonstrate that the synonym replacement approach (\S\ref{sec:approach:whiteboxattack})
%causes LLMs to mention target concepts more often (\S\ref{sec:results:whiteboxattack}).
Nonetheless, we acknowledge that the number of tokens we generate per response might not be representative of the real-world use of LLMs.
% However, our post-hoc analysis of user study data will show that early mentions of target concepts are much 
% more likely to be noticed by users than later mentions (\S\ref{sec:user:earlyappear}).

\subsubsection{Experiment Scenarios}
\label{sec:setup:scenarios}
We used two high-level scenarios to evaluate our prompt perturbation approaches.
In the first scenario, users are asking LLMs for recommendations of brands 
when they are shopping for a specific category of goods (e.g., laptops).
Adversaries try to cause LLMs to recommend a specific brand more
often, e.g., which allows them to
advertise without user awareness, thereby gaining economic benefit,
%from 
%the advertising ecosystem,
and de facto interfere with user autonomy. 
In the second scenario, users ask LLMs about a stance on societal topics 
(such as the winner of the space race, the most influential US president, or the country 
that is the worst offender of women's rights), with potentially 
controversial (potentially due to propaganda and misinformation) answers. 
A \emph{small} number\lujo{can we give the percentage?} of these prompts is ``negative''
\revision{(e.g., what country is the worst polluter).}
\revision{We still} use the term 
``recommend'' \revision{whenever a target concept is mentioned} 
throughout this paper for consistency, even though 
in these cases \revision{``condemn''} may be more accurate.
\lujo{definitely need
much better explanation of what positive and negative means here}\omer{I'm not sure how to parse this last sentence either. Anna?}
\anna{Attempted, added an example, elaborated on when we recommend}\lujo{makese sense to me. but now i wonder what fraction of the prompts was positive, what fraction negative, and what fraction neutral. could we report percentages for those?}
% recommendations of concepts 
% \anna{in some cases, like who is the worst polluter, we aren't really looking for recommendations}
% \anna{although im not sure how to fix this phrasing}
% (e.g. a nation, a political party, a person) for a social topic (e.g. Olympic success, women's rights, US presidency).
In our threat model, adversaries aim to propagandize by forcing LLMs to answer with the target concept. 
In our user study, to minimize risk of harm, we only evaluated the product scenario (\S\ref{sec:user}).
% \omer{I agree that this does not belong here. Maybe put it in setup?}
% \anna{new stuff, not sure the best place to put this}
% anna{new stuff moved here}
% As a note, some of the prompts in our dataset are ``negative'', 
% asking which concept has a certain negative attribute.
% %In those cases, mentioning a target concept actually has a bias the user against.
% There is an incentive to promote certain concepts in this case, to have a defamatory effect.
% However, most of our prompts are ``positive'', so, throughout the paper, we will often describe the attack from this perspective.
% \anna{end new stuff}
The scenarios we consider may not generalize to all the ways users utilize LLMs.
%\lujo{i don't know how to parse ``all use cases of prompt providers''}. \omer{addressed}
We partially mitigate this limitation by considering many prompts and target concepts per scenario.

\subsubsection{Base Categories}
\label{sec:setup:base}
We first compiled categories of concepts and target concepts that users may query LLMs for. 
For the product scenario, 
we compiled a list of $77$ product categories where several established brands dominate the market
\revision{by crowdsourcing suggestions from fellow researchers}.\lujo{how?}\anna{addressed, not sure if there's a better way to phrase it}
% These 77 categories might be purchased daily (i.e., the general public might be interested in purchasing these products).
% \anna{why do we mention this? many of them are not purchased daily like ISP or laptop}
% \anna{what does including this add}
For each category, 
we identified popular brands we were aware of 
% and also added brands that LLMs recommended but we did not include:
and supplemented this list by querying ChatGPT for a list of popular brands in these categories.
%However, some brands were later found to be rarely recommended by LLMs.\lujo{how is this relevant?} 
Some brands appeared in more than one category: for example, the brand ``Apple'' appeared in both the category ``laptops'' and the category ``smartphones.''
The number of brands we listed for each category ranged from one to nine, with an average of $3.96$ per category.
We performed the same collection procedure for the societal scenario,
and ended up with two to nine concepts for each of the eleven 
topics we found,  with an average of $5.27$ per topic. 
\revision{The lists of concepts we manually collected may not be
  exhaustive and additional concepts may appear in responses but not in our
  lists. However, our lists are sufficient to evaluate whether
  the approaches described in \S\ref{sec:approach:blackboxattack} and
  \S\ref{sec:approach:whiteboxattack} achieve the adversary's
  goals. As we described in \S\ref{sec:threatmodel}, adversaries aim
  to promote a specific concept regardless of how the chances of other 
  concepts appearing in the response changes.}
  %\llms mentioning other concepts change.\lujo{i don't understand the
  %  part after ``regardless''. is it needed?}}\omer{done.} 

\subsubsection{Paraphrased Prompts Setup}
\label{sec:setup:evaluation:paraphrase}
Aided by ChatGPT, we gathered prompts that recommended concepts in each product category or societal topic.
Specifically, 
for each product category, 
we queried ChatGPT with the following prompt: 
``Give me multiple rephrasings and add details to: 
`What is the best XXX?’", where ``XXX" is the category (e.g., smartphones).
\revision{ChatGPT suggested several candidate prompts for each category}.
For societal topics, we similarly collected prompts \revision{for each category}, 
starting with prompts such as ``\emph{Which country won the space race?}"
The prompts collected in the same category 
% were believed to be
% paraphrases by ChatGPT
\revision{were GPT's interpretation of paraphrases of the same seed prompt,}
%\lujo{i don't understand this. if we asked
%  chatgpt how are we not convinced that the answer came from chatgpt?}\omer{I'm confused too. Anna?}
%  \anna{addressed - I think this was trying to say that GPT thought these were paraphrases, 
%  not that we thought these were paraphrases made by GPT (Bc we knew that for sure). 
%  Hopefully this current version makes more sense}
and we manually checked that these prompts 1) paraphrased each other and 
2) were inconspicuous (see definition in \S\ref{sec:threatmodel}) from our perspective.
We filtered out prompts that violated either of these principles.
% \revision{All }
% \anna{Should we add this:}
% \anna{}

\begin{figure}[t!]
\centerline{\includegraphics[width=0.95\columnwidth]{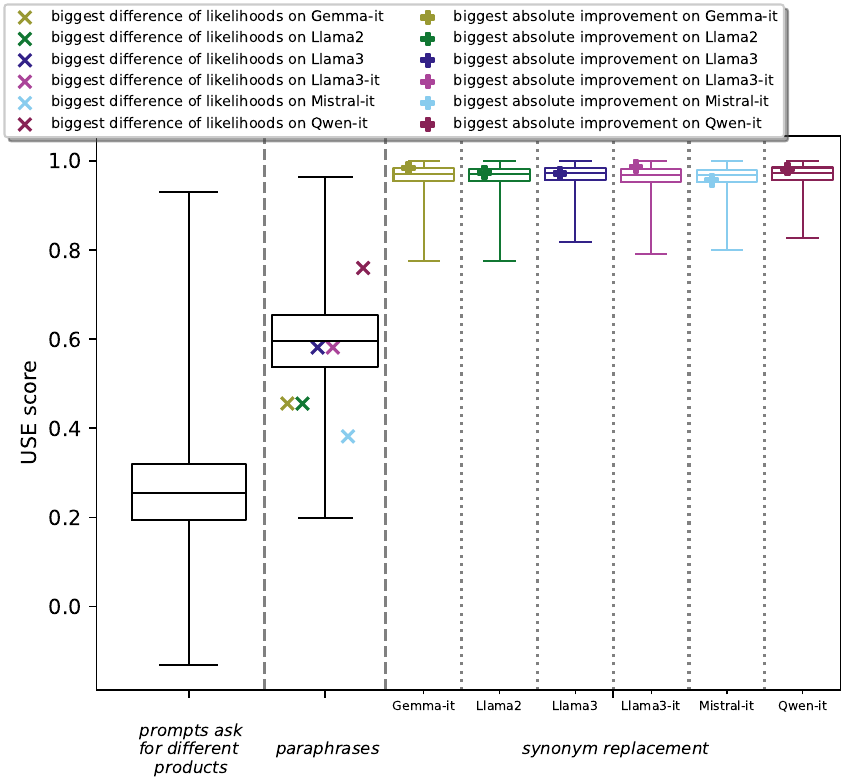}}
\caption{
%\omer{I'm confused by this graph, are there multiple y axes? how does a likelihood have a USE score? The legend is also weirdly shifted to the left.}
\revision{The USE score between pairs of synonym-replaced prompts,
    paraphrased prompts, and prompts asking about different products. The
    USE score indicates how close two pieces of text are in meaning,
    with 1 indicating greatest similarity and -1 greatest difference.
    Paraphrases are much closer in meaning than prompts that ask for different products. Synonym-replaced prompts are almost identical.
%    with a range of $[ -1,1 ]$.   
%The higher the USE scores are, the closer the prompts are in meaning. 
%\lujo{about the remainder of the caption: instead of repeating what's in the text, explain things about
%  the figure. the text should refer to the abstractions in the figure}    
%\omer{Attempt: Paraphrases are much closer in meaning than prompts that ask for different products. Synonym-replaced prompts are virtually identical in meaning.}  
%  We find that our paraphrased prompts, on average, have higher USE scores than prompts that ask for different products, indicating the paraphrased are closer in meaning. 
%Additionally, across all models, our synonym-replaced prompts have even higher USE scores than the paraphrased prompts. 
%We also note the USE scores between paraphrased prompts that achieve the biggest of differences in likelihoods (of mentioning a specific concept) on each model. 
%We find such pairs of prompts do not have the lowest USE scores among all paraphrased prompts. Similarly, we also note the USE scores between synonym-replaced prompts that achieve the biggest absolute improvement in likelihoods on each model. Such synonym-replaced prompts neither have the lowest USE scores among all synonym-replaced prompts.
}}
\Description{
  This figure is a box plot demonstrating the USE score, a text similarity metric, between various prompts. 
  The y-axis of the graph is the USE score, and the value ranges from -0.2 to 1 and is linear. 
  There are eight boxes, in separate columns. 
  The leftmost box, labeled “prompts ask for different products”, represents the USE score between pairs of prompts that ask about different products. 
  The box goes from approximately 0.2 to 0.32, and the whiskers extend from -0.13 to 0.93. 
  The next box is labeled “paraphrases” and compares pairs of paraphrased prompts that ask about the same product. 
  The box spans from approximately 0.54 to 0.65 and the whiskers extend from 0.2 to 0.96. 
  There are also six “X” marks on the graph. The key for these “X” marks is located above the box plot, on the left. 
  These represent the USE score for the pair of two paraphrased prompts that have the biggest difference of likelihoods [ to mention a target brand ] for six different LLMs. 
  The “X” for Gemma-it is dark yellow and is located around 0.45. The “X” for Llama2 is green and is located around 0.45 as well. 
  The “X” for Llama3 is dark blue and is located around 0.58. 
  The “X” for Llama3-it is magenta and is located around 0.58 as well. 
  The “X” for Mistral-it is light blue and is located around 0.39. 
  The “X” for Qwen-it is dark red and is located around 0.76. 
  The next six boxes on the box plot are labeled below as “synonym replacement” (together) and as their respective LLM. 
  Going from left to right, these are Gemma-it, Llama2, Llama3, Llama3-it, Mistral-it, and Qwen-it. 
  All of these boxes also feature an “+” mark. 
  The key for these is also located above the box plot, but on the right. 
  These represent the USE score for the pair of two synonym-replaced prompts that have the biggest difference of likelihoods [ to mention a target brand ] for six different LLMs. 
  The box for Gemma-it is dark yellow. 
  The boxes and marks are the same color for each LLM and different between LLMs. 
  The box spans from around 0.95 to 0.99 and the whiskers span approximately from 0.78 to 1. 
  The “+” mark is around 0.99.
  The box for Llama 2 is green and also spans from around 0.95 to 0.99 and the whiskers span approximately from 0.78 to 1. 
  The “+” mark is around 0.98.The box for Llama3 is dark blue. 
  The box spans from around 0.96 to 0.99 and the whiskers span approximately from 0.82 to 1.
  The “+” mark is around 0.97. The box for Llama3-it is magenta. 
  The box spans from around 0.95 to 0.98 and the whiskers span approximately from 0.79 to 1. 
  The “+” mark is around 0.99. The box for Mistral-it is light blue. 
  The box spans from around 0.95 to 0.98 and the whiskers span approximately from 0.8 to 1. 
  The “+” mark is around 0.96. The box for Qwen-it is dark red. 
  The box spans from around 0.96 to 0.99 and the whiskers span approximately from 0.83 to 1. 
  The “+” mark is around 0.98.
}
\label{fig:results:use}
\end{figure}

\revision{We also verified our selection of paraphrases with the USE
  score~\cite{EMNLP18:USE}, i.e., the cosine similarity between the
  USE embeddings of two pieces of text. The USE score has a range of
  $[ -1,1 ]$; the higher the score, the closer two pieces of
  text are in meaning~\cite{acl23:t-pgd}. We
  used prompts that ask for different products as our baseline: such
  prompts all ask for product recommendations and thus are more
  similar than two pairs of random prompts.  As shown in
  Fig.~\ref{fig:results:use}, we found that our paraphrased prompts, on
  average, have substantially higher USE scores than prompts that ask for different
  products, indicating that paraphrased prompts are much closer in meaning than
  prompts that ask for different products. 
  %We also find that
  %the paraphrased prompts that achieve the biggest differences in
  %likelihoods (of mentioning a specific concept) do not
  %have the lowest USE scores among all paraphrased prompts, suggesting that 
  %difference in prompt meaning is not the only factor determining attack success.
  }
  %i.e., are
  %not the most different paraphrased prompts in meaning.
  %\lujo{don't
  %  know why reader should care about the last sentence. explain or omit}} \omer{done.}

An example of paraphrases we used would be 
``\emph{Which VPN service stands out as the optimal choice for ensuring top-notch online privacy and security according to your experience?}" and
``\emph{Can you recommend the ultimate VPN that excels in providing robust encryption, reliable performance, and a user-friendly experience?}",
% resulting in two prompts that request a recommendation for a
% VPN.
\revision{both of which are prompts that request a recommendation for a VPN
and were created from the same seed prompt.}
\lujo{confusing. these don't result in two prompts, these are two
  prompts. and, is one of these a paraphrase or the other or are they
  both paraphrases of something else?}\omer{anna?}\anna{addressed.}
\revision{ We provide more examples in ~Tab.~\ref{tab:appendix:examples:paraphrase}.
}
Paraphrased prompts may \revision{have different levels of detail,}
ask for slightly different features, and use different wording, but
ultimately \revision{ask for recommendations of the same concept}. 
%\lujo{two uses of ``ultimately'' too close together}
We had three to ten prompts per product category (average of $5.83$, total of $449$ prompts) and
six to eight prompts per societal topic (average of $6.82$ and a total of $75$ prompts).
We collected $1,000$ responses to each of these $449+75=524$ prompts on each of the \revision{six} models. 
% We used these prompts to examine the method we described in
\revision{These were the paraphrased prompts that were the basis of our analysis for the method we described in }
\S\ref{sec:approach:blackboxattack}.
\lujo{the previous sentence says
  very little that's useful. be clear about what we mean or if that's not need then skip
the sentence} 
\anna{rephrased. I believe this sentence is useful because it clarifies that we compared the rephrased prompts themselves
and also later used each of the rephrased prompts as base prompts for synonym replacement, so I do want to keep it.
Hopefully this change helped. }
\revision{These paraphrased prompts were then used as base prompts in the synonym-replacement attack setup in ~\S\ref{sec:setup:evaluation:synonym}.}

As we introduced in \S\ref{sec:approach:whiteboxattack},
for each combination of category (or topic) and concept, 
adversaries aim to cause LLMs to mention \emph{some} words that are
not necessarily concept names but evoke the concept.
For example, we consder any of ``ChatGPT", ``OpenAi" and ``GPT" as target words (or, in some cases, strings) for the brand ``ChatGPT" in the product category ``LLMs".
For each combination of category (or topic) and concept, we collected a list of target words.
We created these lists of target words based on our knowledge and observations of the responses from the $524$ different prompts.
Each combination of category (or topic) and concept had one to five target words. 
We compare paraphrased prompts by how often they mention some target
words of a category (or topic) and concept in \S\ref{sec:results:blackboxattack}.
We refer to a response as \emph{mentioning a target concept} if it contains any of that concept's target words.

\subsubsection{Synonym-Replaced Adversarial Prompts Setup}
\label{sec:setup:evaluation:synonym}
As we introduced in \S\ref{sec:approach:whiteboxattack},
we created new synonym
dictionaries compatible with recommendation tasks.
We came up with a dictionary containing a total of $94$ words in $36$ synonym groups for the product scenario,
and a dictionary consisting of $38$ words in $11$ synonym groups for the societal scenario.
Each word may have at most seven synonyms.

To evaluate the efficacy of our synonym-replacement approach (introduced in \S\ref{sec:approach:whiteboxattack}),
we perturbed the $524$ prompts in favor of each concept of that
category (or topic). 
We obtained $2,207$ synonym-replaced prompts for each model,
consisting of $1,809$ for the product scenario and $398$ for the
societal scenario. \lujo{all of what we did experimentally should be
  in the past tense. please review and make consistent throughout} \anna{attempted}
We compared the $524$ prompts to the $2,207$ prompts by how often they mention a target concept in \S\ref{sec:results:whiteboxattack}.
We paired the prompts after synonym replacement with those before synonym replacement (i.e., $2,207$ corresponding pairs). 
For each pair, we computed the absolute improvement in the percentage of responses that mention a target brand.
If $20\%$ of responses of the prompt before synonym replacement (base prompt),
and $50\%$ of responses of the prompt after synonym replacement (perturbed prompt) mention the target brand,
the absolute improvement was $50\%-20\%=30\%$.
\revision{We also computed the USE score between each pair as shown in Fig.~\ref{fig:results:use}.
Across all models, our synonym-replaced prompts have much higher (close to one) USE scores than the paraphrased prompts,
suggesting the synonym-replaced prompts are near-identical in
meaning. %\lujo{higher does not mean ``near-identical''!}\omer{}
We also find the USE scores between synonym-replaced 
prompts that achieve the biggest absolute improvement 
in likelihoods do not have the lowest USE scores 
among all synonym-replaced prompts, 
indicating that difference in meaning is not a driving factor 
for higher attack success rate.}%\omer{<- rephrased.}
  %\lujo{this makes more sense than the recent previous description of
  %  the same thing, which is both less good and made redundant. please
  %go back and eliminate redundancy}\omer{done}
%such pairs of prompts are not the 
%most different in meaning among all synonym-replaced prompts.
%}

%%%%%%%%%%%%%%%%%%%%%%%%%%%%%%%%%%%%%%%%%%%%%%%%%%%%%%%%%%%%%%%%%%%%%%%%%%%%%%%%%%%%%%%%%%%%%%%%%%%%%%%%%%%%%%%%%%%%%%%%%%%%%%%%%%%%%%%%%%%%%%%%%%%%%%%%%%%%%%%%%%%%%%%%%%%%%%%%%%%%%%%%%%%%%%%%%%%%%%%%%%%%%%%%%%%%%%%%%%%%%%%%%%%%%%%%%%%%%%%%
%%%%%%%%%%%%%%%%%%%%%%%%%%%%%%%%%%%%%%%%%%%%%%%%%%%%%%%%%%%%%%%%%%%%%%%%%%%%%%%%%%%%%%%%%%%%%%%%%%%%%%%%%%%%%%%%%%%%%%%%%%%%%%%%%%%%%%%%%%%%%%%%%%%%%%%%%%%%%%%%%%%%%%%%%%%%%%%%%%%%%%%%%%%%%%%%%%%%%%%%%%%%%%%%%%%%%%%%%%%%%%%%%%%%%%%%%%%%%%%%
%%%%%%%%%%%%%%%%%%%%%%%%%%%%%%%%%%%%%%%%%%%%%%%%%%%%%%%%%%%%%%%%%%%%%%%%%%%%%%%%%%%%%%%%%%%%%%%%%%%%%%%%%%%%%%%%%%%%%%%%%%%%%%%%%%%%%%%%%%%%%%%%%%%%%%%%%%%%%%%%%%%%%%%%%%%%%%%%%%%%%%%%%%%%%%%%%%%%%%%%%%%%%%%%%%%%%%%%%%%%%%%%%%%%%%%%%%%%%%%%
%%%%%%%%%%%%%%%%%%%%%%%%%%%%%%%%%%%%%%%%%%%%%%%%%%%%%%%%%%%%%%%%%%%%%%%%%%%%%%%%%%%%%%%%%%%%%%%%%%%%%%%%%%%%%%%%%%%%%%%%%%%%%%%%%%%%%%%%%%%%%%%%%%%%%%%%%%%%%%%%%%%%%%%%%%%%%%%%%%%%%%%%%%%%%%%%%%%%%%%%%%%%%%%%%%%%%%%%%%%%%%%%%%%%%%%%%%%%%%%%

\subsection{Results}
\label{sec:results}
In this section, we describe our empirical results. 
%Our evaluation procedure and metrics were described in \S\ref{sec:setup}.
First, in \S\ref{sec:results:blackboxattack}, we describe our observations on paraphrases of
prompts---measuring how pairs of paraphrased
prompts can have significant differences in the probability
that the target concept appears.
%\lujo{be more specific (in
%  general). e.g., we
%  describe the results of our experiments that measure how often
%  different prompts from a seemingly equivalent set yield responses
%  that prefer different brands(?)} \anna{addressed}
% \lujo{globally, ``synonym replacement approach'' to ``synonym-replacement approach''}
Next, in \S\ref{sec:results:whiteboxattack}, we report how often our synonym-replacement approach yields a
perturbed prompt that causes LLMs to mention a targeted concept more often than the base prompt.
%\lujo{with different frequency than the original
%  prompts(?) }\anna{addressed} 
% \lujo{let's be more judicious with the use of ``may'' and
%   ``can''. when we're describing results of experiments, we're
%   describing what \emph{did} happen, and so caveats are in many cases
%   unnecessary and incorrect. the caveats still apply whenever we
%   suggest generalizability, however}
%More details of experiment setups can be found in App.~\ref{sec:setup}
\revision{
More experiments characterizing the attack\lujo{say something about dimensions. just ``more over there'' isn't quite enough} can be found in App.~\ref{sec:additionalresult}.
}
%Then in \S\ref{sec:results:negative}. we discuss how our synonym-replacement approach might also yield perturbed prompts that cause LLMs to mention a target concept less often than the base prompt, mitigating existing bias in the LLMs. 
%Finally, in \S\ref{sec:results:transfer}, we explore whether our synonym-replacement approach
%transfers to GPT models. We show that if the right model pair is found, successful perturbed prompts 
%found on an open-source LLM can increase the probability 
%of the target brand appearing more often in the GPT model's responses.

%found by using our
%synonym-replacement approach that increases the probability of a
%certain target brand being mentioned in the responses of an 
%open-source LLM can do the same for ChatGPT, in
%\S\ref{sec:results:transfer}.
%\lujo{hyphenation: when a multi-word phrase describes a noun that
%  follows it, then the words in that phrase generally need to be
%  hyphenated. e.g., ``open-source LLM'', ``synonym-replacement approach''}
%\anna{addressed}
%Ultimately, we report our human experiment results evaluating whether our synonym replacement approach helps adversaries achieve their attack goals, as described in \S\ref{sec:threatmodel} (\S\ref{sec:results:user}).
% \lujo{good intro}

\subsubsection{Concept Frequency Differences on Paraphrased Prompts}
%\lujo{make the title more informative. e.g.,
%  Brand Frequency Differences on Equivalent Prompts}
%  \anna{addressed, may want to rename 4.1 in that case}\weiran{synchronized}
\label{sec:results:blackboxattack}

\begin{figure}[t!]
\centerline{\includegraphics[width=0.95\columnwidth]{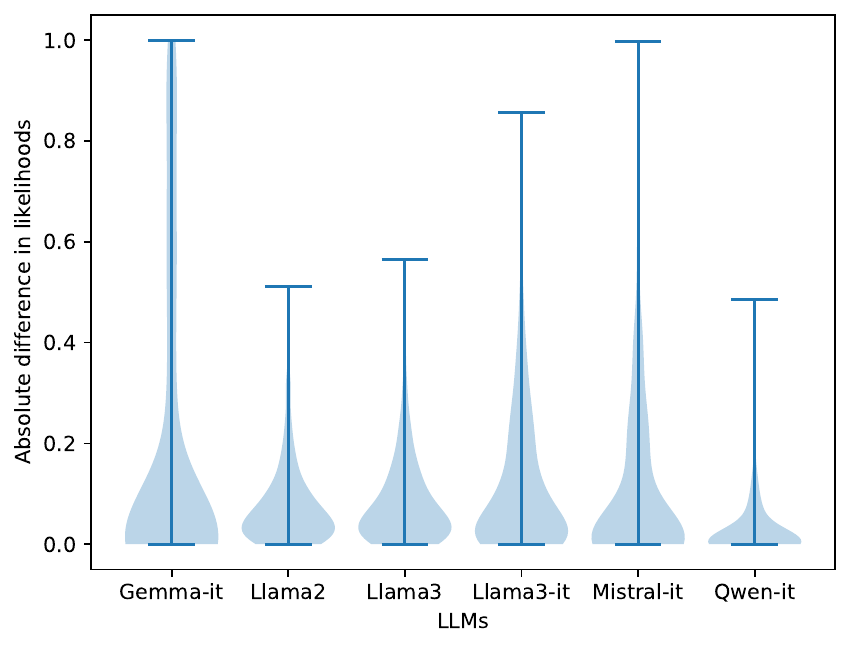}}
\caption{Absolute difference in the likelihoods of responses 
  mentioning a target brand within the first $64$ tokens generated in 
  response to paraphrased prompts.
  Paraphrasing prompts leads to an absolute improvement in
  the likelihood of LLMs mentioning target brands of up to $100\%$
  (i.e., one prompt elicits responses that never mention the target brand while another
  prompt's responses always do).}
\Description{
  This is a violin plot that suggests the range and density of the absolute difference in the likelihoods of 
  responses mentioning a target brand in the shopping scenario within the first 64 tokens generated in response to paraphrased prompts, 
  with respect to six different LLMs. The y axis is labeled with “Absolute difference in likelihoods” with a range from 0 to 1. 
  The x axis has four ticks corresponding to the six LLMs: Gemma-it, Llama2, Llama3, Llama3-it, Mistral-it, and Qwen-it. 
  Each tick has a vertical line segment, which suggests the range of the absolute difference in likelihoods. 
  All line segments start at zero, and end at 1.0 for Gemma-it, about 0.5 for Llama2, about 0.6 for Llama3, over 0.8 for Llama3-it, 1 for Mistral-it, and around 0.5 for Qwen-it. 
  Gemma-it’s absolute difference density is highest at 0, gradually decreases as the difference increases, but is still visible up to 1. 
  The absolute difference density of Llama2 is high at 0, gradually increases as the difference increases with a peak around 0.05, then all gradually decreases as the difference further increases, and ultimately becomes invisible when the difference is higher than 0.2. 
  The absolute difference density of Llama3 is high at 0, gradually increases as the difference increases until a peak around 0.05, then all gradually decreases as the difference further increases, and ultimately becomes invisible when the difference is higher than 0.4. 
  The absolute difference density of Llama3-it is high at 0, gradually increases as the difference increases until a peak around 0.05, then all gradually decreases as the difference further increases, and ultimately becomes invisible when the difference is higher than 0.6. 
The absolute difference density of Mistral-it is high at 0, slightly increases as the difference increases until a peak around 0.05, then all gradually decreases as the difference further increases, and ultimately becomes invisible when the difference is higher than 0.6.
The absolute difference density of Qwen-it is high at 0, and fairly rapidly decreases as the difference further increases, and ultimately becomes invisible when the difference is higher than 0.2. Overall, for each model, the difference has a wide distribution, being overall widest for Gemma-it and most concentrated for Qwen-it.
}
\label{fig:results:blackboxattack}
\end{figure}

\begin{figure}[t!]
\centerline{\includegraphics[width=0.95\columnwidth]{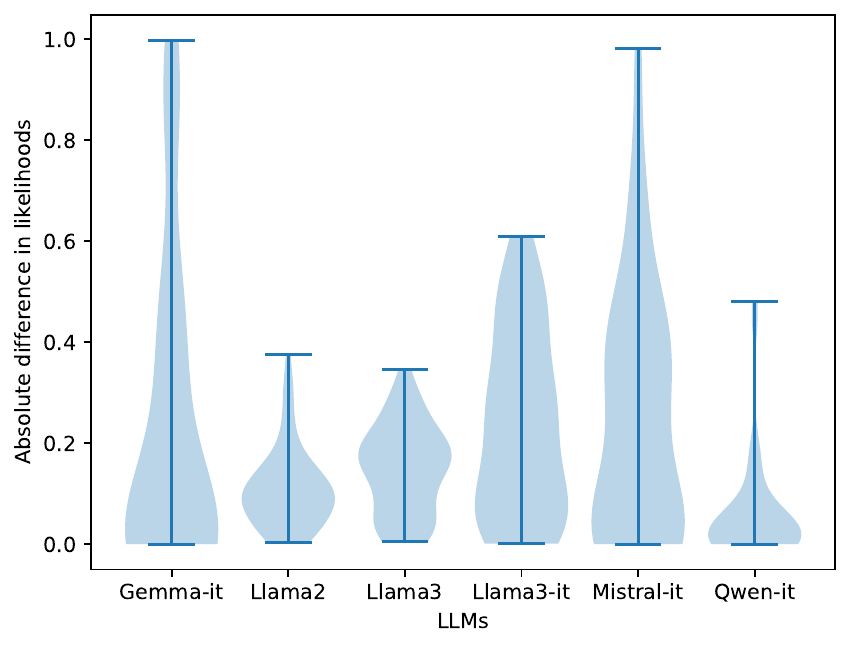}}
\caption{Absolute difference in the likelihoods of responses 
  mentioning a target societal concept within the first $64$ tokens generated in 
  response\lujo{two instances of response meaning different things} to paraphrased prompts.
  Paraphrasing prompts leads to an absolute improvement in
  the likelihood of LLMs mentioning a target societal concept of up to $99.8\%$.}
\Description{
  This is a violin plot that suggests the range and density of the absolute difference in the likelihoods of 
  responses mentioning a target societal concept in the societal scenario within the first 64 tokens generated in response to paraphrased prompts, with respect to six different LLMs. 
  The y axis is labeled with “Absolute difference in likelihoods” with a range from 0 to 1. 
  The x axis has four ticks corresponding to the six LLMs: Gemma-it, Llama2, Llama3, Llama3-it, Mistral-it, and Qwen-it. 
  Each tick has a vertical line segment, which suggests the range of the absolute difference in likelihoods. 
  All line segments start at zero, and end at about 1.0 for Gemma-it, about 0.4 for Llama2, about 0.4 for Llama3, about 0.6 for Llama3-it, about 1 for Mistral-it, and about 0.5 for Qwen-it. 
  Gemma-it’s absolute difference density is highest at 0, gradually decreases as the difference increases, but is still visible up to 1. 
  The absolute difference density of Llama2 is high at 0, gradually increases as the difference increases until a peak around 0.1, then all gradually decreases as the difference further increases, but is still visible up to 0.4. 
  The absolute difference density of Llama3 is high at 0, remains the same as the difference increases up to 0.1, gradually increases as the difference further increases with a peak around 0.2, then all gradually decreases as the difference further increases but is higher compared to Llama2 at the same amount of difference. 
  The absolute difference density of Llama3-it is high at 0, slightly increases as the difference increases with a peak around 0.05, then gradually decreases as the difference further increases, but is higher than all previous three models above 0.05. 
  Mistral-it’s absolute difference density is high at 0, gradually increases up to around 0.05, then gradually decreases until about 0.25, then increases until about 0.4, then gradually increases and is still visible until 1. 
  The absolute difference density of Qwen-it starts high, gradually increases until about 0.05, and then decreases until around 0.1. Overall, for each model, the difference has a wide distribution.
}
\label{fig:results:blackboxattack2}
\end{figure}

%As we described in \S\ref{sec:setup:evaluation:paraphrase}, each of the $77$
%product categories in our dataset have three to ten paraphrased prompts
%and each of the $11$ societal topics had six to eight.
%, which all
%approximately paraphrase each other.  
For each prompt in our dataset \revision{(\S\ref{sec:setup:evaluation:paraphrase})}, we measured the probability that LLMs would mention some 
target concept
%sequences 
% \anna{how do
%   we feel about using "desired sequences" vs "target
%   sequences"?}\lujo{how about just ``string'' or ``token'' or ``token
%   sequences'' related to different brands. i think they're neither
%   ``desired'' nor ``target'' at this point, since we're just measuring
% what happens and we don't have a specific goal in mind}
%related to each brand 
%\lujo{where does the list of brands come from?
%  should this be part of the description of the dataset?}\anna{addressed} 
of a category/topic %, as identified in \S\ref{sec:setup:evaluation}, 
within the first 64 tokens of the response.\lujo{could we please revisit all ``category/topic'' and get rid of the division sign, which is not a grammatically correct way of writing things in english?}
For
each combination of category/topic and concept, we computed the
maximal difference in the probability of responses mentioning the target
concept between paraphrased prompts of the category/topic (within
the first $64$ tokens).  Adversaries may paraphrase a
low-probability base prompt to find the high-probability prompts
to achieve the goals described in \S\ref{sec:threatmodel}.
% \lujo{since this is the first place where we're mentioning 64 tokens
%   in the results section, we should point back to wherever we
%   explained why 64 tokens makes sense} paraphrasing.pdf

We summarize our results in
Fig.~\ref{fig:results:blackboxattack} and Fig.~\ref{fig:results:blackboxattack2}. 
%\lujo{illustrate or summarize?}\anna{changed to summarize} 
\revision{These violin plots illustrate the distribution density of 
absolute difference between likelihoods (of mentioning a 
target concept between paraphrased prompts). The wider 
the violin plot is at a specific height, the more pairs of 
paraphrased prompts have such an absolute difference 
in likelihoods. The height of the violin plots suggest the range the 
absolute difference can be up to. On both plots (i.e., for both 
brands and societal concepts), all violin plots show a 
range of likelihood differences, suggesting it is 
possible for adversaries to manipulate the 
likelihoods by using paraphrased prompts.}
Specifically, on six
open-source LLMs (\S\ref{sec:setup:LLM}), the likelihood of the target concept
being mentioned in the responses can differ between a pair of paraphrased
prompts 
by up to $100.0\%$.
When
comparing responses to ``\emph{I'm curious to know your preference for the pressure
cooker that offers the best combination of cooking performance,
durable construction, and overall convenience in preparing a variety
of dishes.}" with the prompt paraphrased as ``\emph{Can you recommend the ultimate pressure cooker that
excels in providing consistent pressure, user-friendly controls, and
additional features such as multiple cooking presets or a digital
display for precise settings?}", 
the probability of 
Gemma-it mentioning the brand ``InstantPot" (``pressure cooker'' product category)
within the first $64$ tokens
of the response went from $0\%$ to $100.0\%$ (i.e., went from never mentioning ``InstantPot" to always mentioning ``InstantPot").
Among the six models, the average of this absolute difference in likelihood is 
 \revision{$3.6\%$}--$18.6\%$ for brands and \revision{$5.9\%$--$25.2\%$} for societal concepts.

%\lujo{``when we
%  paraphrased'' suggests an attack. perhaps stick to reporting the
%  numbers for the two prompts, e.g., for example, in one category we
%  had the following two prompts: x and y. responses to the first
%  prompt had a 0\% probability of mentioning ..., while responses to
%  the second prompt ...} \anna{addressed}

Our results suggest that while paraphrased prompts appear similar to
humans,
% \anna{removed ``---we evaluate the similarity in in
% \S\ref{sec:results:user}---", we do not}
the
responses to the prompts can differ substantially in how likely they
are to mention
%%se
%%paraphrases that recommend brands for the same category may have a
%%wide range of probabilities of mentioning some
a target 
%sequences related to a specific 
concept (within the first $64$ tokens of the response).
% \lujo{token
% sequences?}  
% \lujo{motivation for switching discussion to talking about adversaries
% is unclear. we may have to remind people about when these adversaries
% are practical} 
Therefore, an adversary wanting to promote a certain concept
can use this to their advantage: 
the adversary may try various paraphrases of prompts,
test the prompts,
and pick the paraphrase that results in the highest probability of 
the target concept being mentioned, ultimately promoting the concept when the perturbed prompt is used 
in the real world.
%mentioning some desired sequences, 
%and thus potentially recommend the
%desired brand more. 
While we generated these paraphrases using ChatGPT, and confirmed that they were valid and reasonable, 
adversaries may also be able to create paraphrases manually or by another method, 
and may be able to test even more paraphrases than we did in our
measurements.  
% \anna{I added this last sentence to emphasize that this is just the
%   method we used for paraphrases, and these observations are meant to
%   be general.}\lujo{makes sense}

%\subsection{Adversarial Prompts Using Synonym Replacement}
\subsubsection{Synonym-Replaced Adversarial Prompts}

%\lujo{again, make the title more descriptive. e.g., ``Impact of
%  Synonym Replacement in Prompts''} \anna{attempted to address}
\label{sec:results:whiteboxattack}

%As described in \S\ref{sec:setup:evaluation:synonym}, we measured and now
%report how
%often our synonym-replacement approach improves the probability of
%a concept being mentioned.
%LLMs mention some desired sequences related to\lujo{sequences related to?}
%\anna{addressed} a brand.
%Rather than using rephrasings
As opposed to paraphrasing, in the synonym-replacement approach, we automatically generated a set of potential candidate
prompts by perturbing a base prompt via synonym replacement,
without needing to confirm whether these perturbed prompts are valid. 
The prompt with the lowest loss was selected, and we assessed how well
these selected prompts increase the probability that 
the target concept is mentioned.
We evaluated the average improvement over 
%multiple models,
%at different response lengths, and for different
%probabilities of a concept 
the probabilities of different concepts
being mentioned in the base prompt.
%In addition, other candidate prompts could also be tested, although we did not take this approach in this paper.
We used the loss as a metric that narrowed down the large set of potential perturbed prompts
we find using synonym replacements to one.
Therefore, we are interested in exploring the \emph{highest} improvement we can achieve
between a base and perturbed score, as adversaries would be able to use a similar method
to \S\ref{sec:results:blackboxattack} and explore multiple \revision{perturbed} prompts.
%and in
%%multiple cases (differing response lengths and 
%%original probabilities of a brand being mentioned),
%as well as
So, we also describe the largest increases in the probability of mentioning a target concept
in the responses to the perturbed prompt we find 
via our synonym-replacement method compared to the base prompt.
% \lujo{here i got confused because i
%   thought this whole paragraph was talking about evaluating synonym
%   replacement. now i don't understand what we were talking about
%   evaluating prior to this sentence}
% \lujo{best-performing measured how?}
Our evaluation shows that 
our synonym-replacement attack increased 
the likelihood of LLMs mentioning a concept on average.
For all evaluations, we focus on the first 64 tokens of the response 
(see~\S\ref{sec:setup:LLM} for details).

% \lujo{i think we want to make the story as follows: (1) unguided
%   synonym replacement successfully generates some prompts that have
%   the desired property of increasing the probability of a brand being
%   mentioned, and (2) using a loss function to guide synonym
%   replacement makes it possible to produce prompts useful to
%   adversaries with one try, although the actual utility of those
%   prompts to the adversary is probably negligible. right now we seem
%   to only sell synonym replacement as something that has to include
%   the loss function, which i think doesn't serve us well because that
%   version of synonmy replacement doesn't work well}

%\lujo{add one-sentence summary of takeaway + describe layout of subsection}
%\anna{addressed summary}
%\anna{addressed layout description}

\begin{figure}[t!]
\centerline{\includegraphics[width=0.95\columnwidth]{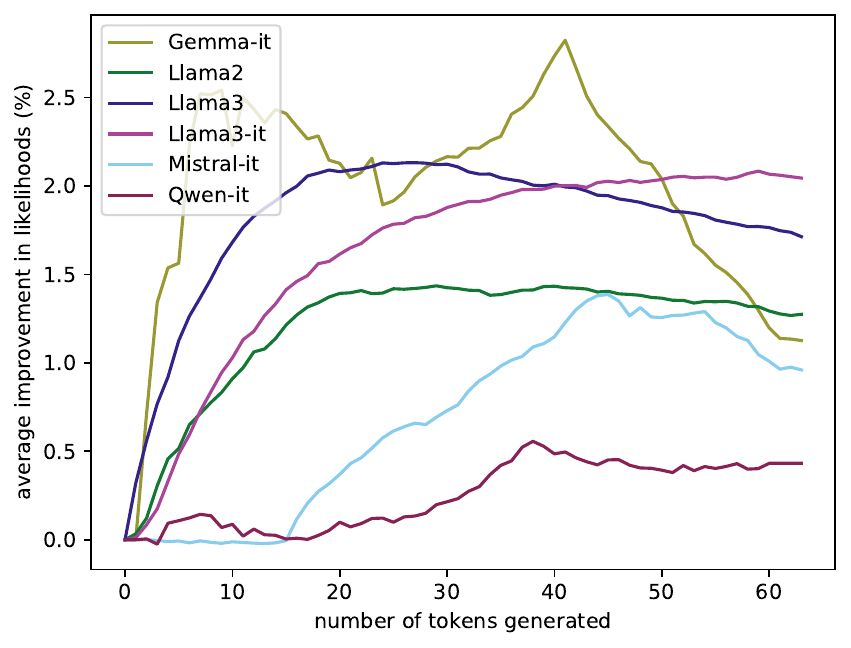}}
\caption{Average absolute improvement in likelihoods that LLMs
  mention a target brand when the base likelihood is within $[ 0.1 \% ,50 \% ]$. Results are presented along number of generated 
  tokens.
  Our synonym-replacement approach
  achieves improvements  in probabilities, which
  verifies the capability of forcing LLMs to mention
  target brands more often.
  }  
\Description{
  This is a line plot suggesting how the average improvement in likelihoods that LLMs mention a target brand change against the number of tokens generated. 
  The y axis is labeled “average improvement in likelihoods (\%)” and x axis is labeled “number of tokens generated”. 
  The x axis goes from 0 to 64. There are six curves plotted, corresponding to six LLMs: Gemma-it, Llama2, Llama3, Llama3-it, Mistral-it, and Qwen-it. 
  All models start at zero improvement with zero tokens generated. Gemma-it reaches around 2.5\% at 10 tokens, goes down to 2\% at 25 tokens, back up to 3\% at 40 tokens, and then decreases to just over 1\%. 
  Llama2 gradually increases to just under 1.5\% at 20 tokens and stays at this level. 
  Llama3 gradually increases to over 2\% at 20 tokens, and then gradually decreases back to around 1.75\%. 
  Llama3-it gradually increases the entire time and stops around 2\%. 
  Mistral-it stays at zero until around 15 tokens, then increases until reaching about 1.5\% around 45 tokens, then decreases back down to around 1\%. 
  Qwen-it stays near zero until about 20 tokens, then increases up to 0.5\% at almost 40, after which it stays at a similar level.
}
\label{fig:results:whitebox:short001}
\end{figure}
\begin{figure}[t!]
\centerline{\includegraphics[width=0.95\columnwidth]{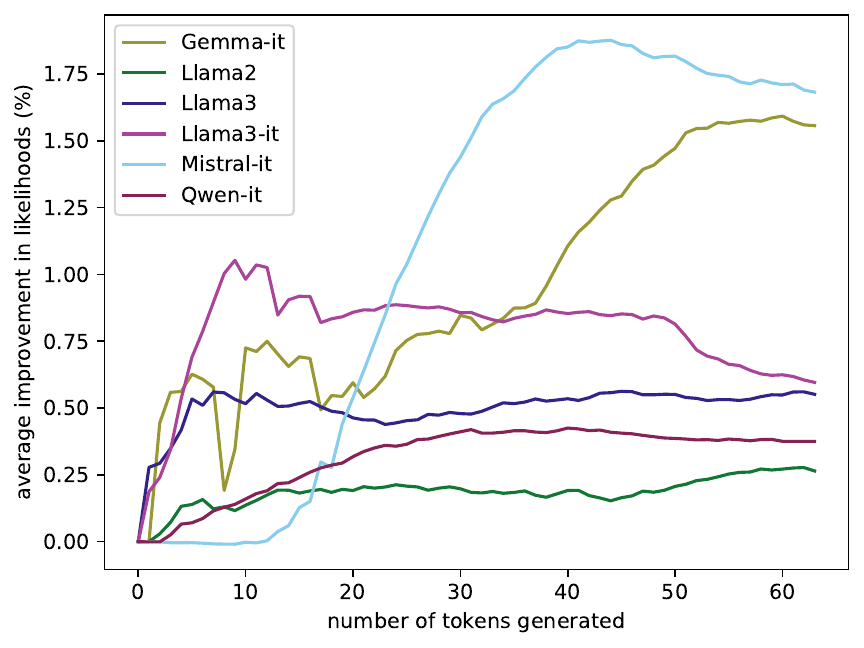}}
\caption{Average absolute improvement in likelihoods that LLMs
  mention a target societal concept when the base likelihood is within $[ 0.1 \% ,50 \% ]$. Results are presented along number of generated 
  tokens.
  Our synonym-replacement approach
  achieves improvements  in probabilities, which
  verifies the capability of forcing LLMs to mention
  target brands more often.
  }  
\Description{
  This is a line plot suggesting how the average improvement in likelihoods that LLMs mention a target societal concept change against the number of tokens generated. 
  The y axis is labeled “average improvement in likelihoods (\%)” and x axis is labeled “number of tokens generated”. 
  The x axis goes from 0 to 64. There are six curves plotted, corresponding to six LLMs: Gemma-it, Llama2, Llama3, Llama3-it, Mistral-it, and Qwen-it. 
  All models start at zero improvement with zero tokens generated. Gemma-it oscillates somewhat between 0 and 0.75\% until 20 tokens, after which it goes up, ending at 1.5\%. 
  Llama2 increases until reaching 0.25\% at 15 tokens, and stays around this level. 
  Llama3 increases until reaching 0.5\% at 5 tokens, and stays around this level. Llama3-it Goes up until reaching 1\% at 10 tokens, then gradually decreases until reaching 0.75\%. 
  Mistral-it stays low until 15 tokens, at which point it begins increasing until reaching about 1.8\% around 40 tokens, then decreases back to around 1.6\%. 
  Qwen-it gradually increases until reaching 0.4\% around 25 tokens, and stays around that level.
}
\label{fig:results:whitebox:short001_other}
\end{figure}

%\omer{I think there's still too much emphasis on the average improvements. What happened to talking about the best averages?}

\paragraph{Average improvement} 
%The results for average improvement in
%likelihoods of a target concept being mentioned
%in the responses to perturbed prompts created via synonym
%replacement compared to in response to the original base prompt, 
%across models, are shown in Fig.~\ref{fig:results:whitebox:short001}, and Fig.~\ref{fig:results:whitebox:short001_other}, corresponding to brands and societal concepts respectively. 
Overall, we 
found that our attack results in an average absolute improvement in all models. 
%\omer{I think we should make this bit more self-contained instead of referring elsewhere. Can we summarize? ->} 
\revision{We took the average over various prompts and target concepts for the absolute improvement (i.e., the difference in likelihoods after synonym replacement, see \S\ref{sec:setup:evaluation:synonym} for more details) on each of the six models.}
%\lujo{definitely need an intro to these results that will prevent readers from thinking that we've completely failed}
%\anna{added an intro, but unsure what to add to discourage readers from thinking we failed :/}
Specifically, for concepts that were mentioned at least once (i.e.,
$0.1\%$ of the time) and at most 500 times (i.e.,
$50\%$ of the time) in responses to the base prompt, 
the six models had an average absolute improvement of 
\revision{$0.43\%$--$2.05\%$} for brands and \revision{$0.26\%$--$1.68\%$} for societal concepts,
as shown in Fig.~\ref{fig:results:whitebox:short001} and Fig.~\ref{fig:results:whitebox:short001_other}. 
%before perturbation,
%Gemma-it had an average absolute improvement
%$0.14\%$ within $64$ tokens and a $1.09\%$ improvement
%within $42$ tokens.
%%\lujo{we need to justify why we're reporting improvement on less than 64 tokens. might make sense to first report performance on 64 tokens for all models and then on the smaller number. might also make sense to summarize the improvement without breaking it down by model in text, since the improvement is small}  
%Llama2 achieves an average absolute improvement of $0.71\%$ within $64$ tokens, and $0.75\%$ within $44$ tokens.
%Llama3 achieves an average absolute improvement of $0.93\%$ within $64$ tokens, and $1.07\%$ within $28$ tokens.
%Llama3-it achieves an average absolute improvement of $1.10\%$ within $64$ tokens, and $1.19\%$ within $60$ tokens.
%The averages of the best absolute improvement in each product category are $24.80\%$, $8.33\%$, $11.23\%$, and $10.17$ on Gemma-it, Llama2, Llama3, and Llama3-it respectively. 
Our synonym replacement approach achieved a positive \revision{average} absolute
improvement in probabilities, which verifies that our approach is
capable of forcing LLMs to mention target concepts more often within
some number of tokens. 
On some models, we saw a bigger absolute
improvement within the first $64$ tokens. 
As we described in \S\ref{sec:approach:whiteboxattack},
our proposed approach might not necessarily be the most effective under the threat model (\S\ref{sec:threatmodel}), 
however, our approach illustrates that attacks under our threat model exist,
\revision{with potentially stronger variations possible.} 
%\anna{I think we may want to add something here that, even with responses len 64 tokens or longer, we do still at times see increases in the probability, even though it decreases w num tokens}
%\anna{We may also want to mention here that it may still be more valuable to have things mentioned earlier}\lujo{agreed. explain practical significance of results for fewer than 64 tokens}  
%
%\omer{these couple sentences I don't understand. }
%Because our loss function only accounts for target brand
%sequences of tokens (representing strings related to a brand)
%appearing in the response, we expect our approach to be more effective with shorter responses.
%So, our approach is more effective at forcing mentions of certain brands \emph{early} in the response,
%which can still provide adversaries an advantage by forcing their brands to appear first. \anna{too speculative?}
%
%While we only collected responses of length 64 tokens for our $2,207$ attacks for each model 
%(see~\autoref{sec:setup:LLM}), we did run a select set of base and perturbed prompt pairs 
%(chosen in \S\ref{Experimental groups})
%to full completion. 
While we did not see improvements in all pairs 
(for example, some base prompts had near $100\%$ probabilities of mentioning certain concepts, not allowing for any improvement),
% (for example, some base prompts had near $100\%$ probabilities of mentioning certain concepts with long completions, not allowing for any improvement),
we still saw pairs with significant increases when considering long responses.
For Llama3-it, the perturbed prompt 
``\emph{Can you recommend the superior video game console that excels in providing top-notch graphics, dissimilar gaming options, and additional features such as online connectivity, appropriate for both casual and hardcore gamers?}'' 
was $55.9\%$ more likely than its base prompt 
``\emph{Can you recommend the ultimate video game console that excels in providing top-notch graphics, diverse gaming options, and additional features such as online connectivity, suitable for both casual and hardcore gamers?}''
(i.e., ``ultimate'' was changed to ``superior'' and ``diverse'' was changed to ``dissimilar'')
to mention Xbox in long completions, 
even more than when completions were only 64 tokens long ($32.9\%$). 
We evaluated if attack objectives are met with full responses in a more realistic setting in \S\ref{sec:results:user}.

\begin{figure}[t!]
\centerline{\includegraphics[width=0.95\columnwidth]{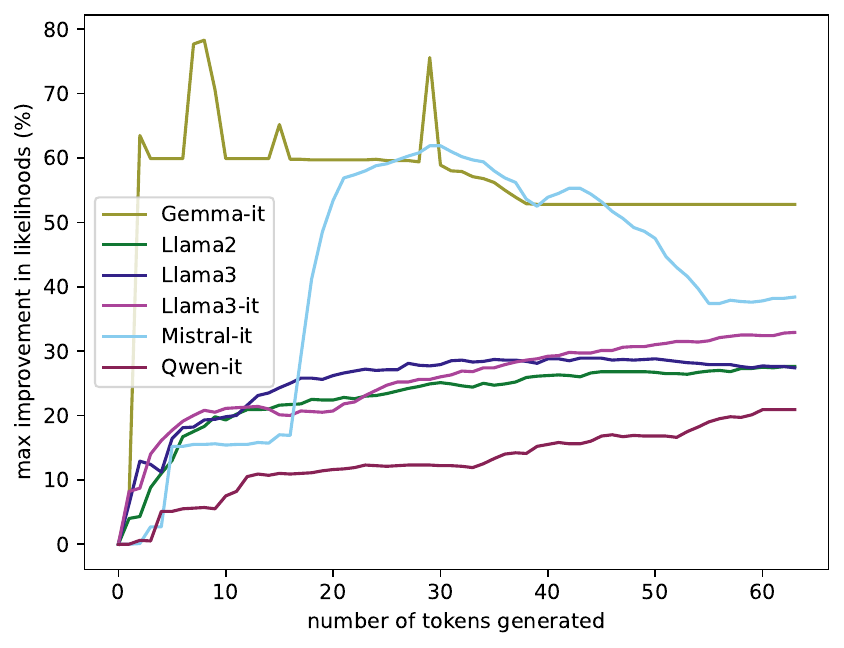}}
\caption{Max absolute improvement in likelihoods that
  LLMs mention 
  a target brand. Results are presented along how many tokens are generated. 
  We achieve a bigger absolute improvement on Gemma-it compared to the
  other three Llama models.
%  , although Gemma-it does not always have the
%  highest average absolute improvement (see
%  Fig.~\ref{fig:results:whitebox:short001})
  }
\Description{
  This is a line plot suggesting how the max absolute improvement in likelihoods that LLMs mention a target brand change against the number of tokens generated. 
  The y axis is labeled “max improvement in likelihoods (\%)” and x axis is labeled “number of tokens generated”. 
  The x axis goes from 0 to 64. There are six curves plotted, corresponding to six LLMs: Gemma-it, Llama2, Llama3, Llama3-it, Mistral-it, and Qwen-it. 
  All models start at zero improvement with zero tokens generated. Gemma-it reaches between 50\% – 6\% after a couple of tokens and stays in the range afterwards, besides a couple of impulses that reach 70\% – 80\% around 10 and 30 tokens generated. 
  The three Llama models reach about 20\% with 10 tokens generated, and then gradually increase towards around 30\% at 64 tokens.
  Mistral-it goes up to around 15\% for 5-20 tokens, after which it rises, with a maximum range of 50-60\% for 20-50 tokens. Mistral-it then decreases to an improvement of 40\% for 55-64 tokens. 
  Qwen-it gradually increases until reaching an improvement of 20\% at the end. 
  Overall, we achieve a significant improvement for all models.
}
\label{fig:results:whitebox:max}
\end{figure}

\begin{figure}[t!]
\centerline{\includegraphics[width=0.95\columnwidth]{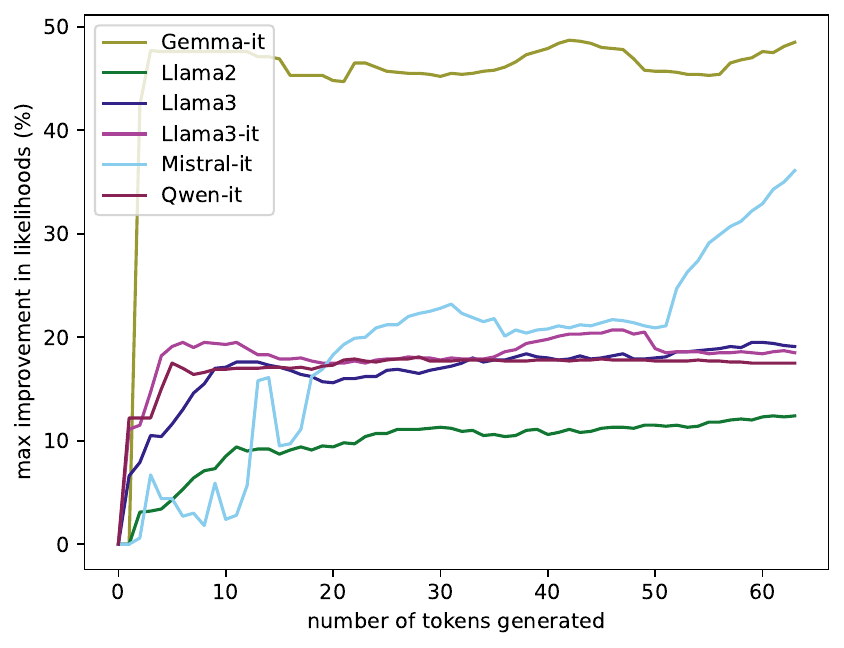}}
\caption{Max absolute improvement in likelihoods that
  LLMs mention 
  a target societal concept. Results are presented along how many tokens are generated. 
  We achieve a bigger absolute improvement on Gemma-it compared to the
  other three Llama models.
%  , although Gemma-it does not always have the
%  highest average absolute improvement (see
%  Fig.~\ref{fig:results:whitebox:short001_other})
  }
\Description{
  This is a line plot suggesting how the max absolute improvement in likelihoods that LLMs mention a target societal concept changes against the number of tokens generated. 
  The y axis is labeled “max improvement in likelihoods (\%)” and x axis is labeled “number of tokens generated”. 
  The x axis goes from 0 to 64. There are six curves plotted, corresponding to six LLMs: Gemma-it, Llama2, Llama3, Llama3-it, Mistral-it, and Qwen-it. 
  All models start at zero improvement with zero tokens generated. Gemma-it almost reaches 50\% within a couple of tokens and stays at the same level afterwards. 
  Llama2 reaches 10\% within ten tokens and stays at the same level afterwards with slight increases. 
  Llama3 almost reaches 20\% within about ten tokens and stays at the same level afterwards with slight increases. 
  Llama3-it and Qwen-it reach 20\% within five tokens and stay at the same level afterwards. 
  Mistral-it oscillates for the first 20 tokens, after which it stays around 20\% until 50 tokens. 
  At 50 tokens, it increases up until reaching around 35\% for 64 tokens. Overall, we achieve a significant improvement for all models.
}
\label{fig:results:whitebox:max_other}
\end{figure}

\lujo{explain that this is practically more meaningful than the
  previous results and why it is meaningful. it might make sense for each bunch of results to
  get its own subsubsection, including because we can then report but
  deemphasize all the results thus far, which aren't very good} 
  \anna{attempted below}

%\omer{I think this paragraph is a bit too long with the added quotes.}

\paragraph{Maximum improvement}
Besides the average absolute improvement, we also explored the maximum
absolute improvement among all combinations of base prompts and concepts,
%we have,
as shown in Fig.~\ref{fig:results:whitebox:max} and  Fig.~\ref{fig:results:whitebox:max_other} for brands and societal concepts respectively.  
While these results do not represent the \emph{expected} improvement using this method,
they do demonstrate that
it is possible to find pairs of prompts with vastly different probabilities of mentioning a certain concept.
In \S\ref{sec:results:blackboxattack} we showed this for prompts that
were manually paraphrased; here, we show that the synonym-replacement
attack can find such alternative prompts automatically.
%%However, with our synonym-replacement attack, these alternative 
%%prompts can be found automatically. 
%without either human-generated or manually checked rephrasings,
Further, the prompts generated by synonym replacement differ from their base prompts minimally 
(at most a seven-synonym difference in our dataset), and are perceptually the same 
along multiple dimensions (see~\S\ref{sec:results:user}).
\revision{We were able to achieve a much higher max absolute improvement compared to the average 
absolute improvement, suggesting the synonym replacement works exceptionally well with specific 
prompts and synonyms. We will explain the implication of this observation in \S\ref{sec:discussion:attributes}.}
\section{User Study: Verifying Practical Attack Success}
\label{sec:user}
%\omer{do a pass to make the overall story fit better.}
%\omer{fix this preamble}
So far we have shown that our synonym-replacement approach
\revision{biases} LLM responses towards a target concept and appears to be 
inconspicuous. However, this does not 
necessarily indicate practical attack success. To evaluate the
attack in a realistic setting, 
\revision{we conduct a user study,} detailing our methods (\S\ref{sec:approach:user}) 
and statistical evaluation  (\S\ref{sec:user:results}). We discuss implications throughout our 
results and takeaways (\S\ref{sec:user:takeaway}).
We find that our attack is indeed successful \revision{in inconspicuously pushing users toward 
chosen concepts.}

All human-subjects procedures were approved by the Carnegie Mellon University institutional review board.
%\subsection{User Study Setup}
%\label{sec:setup:userstudy}
%\anna{do we want this here or in section 7?}
%Among the four open-sourced LLMs we have, we used the two instruction-tuned ones only for user study,
%as we observed that the other two LLMs often have up to more than $4,000$ tokens in their responses which makes the responses too lengthy for human readers. 
%In contrast, the instruction-tuned LLMs usually only have around 400 or 600 tokens and no more than 800 or 1300 tokens for Gemma-it and Llama3-it, respectively.
%For each instruction-tuned LLM, we used four pairs of prompts that
%1) belong to the top $50\%$ product categories that people are interested in according to the pilot study(\S\ref{sec:approach:user}),
%2) belong to the top $50\%$ product categories that people are interested in using LLMs for according to the pilot study(also in \S\ref{sec:approach:user}),
%and 
%3) have the highest improvements in mentioning some desired sequences within complete responses (measured in \S\ref{sec:setup:evalutaion}) among pairs that meet (1) and (2).

\subsection{Methods}%\lujo{can we make the titles more descriptive? both ``methods'' and ``procedures'' seem like they're the same thing. could some adjectives help?}
\label{sec:approach:user}
\revision{This between-subjects study evaluates 
the effectiveness of our proposed attack}
%We performed a between-subjects user study to evaluate 
%whether our synonym-replacement approach can result in prompts that
%advances the adversaries' goals 
(see~\S\ref{sec:threatmodel}), creating inconspicuous prompts 
that trigger inconspicuous responses, making a target brand more noticeable. %\lujo {instead of the opaque ``advances ... goals'', remind us what these are. this is a global coment} 
%\anna{slightly changee phrasing, basically just added `result in a prompt that'} 

\revision{Some of the topics we explored in previous 
sections might be controversial and cause emotional harm. 
As such, we focused the user study 
on the shopping task, a benign topic that users are 
exposed to every day. Similar setups have been 
used in prior work~\cite{CCS16:eyeglasses}.}
\revision{Testing each prompt-and-response pair 
from our previous experimentation would be cost prohibitive. Thus, we limited 
our study to six pairs of prompts, each from a different product category and evenly split between 
two models (see~\S\ref{sec:methods:groups}).}

We tested whether 
users could distinguish between base and perturbed
prompts and responses in multiple dimensions 
(including clarity, likelihood of use, satisfaction, and more).
%\lujo{``, including x, y, and z'' -- otherwise i just get mad because you're teasing me without telling me the answer}. 
Each participant was shown one prompt-and-response pair. 

Fully informed consent was obtained from participants and our procedures 
were approved by the Carnegie Mellon University institutional review board. Following advice from prior work~\cite{chuang2018transparency, cockburn2018prereg}, 
we pre-registered our study.\footnote{\url{https://osf.io/6mycr/?view_only=face90d04806439bb1f69fc110fb9a1e}}
%\anna{I have been using `base prompt' rather than `baseline prompt'}
%\anna{we should decide on one to use throughout}

\subsubsection{Survey procedures}
\label{sec:methods:survey}
We recruited a gender-balanced sample 
from Prolific,\footnote{\url{https://www.prolific.com/}} 
%a crowdsourcing platform commonly used with
%security-relevant user studies (e.g.,~\cite{abrokwa2021comparing}). 
\revision{a commonly used platform for security-relevant user studies~\cite{abrokwa2021comparing}.}
To \revision{reduce} selection bias, 
we avoided mentioning LLMs or 
bias in the study title, ``Chatbot prompting study.'' 
%advertised as  to reduce 
%selection bias %\anna{why does that name reduce selection bias? unclear}. 
%\anna{Maybe something like: ``To avoid selection bias, we avoided mentioning LLMs or bias specifically. Our study was advertised w the title ...''}
Participants had to be in the U.S., be 18 or older, and 
have an approval rate of at least 95\%. \revision{Obeying} Prolific 
guidelines,\footnote{\url{https://researcher-help.prolific.com/hc/en-gb/articles/4407695146002-Prolific-s-payment-principles}}
participants were paid \$1.6 for an estimated 7--9 minute study, \revision{averaging} \$12.45/hour.

After providing consent, participants were first given an overview. 
Next, we asked our participants to imagine that there was a chatbot service 
that was able to recommend prompts appropriate for what users want to 
use the chatbot for. We further instructed participants to imagine that they 
were shopping for a certain product (e.g., laptops) or 
service (e.g., parcel delivery) 
and wanted to use this new chatbot service to help them decide on a brand, 
\revision{similar to the use case of Amazon Rufus~\cite{rufus2024}}.
After a comprehension check on instructions, participants were 
shown a prompt recommendation for the product they were shopping for 
(e.g., ``\emph{Which laptop model do you consider the optimal choice for 
versatile computing, powerful performance, and innovative 
features that enhance your work and entertainment experience?}'') 
and instructed to review it. After a minimum of 10 seconds had passed, 
we asked (1) how likely they were to use this prompt, 
(2) how clear the prompt was, 
(3) was it biased to a certain brand (and which), 
(4) and if anything stood out 
(e.g., unexpected). 

Participants \revision{were then asked to imagine they 
had chosen to use the prompt and presented} a response to 
the prompt \revision{word by word, mimicking chatbots.} 
After a minimum of 20 seconds, they were asked: 
(1) how clear the response was, 
(2) if they were satisfied with the response,
(3) how likely were they to take the recommendation in the response, 
(4) and if anything stood out. In a series of open-ended questions, 
we additionally asked (1) which brand participants would pick based 
on this response, (2) what were all the brands recommended, 
and (3) what the top brand recommendation was. 
These 11 (numbered) questions form the 
practical definition of inconspicuousness and 
increase in target brand perception. 
They form the basis of our statistical analysis~\S\ref{sec:results:user}.

To help with recall, participants could hover over relevant questions 
to reveal the relevant prompts and responses.

%\anna{Possible rephrase:}
%\anna{At the end of the survey participants were asked to answer demographics questions and questions on their their experience with tech, chatbots, and ChatGPT.}.
Participants self-reported how frequently 
they give tech advice and used chatbots, 
if they paid for a chatbot, and their ChatGPT familiarity.
%Participants were then asked to report their frequency of giving tech advice,
%how often they use chatbots, if they paid for a chatbot service, 
%if they heard about and used ChatGPT before.
%~\footnote{These two ChatGPT questions 
%were borrowed from the PEW research \omer{FIX ME}}. 
The survey concluded with demographic 
questions, which we summarize in \autoref{tab:demographics}.

%%%% comp_stats_table.tex starts here %%%%

\aptLtoX{\begin{table*}[t]
	\centering
	\small
	\setlength\tabcolsep{5pt} % default value: 6pt
	\begin{tabular}{r l l l l l l l l l}
\hline  
    & \multicolumn{4}{c}{\textbf{Prompt}} & ~~ & \multicolumn{4}{c}{\textbf{Reponse}}  \\
	\cmidrule{2-5}\cmidrule{7-10}

          %\midrule
& \multicolumn{1}{c}{Clarity (L7)}  & \multicolumn{1}{c}{Use (L7)}  &  \multicolumn{1}{c}{Bias (L5)}  & \multicolumn{1}{c}{Standout (B)}
&& \multicolumn{1}{c}{Clarity (L7)} & \multicolumn{1}{c}{Use (L7)}  & \multicolumn{1}{c}{Satisfied (L7)} & \multicolumn{1}{c}{Standout (B)} \\ %  & \multicolumn{1}{c}{Pick \%} & \multicolumn{1}{c}{Top \%}  
\hline
          \textbf{TV}   & -0.23\dubstar        & -0.15\lonestar   & -0.03\fourstar  & -0.08\fourstar          
                        && ~0.06\fourstar       & ~0.00\lonestar    & 0.05\lonestar   & -0.02\fourstar \\ % & 0.00 & 0.00 & \\

          \textbf{ISP}  & -0.28\dubstar        & -0.04\threestar  & -0.06\fourstar  & -0.07\fourstar              
                        && -0.19\fourstar      & -0.08\lonestar   & -0.15\dubstar  &  -0.10\fourstar \\ % & 0.00 & 0.00 & \\
          \textbf{Parcel delivery} & -0.15\threestar & -0.58 & 0.04\fourstar & -0.06\fourstar 
                                && ~0.04\fourstar & \{0.41\lonestar\} & \{0.55\lonestar\} & -0.06\fourstar \\

          \textbf{Gaming console} & \{-0.8\dubstar\} & -0.10\lonestar & -0.01\fourstar & ~0.00\fourstar 
                                && -0.17\fourstar & -0.21  & -0.21\dubstar & -0.01\fourstar \\
        
          \textbf{Investment plat.} & ~0.01\dubstar & -0.10\threestar & -0.07\fourstar & ~0.00\fourstar 
                                && -0.02\fourstar & ~0.20\dubstar & 0.03\dubstar & -0.02\fourstar  \\
        
          \textbf{Laptop}& -0.04\dubstar & \{0.35\lonestar\} & -0.0\fourstar & ~0.04\fourstar 
                        && ~0.10\fourstar & ~0.20\lonestar & 0.22\dubstar & ~0.04\fourstar  \\
\hline
	\end{tabular}
    \vspace{0.1cm}
	\caption{\small Mean difference between base and perturbed groups among eight questions, four about prompts and four about responses (see~\S\ref{sec:methods:survey} for details). By default test 
    for equivalence is reported (TOST WMU), \{\} indicates test for difference (MWU, CHI$^2$). Higher is better for all differences.
    \lonestar: $p<0.05$, \dubstar : $p<0.01$ \threestar: $p<0.001$, \fourstar: $p<0.0001$. L7: seven-point Likert, L5: five-point Likert, B: binary.}
    \Description{
        This table reports the pairwise test results between perturbed and baseline groups for the eight perceptions questions. 
        The results are reported across multiple dimensions: the product categories, and the individual perceptions questions (four about prompts and four about responses). 
        For each test, up two results can be inferred from the table: 1) results for test of equivalence (TOST MWU p-value, and for effect size mean difference), 
        2) if the test for equivalence is not significant, test for difference (MWU p-value, and and for effect size mean difference).
        From a quick glance, the table shows that out of 48 comparisons reported (6 product categories, 8 questions each), the vast majority (42) were found to be equivalent, 4 different, and 2 neither different nor equivalent.
        Specifically:
        For the clarity question about the prompts, the TV category prompt is found equivalent (p<0.01, mean diff: -0.23).
        For the clarity question about the prompts, the ISP category prompt is found equivalent (p<0.01, mean diff: -0.28).
        For the clarity question about the prompts, the parcel delivery category prompt is found equivalent (p<0.001, mean diff: -0.15).
        For the clarity question about the prompts, the gaming console category prompt is found different (p<0.01, mean diff: -0.15).
        For the clarity question about the prompts, the investment platform category prompt is found equivalent (p<0.001, mean diff: 0.01).
        For the clarity question about the prompts, the laptop category prompt is found equivalent (p<0.001, mean diff: -0.04).
        For the use question about the prompts, the TV category prompt is found equivalent (p<0.05, mean diff: -0.15).
        For the use question about the prompts, the ISP category prompt is found equivalent (p<0.001, mean diff: -0.04).
        For the use question about the prompts, the parcel delivery category prompt is not found equivalent nor different (mean diff: -0.58).
        For the use question about the prompts, the gaming console category prompt is found equivalent (p<0.05, mean diff: -0.10).
        For the use question about the prompts, the investment platform category prompt is found equivalent (p<0.001, mean diff: -0.10).
        For the use question about the prompts, the laptop category prompt is found different (p<0.05, mean diff: 0.35).
        For the bias question about the prompts, the TV category prompt is found equivalent (p<0.0001, mean diff: -0.03).
        For the bias question about the prompts, the ISP category prompt is found equivalent (p<0.0001, mean diff: -0.06).
        For the bias question about the prompts, the parcel delivery category prompt is found equivalent (p<0.0001, mean diff: 0.04).
        For the bias question about the prompts, the gaming console category prompt is found equivalent (p<0.0001, mean diff: -0.01).
        For the bias question about the prompts, the investment platform category prompt is found equivalent (p<0.0001, mean diff: -0.07).
        For the bias question about the prompts, the laptop category prompt is found equivalent (p<0.0001, mean diff: 0.00).
        For the standout question about the prompts, the TV category prompt is found equivalent (p<0.0001, mean diff: -0.08).
        For the standout question about the prompts, the ISP category prompt is found equivalent (p<0.0001, mean diff: -0.07).
        For the standout question about the prompts, the parcel delivery category prompt is found equivalent (p<0.0001, mean diff: 0.06).
        For the standout question about the prompts, the gaming console category prompt is found equivalent (p<0.0001, mean diff: 0.00).
        For the standout question about the prompts, the investment platform category prompt is found equivalent (p<0.0001, mean diff: 0.00).
        For the standout question about the prompts, the laptop category prompt is found equivalent (p<0.0001, mean diff: 0.04).
        For the clarity question about responses, the TV category prompt is found equivalent (p<0.0001, mean diff: 0.06).
        For the clarity question about responses, the ISP category prompt is found equivalent (p<0.0001, mean diff: -0.19).
        For the clarity question about responses, the parcel delivery category prompt is found equivalent (p<0.0001, mean diff: 0.04).
        For the clarity question about responses, the gaming console category prompt is found equivalent (p<0.0001, mean diff: -0.17).
        For the clarity question about responses, the investment platform category prompt is found equivalent (p<0.0001, mean diff: -0.02).
        For the clarity question about responses, the laptop category prompt is found equivalent (p<0.0001, mean diff: 0.10).
        For the use question about responses, the TV category prompt is found equivalent (p<0.05, mean diff: 0.00).
        For the use question about responses, the ISP category prompt is found equivalent (p<0.05, mean diff: -0.08).
        For the use question about responses, the parcel delivery category prompt is found different (p<0.05, mean diff: 0.041).
        For the use question about responses, the gaming console category prompt is not found equivalent nor different (mean diff: -0.21).
        For the use question about responses, the investment platform category prompt is found equivalent (p<0.01, mean diff: 0.20).
        For the use question about responses, the laptop category prompt is found equivalent (p<0.05, mean diff: 0.20).
        For the satisfied question about responses, the TV category prompt is found equivalent (p<0.05, mean diff: 0.05).
        For the satisfied question about responses, the ISP category prompt is found equivalent (p<0.01, mean diff: -0.15).
        For the satisfied question about responses, the parcel delivery category prompt is found different (p<0.05, mean diff: 0.55).
        For the satisfied question about responses, the gaming console category prompt is found equivalent (p<0.01mean diff: -0.21).
        For the satisfied question about responses, the investment platform category prompt is found equivalent (p<0.01, mean diff: 0.03).
        For the satisfied question about responses, the laptop category prompt is found equivalent (p<0.01, mean diff: 0.22).
        For the standout question about responses, the TV category prompt is found equivalent (p<0.0001, mean diff: -0.02).
        For the standout question about responses, the ISP category prompt is found equivalent (p<0.0001, mean diff: -0.10).
        For the standout question about responses, the parcel delivery category prompt is found equivalent (p<0.0001, mean diff: -0.06).
        For the standout question about responses, the gaming console category prompt is found equivalent (p<0.0001, mean diff: -0.01).
        For the standout question about responses, the investment platform category prompt is found equivalent (p<0.0001, mean diff: -0.02).
        For the standout question about responses, the laptop category prompt is found equivalent (p<0.0001, mean diff: 0.04).
    }
  \label{tab:comp_stats}
	
\end{table*}}{\begin{table*}[t]
	\centering
	\small
	\setlength\tabcolsep{5pt} % default value: 6pt
	\begin{tabular}{r l l l l l l l l l}
	\toprule
    \midrule
    
    & \multicolumn{4}{c}{\textbf{Prompt}} & ~~ & \multicolumn{4}{c}{\textbf{Reponse}}  \\
	\cmidrule{2-5}\cmidrule{7-10}

          %\midrule
& \multicolumn{1}{c}{Clarity (L7)}  & \multicolumn{1}{c}{Use (L7)}  &  \multicolumn{1}{c}{Bias (L5)}  & \multicolumn{1}{c}{Standout (B)}
&& \multicolumn{1}{c}{Clarity (L7)} & \multicolumn{1}{c}{Use (L7)}  & \multicolumn{1}{c}{Satisfied (L7)} & \multicolumn{1}{c}{Standout (B)} \\ %  & \multicolumn{1}{c}{Pick \%} & \multicolumn{1}{c}{Top \%}  
\midrule
          \textbf{TV}   & -0.23\dubstar        & -0.15\lonestar   & -0.03\fourstar  & -0.08\fourstar          
                        && ~0.06\fourstar       & ~0.00\lonestar    & 0.05\lonestar   & -0.02\fourstar \\ % & 0.00 & 0.00 & \\

          \textbf{ISP}  & -0.28\dubstar        & -0.04\threestar  & -0.06\fourstar  & -0.07\fourstar              
                        && -0.19\fourstar      & -0.08\lonestar   & -0.15\dubstar  &  -0.10\fourstar \\ % & 0.00 & 0.00 & \\
          \textbf{Parcel delivery} & -0.15\threestar & -0.58 & 0.04\fourstar & -0.06\fourstar 
                                && ~0.04\fourstar & \{0.41\lonestar\} & \{0.55\lonestar\} & -0.06\fourstar \\

          \textbf{Gaming console} & \{-0.8\dubstar\} & -0.10\lonestar & -0.01\fourstar & ~0.00\fourstar 
                                && -0.17\fourstar & -0.21  & -0.21\dubstar & -0.01\fourstar \\
        
          \textbf{Investment plat.} & ~0.01\dubstar & -0.10\threestar & -0.07\fourstar & ~0.00\fourstar 
                                && -0.02\fourstar & ~0.20\dubstar & 0.03\dubstar & -0.02\fourstar  \\
        
          \textbf{Laptop}& -0.04\dubstar & \{0.35\lonestar\} & -0.0\fourstar & ~0.04\fourstar 
                        && ~0.10\fourstar & ~0.20\lonestar & 0.22\dubstar & ~0.04\fourstar  \\

          \midrule
          \bottomrule
	\end{tabular}
    \vspace{0.1cm}
	\caption{\small Mean difference between base and perturbed groups among eight questions, four about prompts and four about responses (see~\S\ref{sec:methods:survey} for details). By default test 
    for equivalence is reported (TOST WMU), \{\} indicates test for difference (MWU, CHI$^2$). Higher is better for all differences.
    \lonestar: $p<0.05$, \dubstar : $p<0.01$ \threestar: $p<0.001$, \fourstar: $p<0.0001$. L7: seven-point Likert, L5: five-point Likert, B: binary.}
    \Description{
        This table reports the pairwise test results between perturbed and baseline groups for the eight perceptions questions. 
        The results are reported across multiple dimensions: the product categories, and the individual perceptions questions (four about prompts and four about responses). 
        For each test, up two results can be inferred from the table: 1) results for test of equivalence (TOST MWU p-value, and for effect size mean difference), 
        2) if the test for equivalence is not significant, test for difference (MWU p-value, and and for effect size mean difference).
        From a quick glance, the table shows that out of 48 comparisons reported (6 product categories, 8 questions each), the vast majority (42) were found to be equivalent, 4 different, and 2 neither different nor equivalent.
        Specifically:
        For the clarity question about the prompts, the TV category prompt is found equivalent (p<0.01, mean diff: -0.23).
        For the clarity question about the prompts, the ISP category prompt is found equivalent (p<0.01, mean diff: -0.28).
        For the clarity question about the prompts, the parcel delivery category prompt is found equivalent (p<0.001, mean diff: -0.15).
        For the clarity question about the prompts, the gaming console category prompt is found different (p<0.01, mean diff: -0.15).
        For the clarity question about the prompts, the investment platform category prompt is found equivalent (p<0.001, mean diff: 0.01).
        For the clarity question about the prompts, the laptop category prompt is found equivalent (p<0.001, mean diff: -0.04).
        For the use question about the prompts, the TV category prompt is found equivalent (p<0.05, mean diff: -0.15).
        For the use question about the prompts, the ISP category prompt is found equivalent (p<0.001, mean diff: -0.04).
        For the use question about the prompts, the parcel delivery category prompt is not found equivalent nor different (mean diff: -0.58).
        For the use question about the prompts, the gaming console category prompt is found equivalent (p<0.05, mean diff: -0.10).
        For the use question about the prompts, the investment platform category prompt is found equivalent (p<0.001, mean diff: -0.10).
        For the use question about the prompts, the laptop category prompt is found different (p<0.05, mean diff: 0.35).
        For the bias question about the prompts, the TV category prompt is found equivalent (p<0.0001, mean diff: -0.03).
        For the bias question about the prompts, the ISP category prompt is found equivalent (p<0.0001, mean diff: -0.06).
        For the bias question about the prompts, the parcel delivery category prompt is found equivalent (p<0.0001, mean diff: 0.04).
        For the bias question about the prompts, the gaming console category prompt is found equivalent (p<0.0001, mean diff: -0.01).
        For the bias question about the prompts, the investment platform category prompt is found equivalent (p<0.0001, mean diff: -0.07).
        For the bias question about the prompts, the laptop category prompt is found equivalent (p<0.0001, mean diff: 0.00).
        For the standout question about the prompts, the TV category prompt is found equivalent (p<0.0001, mean diff: -0.08).
        For the standout question about the prompts, the ISP category prompt is found equivalent (p<0.0001, mean diff: -0.07).
        For the standout question about the prompts, the parcel delivery category prompt is found equivalent (p<0.0001, mean diff: 0.06).
        For the standout question about the prompts, the gaming console category prompt is found equivalent (p<0.0001, mean diff: 0.00).
        For the standout question about the prompts, the investment platform category prompt is found equivalent (p<0.0001, mean diff: 0.00).
        For the standout question about the prompts, the laptop category prompt is found equivalent (p<0.0001, mean diff: 0.04).
        For the clarity question about responses, the TV category prompt is found equivalent (p<0.0001, mean diff: 0.06).
        For the clarity question about responses, the ISP category prompt is found equivalent (p<0.0001, mean diff: -0.19).
        For the clarity question about responses, the parcel delivery category prompt is found equivalent (p<0.0001, mean diff: 0.04).
        For the clarity question about responses, the gaming console category prompt is found equivalent (p<0.0001, mean diff: -0.17).
        For the clarity question about responses, the investment platform category prompt is found equivalent (p<0.0001, mean diff: -0.02).
        For the clarity question about responses, the laptop category prompt is found equivalent (p<0.0001, mean diff: 0.10).
        For the use question about responses, the TV category prompt is found equivalent (p<0.05, mean diff: 0.00).
        For the use question about responses, the ISP category prompt is found equivalent (p<0.05, mean diff: -0.08).
        For the use question about responses, the parcel delivery category prompt is found different (p<0.05, mean diff: 0.041).
        For the use question about responses, the gaming console category prompt is not found equivalent nor different (mean diff: -0.21).
        For the use question about responses, the investment platform category prompt is found equivalent (p<0.01, mean diff: 0.20).
        For the use question about responses, the laptop category prompt is found equivalent (p<0.05, mean diff: 0.20).
        For the satisfied question about responses, the TV category prompt is found equivalent (p<0.05, mean diff: 0.05).
        For the satisfied question about responses, the ISP category prompt is found equivalent (p<0.01, mean diff: -0.15).
        For the satisfied question about responses, the parcel delivery category prompt is found different (p<0.05, mean diff: 0.55).
        For the satisfied question about responses, the gaming console category prompt is found equivalent (p<0.01mean diff: -0.21).
        For the satisfied question about responses, the investment platform category prompt is found equivalent (p<0.01, mean diff: 0.03).
        For the satisfied question about responses, the laptop category prompt is found equivalent (p<0.01, mean diff: 0.22).
        For the standout question about responses, the TV category prompt is found equivalent (p<0.0001, mean diff: -0.02).
        For the standout question about responses, the ISP category prompt is found equivalent (p<0.0001, mean diff: -0.10).
        For the standout question about responses, the parcel delivery category prompt is found equivalent (p<0.0001, mean diff: -0.06).
        For the standout question about responses, the gaming console category prompt is found equivalent (p<0.0001, mean diff: -0.01).
        For the standout question about responses, the investment platform category prompt is found equivalent (p<0.0001, mean diff: -0.02).
        For the standout question about responses, the laptop category prompt is found equivalent (p<0.0001, mean diff: 0.04).
    }
  \label{tab:comp_stats}
	
\end{table*}}

%%%% comp_stats_table.tex ends here %%%%

%%%% brand_stats_table.tex starts here %%%%

\aptLtoX{\begin{table*}[t]
	\centering
	\small
	\setlength\tabcolsep{2pt} % default value: 6pt
	\begin{tabular}{r  lll c lll c lll c lll c lll c lll}
\hline
  
     \multicolumn{11}{c}{\textbf{Gemma-it}} & ~~ &  \multicolumn{11}{c}{\textbf{LLama3-it}}\\
	\cmidrule{2-12}\cmidrule{14-24}

	 & \multicolumn{3}{c}{\textbf{TV}} & ~~ & 
	       \multicolumn{3}{c}{\textbf{ISP}} & ~~ &
	       \multicolumn{3}{c}{\textbf{Parcel delivery}} & ~~ &
	       \multicolumn{3}{c}{\textbf{Gaming console}} & ~~ &
	       \multicolumn{3}{c}{\textbf{Invesment plat.}} & ~~ &
	       \multicolumn{3}{c}{\textbf{Laptop}} \\
          %\midrule
\cmidrule{2-4}\cmidrule{6-8}\cmidrule{10-12}\cmidrule{14-16}\cmidrule{18-20}\cmidrule{22-24}
& \multicolumn{1}{c}{P\%} & \multicolumn{1}{c}{A\%} &  \multicolumn{1}{c}{T\%}  
&& \multicolumn{1}{c}{P\%} & \multicolumn{1}{c}{A\%} &  \multicolumn{1}{c}{T\%}  
&& \multicolumn{1}{c}{P\%} & \multicolumn{1}{c}{A\%} &  \multicolumn{1}{c}{T\%}  
&& \multicolumn{1}{c}{P\%} & \multicolumn{1}{c}{A\%} &  \multicolumn{1}{c}{T\%}  
&& \multicolumn{1}{c}{P\%} & \multicolumn{1}{c}{A\%} &  \multicolumn{1}{c}{T\%}  
&& \multicolumn{1}{c}{P\%} & \multicolumn{1}{c}{A\%} &  \multicolumn{1}{c}{T\%}   

\\
\midrule
          \textbf{Base}  & 25.0~       & 33.8          & 35.3~           && 23.9 & 46.2            &  34.3    && 10.6           & 22.7           & 13.6~                 &&         ~0.0          & 18.3          & 0.0  && 14.1          & 38.0          & 14.1           && 5.3 & 61.3 & 1.3 \\
          \textbf{Pert}  & 57.1~        & 80.0          & 82.9~           && 39.1 & 78.2            &  50.7    && 41.4           & 58.6           & 44.3~                 &&        11.2          & 64.8          & 1.4  && 40.3           & 58.4          & 42.9           && 7.1 & 74.3 & 0.0 \\
		\midrule
		  \textbf{Diff} & 32.1\threestar & 46.2\fourstar & 47.6\fourstar   && 15.2 & 31.9\threestar  &  16.4    && 30.8\threestar & 35.8\threestar  & 30.6\threestar       &&        11.2\lonestar & 46.5\fourstar & 1.4  && 26.1\dubstar   & 
20.4\lonestar & 28.8\threestar && 1.8 & 12.9 & -1.3\\ 
\hline
	\end{tabular}
    \vspace{0.1cm}
	\caption{\small \% of responses mentioning the targeted brand. 
	P: What brand the participant (p)icked given a response, 
	A: What are (a)ll recommended brands the participant found, 
	T: What the (t)op recommended brand is to the participant. 
	\lonestar: $p<0.05$, \dubstar : $p<0.01$ \threestar: $p<0.001$, \fourstar: $p<0.0001$.}
	\Description{
		This table reports the percentage of participants that noticed the target brand being promoted. 
		The results are broken down in three metrics of noticing a brand, P: What brand is (p)icked by the user given the response, 
		A:What are (a)ll of the brands recommended in the response, 
		T: What is the (t)op brand recommended in the response. 
		The percentage of participants noticing the brand in the base condition, the perturbed condition, and the difference between the two conditions is given. A test showing if the difference is significant is also reported.
		A quick glance shows that in every condition except ISP and P and T; Gaming console and T; and Laptop and P, A, and T are significantly different (12 out of 18 comparisons). 
		All of the differences are in the expected direction (perturbed prompts lead to more people noticing the target brand) except with Laptop and T.
		Specifically the results are:
		For TV, P base is 25.0\%, perturbed is 57.1\%, the difference is 32.1\% and is significant (p<0.001).
		For TV, A base is 33.8\%, perturbed is 80.0\%, the difference is 42.6\% and is significant (p<0.0001).
		For TV, T base is 35.3\%, perturbed is 82.9\%, the difference is 47.6\% and is significant (p<0.0001).
		For ISP, P base is 23.9\%, perturbed is 39.1\%, the difference is 15.9\% and is not significant.
		For ISP, A base is 46.2\%, perturbed is 78.2\%, the difference is 31.9\% and is significant (p<0.001).
		For ISP, T base is 34.3\%, perturbed is 31.9\%, the difference is 16.4\% and is not significant.
		For parcel delivery, P base is 10.6\%, perturbed is 41.4\%, the difference is 30.8\% and is significant (p<0.001).
		For parcel delivery, A base is 22.7\%, perturbed is 58.6\%, the difference is 35.8\% and is significant (p<0.001).
		For parcel delivery, T base is 13.6\%, perturbed is 44.3\%, the difference is 30.6\% and is significant (p<0.001).
		For gaming console, P base is  0.0\%, perturbed is 11.2\%, the difference is 11.2\% and is significant (p<0.05).
		For gaming console, A base is 18.3\%, perturbed is 64.8\%, the difference is 46.5\% and is significant (p<0.0001).
		For gaming console, T base is  0.0\%, perturbed is  1.4\%, the difference is  1.4\% and is not significant.
		For investment platform, P base is 14.1\%, perturbed is 40.3\%, the difference is 26.1\% and is significant (p<0.01).
		For investment platform, A base is 38.0\%, perturbed is 58.4\%, the difference is 20.4\% and is significant (p<0.05).
		For investment platform, T base is 14.1\%, perturbed is 42.9\%, the difference is 28.8\% and is significant (p<0.001).
		For laptop, P base is  5.3\%, perturbed is  7.1\%, the difference is  1.8\% and is not significant.
		For laptop, A base is 61.3\%, perturbed is 74.3\%, the difference is 12.9\% and is not significant.
		For laptop, T base is  1.3\%, perturbed is  0.0\%, the difference is -1.3\% and is not significant.

	}
  \label{tab:brand_stats}
	
\end{table*}}{\begin{table*}[t]
	\centering
	\small
	\setlength\tabcolsep{2pt} % default value: 6pt
	\begin{tabular}{r  lll c lll c lll c lll c lll c lll}
	\toprule
	\midrule
%	\multirow{2}{*}{\backslashbox{Capability}{Adversary}} &&&&&&&&\multicolumn{2}{c}{\textbf{Server}}\\
    
     \multicolumn{11}{c}{\textbf{Gemma-it}} & ~~ &  \multicolumn{11}{c}{\textbf{LLama3-it}}\\
	\cmidrule{2-12}\cmidrule{14-24}

	 & \multicolumn{3}{c}{\textbf{TV}} & ~~ & 
	       \multicolumn{3}{c}{\textbf{ISP}} & ~~ &
	       \multicolumn{3}{c}{\textbf{Parcel delivery}} & ~~ &
	       \multicolumn{3}{c}{\textbf{Gaming console}} & ~~ &
	       \multicolumn{3}{c}{\textbf{Invesment plat.}} & ~~ &
	       \multicolumn{3}{c}{\textbf{Laptop}} \\
          %\midrule
\cmidrule{2-4}\cmidrule{6-8}\cmidrule{10-12}\cmidrule{14-16}\cmidrule{18-20}\cmidrule{22-24}
& \multicolumn{1}{c}{P\%} & \multicolumn{1}{c}{A\%} &  \multicolumn{1}{c}{T\%}  
&& \multicolumn{1}{c}{P\%} & \multicolumn{1}{c}{A\%} &  \multicolumn{1}{c}{T\%}  
&& \multicolumn{1}{c}{P\%} & \multicolumn{1}{c}{A\%} &  \multicolumn{1}{c}{T\%}  
&& \multicolumn{1}{c}{P\%} & \multicolumn{1}{c}{A\%} &  \multicolumn{1}{c}{T\%}  
&& \multicolumn{1}{c}{P\%} & \multicolumn{1}{c}{A\%} &  \multicolumn{1}{c}{T\%}  
&& \multicolumn{1}{c}{P\%} & \multicolumn{1}{c}{A\%} &  \multicolumn{1}{c}{T\%}   

\\
\midrule
          \textbf{Base}  & 25.0~       & 33.8          & 35.3~           && 23.9 & 46.2            &  34.3    && 10.6           & 22.7           & 13.6~                 &&         ~0.0          & 18.3          & 0.0  && 14.1          & 38.0          & 14.1           && 5.3 & 61.3 & 1.3 \\
          \textbf{Pert}  & 57.1~        & 80.0          & 82.9~           && 39.1 & 78.2            &  50.7    && 41.4           & 58.6           & 44.3~                 &&        11.2          & 64.8          & 1.4  && 40.3           & 58.4          & 42.9           && 7.1 & 74.3 & 0.0 \\
		\midrule
		  \textbf{Diff} & 32.1\threestar & 46.2\fourstar & 47.6\fourstar   && 15.2 & 31.9\threestar  &  16.4    && 30.8\threestar & 35.8\threestar  & 30.6\threestar       &&        11.2\lonestar & 46.5\fourstar & 1.4  && 26.1\dubstar   & 20.4\lonestar & 28.8\threestar && 1.8 & 12.9 & -1.3\\ 
		\midrule
		\bottomrule
	\end{tabular}
    \vspace{0.1cm}
	\caption{\small \% of responses mentioning the targeted brand. 
	P: What brand the participant (p)icked given a response, 
	A: What are (a)ll recommended brands the participant found, 
	T: What the (t)op recommended brand is to the participant. 
	\lonestar: $p<0.05$, \dubstar : $p<0.01$ \threestar: $p<0.001$, \fourstar: $p<0.0001$.}
	\Description{
		This table reports the percentage of participants that noticed the target brand being promoted. 
		The results are broken down in three metrics of noticing a brand, P: What brand is (p)icked by the user given the response, 
		A:What are (a)ll of the brands recommended in the response, 
		T: What is the (t)op brand recommended in the response. 
		The percentage of participants noticing the brand in the base condition, the perturbed condition, and the difference between the two conditions is given. A test showing if the difference is significant is also reported.
		A quick glance shows that in every condition except ISP and P and T; Gaming console and T; and Laptop and P, A, and T are significantly different (12 out of 18 comparisons). 
		All of the differences are in the expected direction (perturbed prompts lead to more people noticing the target brand) except with Laptop and T.
		Specifically the results are:
		For TV, P base is 25.0\%, perturbed is 57.1\%, the difference is 32.1\% and is significant (p<0.001).
		For TV, A base is 33.8\%, perturbed is 80.0\%, the difference is 42.6\% and is significant (p<0.0001).
		For TV, T base is 35.3\%, perturbed is 82.9\%, the difference is 47.6\% and is significant (p<0.0001).
		For ISP, P base is 23.9\%, perturbed is 39.1\%, the difference is 15.9\% and is not significant.
		For ISP, A base is 46.2\%, perturbed is 78.2\%, the difference is 31.9\% and is significant (p<0.001).
		For ISP, T base is 34.3\%, perturbed is 31.9\%, the difference is 16.4\% and is not significant.
		For parcel delivery, P base is 10.6\%, perturbed is 41.4\%, the difference is 30.8\% and is significant (p<0.001).
		For parcel delivery, A base is 22.7\%, perturbed is 58.6\%, the difference is 35.8\% and is significant (p<0.001).
		For parcel delivery, T base is 13.6\%, perturbed is 44.3\%, the difference is 30.6\% and is significant (p<0.001).
		For gaming console, P base is  0.0\%, perturbed is 11.2\%, the difference is 11.2\% and is significant (p<0.05).
		For gaming console, A base is 18.3\%, perturbed is 64.8\%, the difference is 46.5\% and is significant (p<0.0001).
		For gaming console, T base is  0.0\%, perturbed is  1.4\%, the difference is  1.4\% and is not significant.
		For investment platform, P base is 14.1\%, perturbed is 40.3\%, the difference is 26.1\% and is significant (p<0.01).
		For investment platform, A base is 38.0\%, perturbed is 58.4\%, the difference is 20.4\% and is significant (p<0.05).
		For investment platform, T base is 14.1\%, perturbed is 42.9\%, the difference is 28.8\% and is significant (p<0.001).
		For laptop, P base is  5.3\%, perturbed is  7.1\%, the difference is  1.8\% and is not significant.
		For laptop, A base is 61.3\%, perturbed is 74.3\%, the difference is 12.9\% and is not significant.
		For laptop, T base is  1.3\%, perturbed is  0.0\%, the difference is -1.3\% and is not significant.

	}
  \label{tab:brand_stats}
	
\end{table*}}

%%%% brand_stats_table.tex ends here %%%%

%%%% demo_table.tex starts here %%%%

\aptLtoX{\begin{table}[h!]
    \centering
    \small
    \begin{tabular}{p{2cm} p{3.5cm} p{1cm} p{1cm}}
\hline
    \textbf{Gender}         & Male                    & 48.2 \\
                            & Female                  & 50.3 \\
                            & Self-described          & 0.9 \\
    \midrule
    \textbf{Age}            & 18-25                   & 16.7 \\
                            & 26-35                   & 36.0 \\
                            & 36-45                   & 21.5 \\
                            & 46-60                   & 18.2 \\
                            & 61+                     & 5.2 \\
    \midrule
    \textbf{Ethnicity}      & White                   & 63.0 \\
                            & Black or African Am.    & 12.1 \\
                            & Asian                   & 9.8 \\
                            & Hispanic or Latino      & 5.6 \\
                            & Other or mixed          & 12.7 \\
    \midrule
    \textbf{Education}      & Completed H.S. or below  & 10.4 \\
                            & Some college, no degree  & 18.0 \\
                            & Trade or vocational      & 2.5 \\
                            %& Associate's or Bachelor's & 50.2 \\
                            & Associate's degree       & 11.1 \\
                            & Bachelor's degree        & 39.1 \\
                            & Master's or higher       & 18.5 \\
    \midrule
    \textbf{Chatbot usage} & Daily or more freq.     & 16.6 \\
     \textbf{frequency}    & Daily to monthly        & 49.7 \\
                            & Monthly or less freq.  & 33.7 \\
    \midrule
    \textbf{ChatGPT}     & A lot          & 57.2 \\
    \textbf{familiarity} & A little       & 40.7 \\
                         & Nothing at all & 2.1 \\
    \hline
    \end{tabular}
    \vspace{0.1cm}
    \caption{\small Demographics. 
    May not total 100\% (rounding, opt-outs).}    
    \Description{
        This table reports the demographics of the user study participants. 
        It reports the distribution of Gender (48.2\% Male, 50.3\% Female, and 0.9\% Self-described), 
        Age (16.7\% 18-25, 36.0\% 26-35, 21.5\% 36-45, 18.2\% 46-60, and 5.2\% 61 and older), 
        Ethnicity (63\% white, 12.1\% black or african american, 9.8\% asian, 5.6\% hispanic or latino, and 12.7\% other or mixed), 
        Education attainment (10.4\% completed high school. or below, 18.0\% some college but no degree, 2.5\% trade or vocational, 11.1\% associate’s degree, 39.1\% bachelor’s degree, and 18.5\% master’s or higher), 
        Chatbot usage frequency (16.6\% daily or more frequent, 49.7\% daily to monthly, 33.7\% monthly or less frequent), and ChatGPT familiarity (57.2\% a lot, 40.7\% a little, and 2.1\% nothing at all).
    }
    \label{tab:demographics}
\end{table}}{\begin{table}[h!]
    \centering
    \small
    \begin{tabular}{p{2cm} p{3.5cm} p{1cm} p{1cm}}
    \toprule
    \midrule
    \textbf{Gender}         & Male                    & 48.2 \\
                            & Female                  & 50.3 \\
                            & Self-described          & 0.9 \\
    \midrule
    \textbf{Age}            & 18-25                   & 16.7 \\
                            & 26-35                   & 36.0 \\
                            & 36-45                   & 21.5 \\
                            & 46-60                   & 18.2 \\
                            & 61+                     & 5.2 \\
    \midrule
    \textbf{Ethnicity}      & White                   & 63.0 \\
                            & Black or African Am.    & 12.1 \\
                            & Asian                   & 9.8 \\
                            & Hispanic or Latino      & 5.6 \\
                            & Other or mixed          & 12.7 \\
    \midrule
    \textbf{Education}      & Completed H.S. or below  & 10.4 \\
                            & Some college, no degree  & 18.0 \\
                            & Trade or vocational      & 2.5 \\
                            %& Associate's or Bachelor's & 50.2 \\
                            & Associate's degree       & 11.1 \\
                            & Bachelor's degree        & 39.1 \\
                            & Master's or higher       & 18.5 \\
    \midrule
    \textbf{Chatbot usage} & Daily or more freq.     & 16.6 \\
     \textbf{frequency}    & Daily to monthly        & 49.7 \\
                            & Monthly or less freq.  & 33.7 \\
    \midrule
    \textbf{ChatGPT}     & A lot          & 57.2 \\
    \textbf{familiarity} & A little       & 40.7 \\
                         & Nothing at all & 2.1 \\
    \hline
    \bottomrule
    \end{tabular}
    \vspace{0.1cm}
    \caption{\small Demographics. 
    May not total 100\% (rounding, opt-outs).}    
    \Description{
        This table reports the demographics of the user study participants. 
        It reports the distribution of Gender (48.2\% Male, 50.3\% Female, and 0.9\% Self-described), 
        Age (16.7\% 18-25, 36.0\% 26-35, 21.5\% 36-45, 18.2\% 46-60, and 5.2\% 61 and older), 
        Ethnicity (63\% white, 12.1\% black or african american, 9.8\% asian, 5.6\% hispanic or latino, and 12.7\% other or mixed), 
        Education attainment (10.4\% completed high school. or below, 18.0\% some college but no degree, 2.5\% trade or vocational, 11.1\% associate’s degree, 39.1\% bachelor’s degree, and 18.5\% master’s or higher), 
        Chatbot usage frequency (16.6\% daily or more frequent, 49.7\% daily to monthly, 33.7\% monthly or less frequent), and ChatGPT familiarity (57.2\% a lot, 40.7\% a little, and 2.1\% nothing at all).
    }
    \label{tab:demographics}
\end{table}}

%%%% demo_table.tex ends here %%%%

\subsubsection{Experimental groups}
\label{sec:methods:groups}
Our overall goal was to find whether people notice differences %\lujo{whether people perceived any difference(?)}
between our perturbed prompts 
\revision{(and corresponding responses)} compared to 
base prompts. 
%For each of our 447 base product recommendation prompts, 
%we found a perturbed prompt for each possible target 
%brand using the loss of each of our models. 
%This resulted in 
\revision{
Due to prohibitive cost, we 
could not test all of our 1,809 base and perturbed
(brand) prompt pairs from ~\S\ref{sec:measurement} in the user study.}
%Because of the large number of prompt pairs, 
%we could not test all of them in the user study.
\revision{Instead}, 
we selected six pairs of 
prompts to use in the user study,
each pair from a different product category, giving us 12 prompts total.\footnote{\revision{We aimed 
to detect ``medium'' effect sizes with 80\% power 
at $\alpha=0.05$ for two tailed Mann-Whitney U,
requiring $\sim$75 participants per group.}}
Because each pair of prompts belonged to a unique product category, 
we refer to them by their product categories in the results.
% However, our prompts have to exist in the context of 
% product categories. Thus we chose six product categories
% to test in our user study and selected 
% one perturbed and baseline prompt for each category, creating a total of 12 study groups.
%\anna{I commented out three four lines in the tex file - i don't think they were clear enough and i feel like the focused too heavily on the product category.}
%\anna{I was ucnlear on what ``However, our prompts have to exist in the context of product categories'' means.}
The prompt pairs were split between two models, 
three pairs for Llama3-it and three for Gemma-it.
We focused on these more user-friendly 
instruction-tuned models since \revision{they 
display chatbot-like behavior}~\cite{ouyang2022training}.
%The pairs selected met the following criteria: %\anna{changed product categories to pairs, bc third item does not apply to product categories}: 
\revision{To increase the realism of the study 
we set the following criteria when picking which prompt pairs 
to use:}
\begin{itemize}[itemsep=1pt, left=0pt]
    \item \revision{Prompts pairs from product categories} among the top 50\% 
    of product categories that participants reported caring about. 
    \revision{This increases the chances that participants would have 
    shopped for the product outside an experimental environment.}
    \item \revision{Prompts pairs from product categories} 
    among the top 50\% of product categories that participants 
    reported they might use an LLM when shopping for. 
    \revision{This increases the chances that participants 
    would have used an LLM when shopping for the product. 
    Combined with the previous criteria, we selected the product 
    categories that are most likely to be investigated with an 
    LLM in a real-world scenario.}
    \item \revision{The prompt pairs} had the highest probability increase that the 
    target brand is mentioned with the attack, as measured 
    in~\S\ref{sec:results:blackboxattack}, 
    \revision{mimicking the type of prompts an attacker 
    might choose to deploy for the highest effectiveness.}
    %\item they were not repetitive - i.e., we filtered out baseline/perturbed prompt pairs even if they met the above criteria if another pair that belonged to their same category was being used \anna{maybe this is not necessary}
    % \item the most successful categories our perturbed prompts were able to attack (as measured in~\omer{FILL IN!}).
\end{itemize}

% For each product category/prompts pair (i.e., study group), 
% the ratio in our larger experiments. 
%%For each participant within a study group, the prompt 
%%\anna{changed `product category and prompts' to `prompt' bc each prompt is tied to its category} 
%%remained the same\lujo{the same as what?}, but the response was unique.
All participants within a group saw the same prompts, but,
mirroring real-world chatbots, each
participant was shown a unique full-length response.
\revision{For each of the 12 prompts,
we obtained a random sample of $\sim$75
model responses from our earlier experiments
with the models (\S\ref{sec:results:whiteboxattack}). 
This sample was stratified to ensure the 
ratio of responses that mentioned the target was the same 
in this set of $\sim$75 
as it was in the overall set of 1000.} %As opposed to~\S\ref{sec:results}, 
%\revision{Participants} saw full-length responses.%\lujo{is my rephrasing correct?}  
%We chose this way of  to 
%realistically mirror the nature of how our distributional attack
%works.\lujo{no longer clear what ``this approach'' encompasses. avoid
%  ``this'' in favor of writing down what ``this'' actually refers to}

\subsubsection{Piloting and preliminary data collection}
% label
We piloted our study extensively. % before the main study. 
We ran a series of preliminary studies to determine whether 
the study design was feasible and whether participants would encounter
issues. \revision{To detect potential problems,}
questions were timed and augmented with meta-questions on how clear 
the main questions were.
We also collected participants' interest in product categories and 
their likelihood of using chatbots when shopping for these 
categories. In total, we collected responses from 90 Prolific 
participants for piloting and 63 for product category preferences.
%\lujo{unclear if ``this'' purpose is
%  just the interest in categories and chatbots or also the pilots}
We further ran pilots with two HCI researchers, asking them to 
review and criticize our study.
%\lujo{it is standard to write out
%  numbers smaller than 10 (or some arbirary other number, like 12)}

Based on responses, we made the study more concise with 
clearer instructions. Mimicking chatbots, we 
changed how prompts and responses were 
displayed. % to be more realistic of how chatbots are used in the wild.
Attention and comprehension checks were added to ensure high data 
quality.

\subsubsection{Statistical analysis}
For each of the six base and perturbed prompt pairs, 
% For each category,
we analyzed the difference between base and perturbed group responses
% for each baseline and perturbed prompt pair
with a series of non-parametric tests on our 11 main measurement variables: two-tailed 
Mann-Whitney U tests for Likert data and chi-squared tests for binary data.
%Each question asked about prompts and responses were analyzed
%separately.\lujo{can't parse. what was analyzed separately?} 
To understand whether the base and perturbed groups are equivalent, 
we replaced non-significant tests for difference with tests of equivalence. 
To establish equivalence, we used the two one-sided
tests procedure (TOST) and set our equivalence 
margin to be $\Delta=0.5$ for 80\% 
power~\cite{lakens2017equivalence}.\footnote{Testing 
for equivalence is a deviation from the pre-registration; 
however, we believe this approach paints a more complete picture.}
Brand-recall questions (e.g., brand 
participants pick based on the response), 
were coded into binary categories: was the targeted brand reported or not. %(products of the target brand were also included, 
Hesitant or unclear responses did not count as matches.

\revision{Because of our extensive testing (11 tests between the base 
and perturbed prompt per product category), 
we controlled our false-discovery rate per product} 
category with the Benjamini-Hochberg procedure~\cite{benjamini1995controlling}.

\subsection{Results}
\label{sec:user:results}
\revision{Our data show} that, under most measures, 
perturbed prompts and corresponding responses are inconspicuous.
Further, these prompts successfully nudge more users 
into noticing the target brand in most measures, \revision{fulfilling the adversarial objectives.} %\lujo{what's the two sentence takeaway?}

\subsubsection{Participants}
We recruited 845 participants and evenly distributed them to 12 groups, 
each group defined by the product category and prompt type (base or
perturbed).
%\lujo{this is confusing. one recruits participants across
%  age groups, for example. this sounds like we recruited XXX and
%  assigned each participant to one of 12 groups. correct?}
Product categories were split between Llama3-it and Gemma-it. 
Our participants were more educated, younger, 
less Hispanic, and had more familiarity 
with ChatGPT than the national 
average~\cite{uscensus2020,pew2024views}. 
\revision{\autoref{tab:demographics} 
summarizes the demographics of our participants.}

\subsubsection{Statistical Evaluation}
\label{sec:results:user}

We asked \revision{11 core} questions about the prompts and responses. 
Four of these were to test the inconspicuousness of the perturbed prompts. 
Another four were to test if users were significantly more 
dissatisfied with the perturbed responses. We further asked 
three questions to test if \revision{our perturbations} 
made the targeted brand more perceptible.
\autoref{tab:comp_stats} and~\autoref{tab:brand_stats} show the results.
%\lujo{do we include our study script somewhere? if
%  not, we should (or offer to make it available if we don't have space
%to include it)} \omer{we should probably drop it in osf, but I don't know if 
%we have the time today.}

\paragraph{Equivalence} 
\revision{From participants' perspectives,}
nearly all perturbed prompts 
and responses were equivalent to corresponding base prompts and responses 
in terms of variables measured (42/48 comparisons were equivalent, $p<0.05$) 
with the following exceptions: 
statistical tests showed no difference or equivalence
%\lujo{which one of these two? do you mean
%  ``statistical tests showed neither difference nor equivalence''?} 
for the likelihood of using the parcel delivery prompts ($p>0.05$); 
responses to the parcel delivery perturbed prompt were more satisfactory and more 
likely to be used (all $p<0.05$); the perturbed gaming console prompt
was less clear than the base prompt ($p<0.05$);
no difference nor equivalence
%\lujo{same comment as for first} 
was found for the 
likelihood of using the response for gaming platforms ($p>0.05$);
the laptop perturbed prompt was more likely to be used ($p<0.05$). 
These results suggest that not only is our 
attack imperceptible to users, 
but in a few cases the perturbed prompts 
and responses might be preferable.

\paragraph{Attack success} In order to measure attack success, we recorded 
the percentage of participants who \emph{would pick} the targeted 
brand given the response, the percentage of participants who 
\emph{noticed} the brand in the response, and the percentage of participants who
said the targeted brand \emph{was the top recommendation}. 
%We see \lujo{avoid using "we see" and similar when refering to tables
%and figures. what _we_ see is irrelevant, and what readers see we
%have no control over. we only control what the figures and tables _show_}
As shown in
\autoref{tab:brand_stats} \anna{removed parenthesis?}, in five out of six 
categories our attacks were able to increase the prominence of the
target brand in at least 
one of the three questions:
in four categories participants 
were more likely to pick the targeted brand when given the perturbed prompts, 
in five categories participants were more likely to notice the targeted brand, 
and in three categories participants were more 
likely to say the targeted brand was the top recommendation.

\revision{
\paragraph{Effect of earlier brand appearance}
Attack success is likely dependent on various factors, 
including how early the target is mentioned in the response. 
We made this assumption when generating perturbed 
prompts through synonym replacement. 
%we assumed that earlier mention of a target concept would 
%be more likely to influence users. 
Thus, we used a loss function that increases the probability 
of a target concept being mentioned at the beginning of the response. 
Here, we run an exploratory (not pre-registered) analysis on the user study 
data to test this assumption.}

\revision{
Specifically, we ran a series of logistic regressions
to predict whether participants would notice the target brands based on 
how soon the target brand was mentioned in the response. 
We formulated the model where 
correctly noticing the target brand is the dependent variable 
and the position of the target brand 
in the response is the independent variable 
(if ``Verizon'' is the 15th word, the position would be 15). 
We included a random effect for the prompt used, since product 
categories and base prompt phrasing might have different baseline effects.
We find that, later mentions of target brands in 
responses are less likely to be noticed by participants. 
This observation remains true for all three measurements 
of attack success: what brand is picked, what the top 
perceived recommendation is,
and if the brand is perceived as recommended by the LLM. 
A 10-word increase in the position of the target brand 
reduces the chances of users noticing the target brand 
by 1.7\%, 5.9\%, and 0.6\% respectively (all $p<0.0001$).
}

\revision{
\subsubsection{Takeaways}
\label{sec:user:takeaway} We show that our synonym-based attack can 
practically shift users towards a concept of the attacker's choosing. 
In nearly all measures, perturbed prompts and responses were statistically 
indistinguishable—or even occasionally preferred—compared to their 
base counterparts. Meanwhile, in five of six product categories, 
participants were more likely to notice or choose the attacker's target 
brand when given perturbed prompts. Taken together, this study verifies 
the assumptions and results of our earlier experiments.
Though we only tested this attack on a benign case (brands) to avoid harm, 
the results show potential for harm and serious 
implications on user autonomy, which we discuss next.
}

%%%% user_study.tex ends here %%%%

%%%% conclusion.tex starts here %%%%

\section{Discussion}
\label{sec:discussion}

We identify a novel threat model in which an adversary
aims to induce biases in LLM responses users receive 
by using adversarially crafted but innocuous-seeming prompts. We 
empirically show that, when generated with the right method, 
users do not notice that these prompts are adversarial 
but are influenced by the biased responses that the prompts result in. Here, we first discuss the 
implications of our findings (\S\ref{sec:discussion:implications}) and suggest potential defenses (\S\ref{sec:discussion:defenses}).
\revision{Then we attempt to explain why our attacks work (\S\ref{sec:discussion:attributes}).
%  \lujo{the word ``attributes''
%    here is insufficient. i haven't the slightest idea what it refers to} of the attacks.
Finally, we end with an economic analysis in \S\ref{sec:discussion:costs}.}

\subsection{Implications}
\label{sec:discussion:implications}
This risk we highlight in this work stems from 
real-world deployments of LLMs in chatbots and 
the accompanying prompt providers, \revision{e.g., Amazon Rufus \cite{rufus2024}}. As such, 
it has real-world implications. Users may be 
manipulated into thinking a certain way, 
while being under the impression that they are receiving 
unbiased advice. Prompt providers 
could be giving the impression of 
a personalized experience, while subtly
undermining user autonomy.
This risk can fundamentally be thought of as 
a risk of intentional algorithmic bias, which is well understood to be particularly
dangerous in the context of political and social
issues~\cite{blodgett2022responsible, weidinger2021ethicalsocialrisksharm}, 
but also concerning in commercial 
contexts~\cite{akter2022algorithmic}. As outlined in prior 
work, this type of misuse can produce effective propaganda 
and misinformation, resulting in emotional and financial 
distress~\cite{weidinger2021ethicalsocialrisksharm}. Notably, 
LLMs already disseminate propaganda~\cite{myers2025deepseek}--prompt 
providers may follow suit at a lower cost.
We argue that the seriousness of this risk necessitates 
defensive measures, likely requiring an ensemble 
of defenses to be effective.

\subsection{Defenses: Multi-Pronged}
\label{sec:discussion:defenses}
We identify a series of complementary defenses---with various 
tradeoffs between effectiveness and deployment cost---that different 
stakeholders can employ to mitigate the risk of adversarial prompts.
%As such, we base some of our suggestions 
%on the defense mechanisms from this domain.\lujo{clarify ``this'' domain. stay away from ``this'' when possible} 
%The suggested defenses are complementary, and can, and we believe should be, used in conjunction.

\paragraph{User warnings, labeling, and education}
Using prompts from untrusted sources is akin to running 
code copied from untrusted sources, a well-studied problem in computer
security~\cite{bai2019qualitative, fischer2017stack}. \omer{more cites, not entirely sure
  if this connection is worth making} \anna{I like it, but it does sound a bit strong}  
Thus, similar protection mechanisms, like warnings~\cite{fischer2019stack}, might be effective.
However, unlike untrusted code, inconspicuous attacks are difficult to detect and therefore might 
require more invasive warnings.
Security warnings have a long history in
human-centered design~\cite{reeder2018experience, zhang2014effects, gorski2020listen, bravo2013your},
showing they can be integrated into the user interface to possibly help users make more informed decisions. 
These warnings could appear in various locations depending on the owner of the prompt provider. If the prompt provider is not the 
chatbot owner, the chatbot could warn users about the potential risks of using prompts from untrusted sources. 
Notably, certain chatbots already warn users, e.g., ChatGPT warns, ``ChatGPT can make mistakes. Check important
info.''~\cite{verge2024}. Further, prompt libraries can warn users about the potential risks of using prompts from other users, 
similar to proposed warnings in programming forums~\cite{fischer2019stack}.

Warnings tend to be reactive, appearing only when an emergency arises. In contrast, users 
can be proactively educated with ``nutrition'' labels~\cite{kelley2009nutrition}, 
a standardized summary of what users need to pay attention to (e.g., privacy practices) before using 
a product. Notably, privacy-relevant nutrition labels have had widespread 
adoption\footnote{Google Play and Apple App Store made privacy nutrition labels mandatory for apps.}
and has been shown to be effective in 
multiple computing contexts, including mobile applications and IoT devices~\cite{pardis2022nutrition, balash2024would}. 
Similar labels for the capabilities and pitfalls of LLMs could be developed and publicized.

Labels and warnings, however, are likely only part of the solution. More involved 
user education campaigns on how to use LLMs and chatbots safely 
and effectively might need to be developed~\cite{zamfirescu2023johnny}.

%\lujo{probably hard to avoid talking about user education and
%  warnings, but it's not very inspiring}

\paragraph{Robust models}
Our attacks fundamentally rely on LLMs (like other deep neural networks) being
brittle, a fact we also observe in our work (\S\ref{sec:results:blackboxattack}).
%\lujo{i'm not sure we should call them brittle. maybe avoid
%negative judgments about LLMs that aren't supported by our work and
%instead just focus on the facts} 
They respond to small changes in prompts in a way that is 
unexpected and can be manipulated. Despite the claims of robustness by many LLM providers, our work (and many others) show 
that there is still much work to be done. 
Existing robust LLMs focused on the correctness of LLM answers~\cite{ndss24:LMSanitator,ndss24:TextGuard,iclr24:certifying,arXiv24:AutoDefense,arXiv24:llamaguard,emnlp23:nemo}.
To the best of our knowledge, we are the first to explore how this model fragility
allows slight (inconspicuous) perturbation in LLM prompts to vastly 
different probabilities of LLMs mentioning target concepts.
%\lujo{i think inconspicuousness is
%key, because many attacks can be argued to involve ``slight''
%  changes to prompts}
Such differences may lead to bias and risks that hurt user autonomy, including but not limited to 
advertising without user awareness, misinformation, and propaganda, as we showed in this paper. 
We suggest LLM providers emphasize robustness against these defenses. 
However, given the rapid---perhaps rushed---deployment of LLMs in the wild (e.g.,~\cite{verge2024gemini}), 
it might be up to regulators to enforce robustness requirements~\cite{WhiteHouse2023}.

\paragraph{New bias metrics}
Our work does not use prior definitions of bias~\cite{FAccT21:definition,  FAccT22:definition} during evaluation. 
This is intentional: existing bias metrics focus on discrimination, hate speech, and exclusion,
 but do not capture the type of bias we introduce in this work.
For instance, many brands could be mentioned in an LLM response, and
an ideal bias metric should be able to capture how much each one can be biased.
This suggests that new metrics need to be developed to capture this type of bias, 
enrich the definition of LLM robustness,
and 
eventually, make testing meaningful.

\paragraph{\revision{Continuous audits}}
In the absence of guaranteed robustness, an elusive target, frequent testing of models for bias and other risks is essential.
We advocate for systematic testing of prompt providers for bias, including the type of bias we introduce in this work.
Such a system could regularly check if prompts are biased towards certain concepts, and if so, 
alert users. The effectiveness of such a system would fundamentally depend on the quality of 
the bias detection tests. 
% This brings us to our next suggestion.

We further find that running measurements on LLMs is difficult due to the variability in responses.
While this is a feature that LLM developers intentionally build, 
it also complicates the measurement of bias. How can we measure the underlying 
bias in the model? Is simply averaging numerous responses enough? We advocate for 
future research to explore this question.

\subsection{\revision{Why Does the Attack Work?}}
\label{sec:discussion:attributes}
\revision{ As we illustrate in \S\ref{sec:results:whiteboxattack},
  our synonym replacement method has a much higher max absolute improvement
  compared to the average absolute improvement, suggesting the
  approach works particularly well with specific prompts and
  synonyms.
  We hypothesize that one reason this may be caused by the synonym occurring in
  close proximity to the promoted concept
  in training instances.
  %% This may correspond to similarly seen training instances
  %% where the synonym and the promoted concept might have appeared
  %% closer in proximity.
  For example, when the prompt includes the word
  `\textit{reliable}' and asks \llms to recommend a streaming service,
  \llms almost always recommend the brand ``'\textit{Netflix}'', and
  `\textit{Netflix}' is commonly associated with `\textit{reliable}'
  in online text,\footnote{E.g.,
  \url{https://netflixtechblog.com/keeping-netflix-reliable-using-prioritized-load-shedding-6cc827b02f94},
  \url{https://www.forbes.com/sites/rosaescandon/2020/05/19/netflix-is-the-most-reliable-streaming-service-new-survey-shows},
  and
  \url{https://www.infoq.com/articles/netflix-highly-reliable-stateful-systems/}. All
  visited on Dec 3, 2024.} while other streaming services are less
  often. However, we are not able to identify such words (or word
  combinations) for every prompt where our synonym replacement works
  particularly well, suggesting that there may be other factors at play. 
  Our attacks can be thought of as adversarial examples, which are not fully understood.}

\subsection{\revision{Economic Analysis}}
\label{sec:discussion:costs}
%\lujo{i don't think this is related to ecological validity. i think
%  this is an economic analysis of synonym-replacement compared to
%  paraphrasing via LLMs}\weiran{Addressed}

\revision{
As we described in \S\ref{sec:threatmodel}, adversaries may run 
attacks we discuss in this paper to perform 
advertising without user awareness.
%These adversaries may be
%e-commerce service providers, who also
%develop side products such as prompt-optimization engines.
Prompt-optimization engines may advertise products on behalf of others
%Such attackers may hope that their prompt-optimization engines, which are open to
%customers, may act as recommendation engines and 
to generate revenue.
In the following paragraphs, we briefly analyze the potential economic
incentives of such attacks, 
we list benefits and current costs to run 
attacks (i.e., synonym replacement and paraphrasing) 
compared to more traditional advertising methods.
%according to current prices of advertising and 
%running \llms. 
We also compute the number of 
increased mentions of the target concept needed for 
adversaries to make profits. 
}

\revision{
%\lujo{at the start, expalain what you're trying to show, and
%   tell us who needs to pay what under what
%  circumstances, for some case A and some case B. only later work out the math} 
%  \weiran{Addressed}
  At the time of writing, the cost per mille (CPM),
  i.e., price to show an ad $1,000$ times 
  %(regardless of clicks) 
  is $\$4-\$10$ on Meta, Youtube, Snapchat, and TikTok\footnote{According to
  \url{https://www.guptamedia.com/social-media-ads-cost}, visited on
  Dec 2 2024.}, making the price to show a single ad
  $\$0.004-\$0.01$. %\lujo{in the US, \$ goes before the value.}
  %\lujo{switch the ranges to be formatted as i did earlier in this paragraph}\weiran{Addressed}
  The price for running an LLMs is
  $\$0.04 - \$30$ for one million input tokens and $\$0 - \$75$ for
  one million output tokens\footnote{According to
  \url{https://llm-price.com/}, visited on Dec 2 2024.}. 
  %For
  %example, the cost to run Mistral (7B), an open-sourced model, on
  %AWS, is $\$0.15$ for one million input tokens and $\$0.20$ for one
  %million output tokens.  
%
%  The cost to run GPT4 is $\$30$ for one
%  million input tokens and $\$60$ for one million output
%  tokens\footnote{Double checked at
%  \url{https://help.openai.com/en/articles/7127956-how-much-does-gpt-4-cost},
%  visited on Dec 2 2024.}.\lujo{don't need this second sentence --
%    just provide both citations for the first sentence, or integrate
%    some example from the second into the first}  
  Adversaries need to pay the cost for
  generating the adversarial prompt (either by paraphrasing or synonym
  replacement).  The victims will take the adversarial prompt and run
  on \llms by themselves (as we defined in \S\ref{sec:threatmodel})
  so there is no additional cost after the prompt is released, 
  regardless of how many times the adversarial prompts
  are used.  }

\revision{ Our prompts 
  %any unperturbed, paraphrased, or
  %synonym-replaced prompts, 
  are no longer than $400$ tokens. The
  synonym replacement (\S\ref{sec:approach:whiteboxattack} and
  \S\ref{sec:results:whiteboxattack}) only needs to perform inference
  once (i.e., compute one output token), and therefore, the cost can be up to
  $\$0.15/1,000,000*400+0.20/1,000,000=0.0000602$ for Mistral and
  $\$30/1,000,000*400+75/1,000,000=0.012075$ with the most expensive token prices.
  %although synonym replacement may not apply to close-sourced API-only
  %models as we explained in \S\ref{sec:setup:LLM} 
  Adversaries only
  need one more mention of the target concept (as $\$0.0000602$ is
  less than $\$0.004$) on Mistral or up to $0.012075/0.004 \approx 3$
  more mentions of the target concept for any other models to have
  more benefits than cost. The synonym replacement achieves more than
  $0.3\%$ average absolute improvement on each model we used
  (Fig.~\ref{fig:results:whitebox:short001}), i.e., more than three more
  mentions of the target concepts within $1,000$ uses of the prompt.
 Thus, adversaries can run much more efficient advertising 
 using our synonym replacement approach.}

\revision{ In contrast, if adversaries only use paraphrasing
  (\S\ref{sec:approach:blackboxattack}), 
  %albeit they might find more
  %attacks on specific prompts (\S\ref{sec:results:blackboxattack}),
  the cost is much higher: with the same setup we used
  (\S\ref{sec:setup:evaluation:paraphrase}), adversaries may need to
  generate up to $64$ tokens $1,000$ times for an average of $2,207/524
  \approx 4.21$ prompts (including the unperturbed prompts and
  paraphrases). The cost can be up to
  $\$(0.15/1,000,000*400+0.20/1,000,000*64)*1,000*4.21=0.306488$ on Mistral
  and $\$(30/1,000,000*400+75/1,000,000*64)*1,000*4.21=70.728$ in the worst
  case scenario. Up to $77$ more mentions on Mistral and $17,682$ more
  mentions of the target concept in the worst case on other models are
  needed for the adversaries to have more benefits than costs. As we
  suggested in \S\ref{sec:approach:blackboxattack}, this makes 
  paraphrasing more time- and cost-intensive than 
  synonym replacement.} 
  \revision{The number of more mentions 
  we compute are all upper bounds, and whether
  adversaries can make profits by only paraphrasing depends on
  specific prompts, models used, and the number of times that the
  prompts will be used.}

%\lujo{how about a tool that you could run that would take a prompt as
%  input and suggest alternative prompts (that would be rephrased
%  without trying to achieve a specific objective)?}

%\lujo{we could suggest auditing services that test the prompts
%  suggested by prompting services for the kinds of biases we
%  detect. the auditing services would check that the specific prompts
%  aren't outliers in terms of their answers compared to rephrasings of
%those prompts} \omer{I like this idea}

%\lujo{maybe prompting services should always suggest multiple variants
%of a prompt and let people choose one, or maybe they should allow
%users to rephrase a prompt as many times as they want}

%\lujo{i think our work also points to the need to better understand
%  the distributions of responses LLM give on a broad range of
%  topics. people talk about over bias a lot, but our work suggests
%  that LLM reponses in general can't be trusted to accurately reflect
%  the underlying data in a manner consistent to human intuition}
%\new{ideas: 
%Measuring bias is hard with LLM because they produce something different every time (sometimes small diffs sometimes large). The input and output cutoff are params 
%that complicate this problem. This is a feature of LLMs that ends up being a problem for measurement.}

%\lujo{maybe a point to bring out in this paragraph that our work
%  highlights the need for yet broader definitions of robustness in the
%efforts that seek to make LLMs robust}
%\new{existing metrics don't work to measure this type of bias. We need to develop new metrics.}

\section{Limitations}
\label{sec:limitations}
We explore methods to make LLMs mention a specific concept while remaining inconspicuous to users and our results may depend on many factors.
The concepts and brands we chose might not be realistic of real world use-cases. To address this, we make the case that the use of LLMs when shopping 
is already a reality and collect a list of products that users had the highest likelihood of shopping for with LLMs (\S\ref{sec:methods:groups}). 

Our LLM use is bound to specific temperature settings. We address our temperature choice in \S\ref{sec:setup:LLM} and explore 
the effect of different temperature settings in App.~\ref{sec:results:temperature}.

We also were only able to test our attack on a small set of LLMs, all of which are open-source.
This means we cannot evaluate the success of our attack on all LLMs, including popular closed-source models.
We address this by exploring transferability in \S\ref{sec:results:transfer}.

Our assumptions of what change in resulting prompts would influence users might be flawed. We test (some of) our assumptions in the 
user study (\S\ref{sec:user}) and find them to be reasonable.

Like all user studies, ours is limited in a multitude of ways. Our sample might not be representative of 
general LLM users. However, we only find a small minority of our participants to not be familiar with chatbots (only 2.1\% knows nothing about ChatGPT). 
Though we phrased the initial study advertisement generically, participants might 
have self-selected themselves, limiting generalizability. Our study was run on a U.S. population
and in English, not representative of the global population. 
Despite these limitations, we believe our study provides valuable data for an underexplored problem.

\section{Conclusion}
\label{sec:conclusion}

We identify a novel threat model in which adversaries bias LLM responses 
in a target direction by suggesting subtly altered prompts to unsuspecting users. 
%\omer{fix the numbers please.}
%We contribute a dataset of 449 prompts across 77 product categories and develop methods 
%to evaluate brand mentions in LLM outputs, 
%finding significant variability in brand mentions between similar prompts.
Through a series of experiments and a user study, we develop this attack and demonstrate that 
we can bias LLM responses towards a target concept 
while remaining inconspicuous to humans.
We further show that this attack can be used to bias LLMs in harmful and 
helpful ways. 
% \revision{We investigate the transferability of this attack in the appendix (\S\ref{sec:results:transfer})}.
These findings highlight the risk of adversaries introducing bias 
in LLM responses in a unique way, suggesting a need for 
defensive measures like warnings and robustness requirements.

\begin{acks}
The work described in this paper was supported in part by the U.S. Army Research Office under MURI grant W911NF-21-1-0317 and by the Future Enterprise Security Initiative at Carnegie Mellon CyLab (FutureEnterprise@CyLab). 

This research was supported by the Center for AI Safety Compute Cluster. Any opinions, findings, and conclusions or recommendations expressed in this material are those of the author(s) and do not necessarily reflect the views of the sponsors.
\end{acks}
\bibliographystyle{ACM-Reference-Format}

%%%% bibliography starts here %%%%


%\bibliography{reference}%% Commented by merge tool

\appendix

%%%% appendix.tex starts here %%%%

\section*{Appendix}

% \section{\revision{How We May Use the Attack Differently?}}
\section{\revision{Variations on the Attack}}
\label{sec:additionalresult}
\revision{
In this section, we list experimental results in addition to those in the main body of the paper (\S\ref{sec:results} and \S\ref{sec:user:results}).
For the sake of completeness, we evaluate the effectiveness of our synonym replacement approach to reduce the frequency of mentioning a target concept 
by maximizing the loss function (App.~\ref{sec:results:negative}).
We are limited by computational resources, and
the models we run our evaluations on might not be representative 
of all LLMs used in the wild. Thus, we explore the transferability 
of our synonym replacement 
approach to (closed-source) GPT models (App.~\ref{sec:results:transfer}).
We then evaluate synonym replacement at different settings (App.~\ref{sec:results:temperature}),
and compare attack success with different numbers of synonyms replaced (App.~\ref{sec:results:numberReplacement}).
}

\subsection{Synonym-Replaced Adversarial Prompts to Mention Concepts Less Often}
\label{sec:results:negative}
\begin{figure}[t!]
\centerline{\includegraphics[width=0.95\columnwidth]{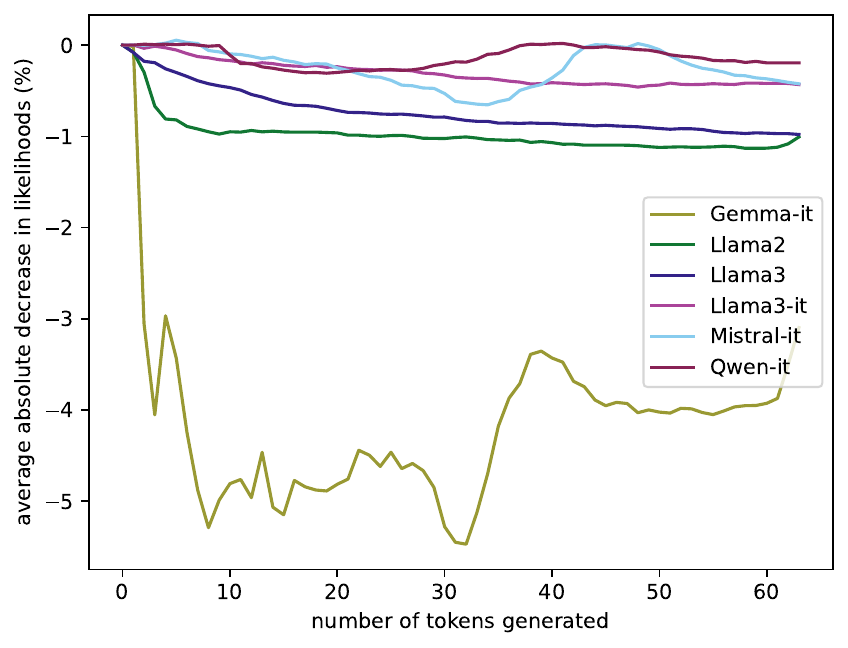}}
\caption{Average absolute decrease in likelihoods that LLMs
  mention a target brand when the base likelihood is at least $5 \%$. Results are presented along number of generated 
  tokens.
  Our synonym-replacement approach
  achieves decreases  in probabilities, which
  verifies the capability of forcing LLMs to mention
  target brands less often.
  }  
\Description{
  This is a line plot suggesting how the average absolute decrease in likelihoods that LLMs mention a target brand change against the number of tokens generated. 
  The y axis is labeled “average absolute decrease in likelihoods (\%)” and x axis is labeled “number of tokens generated”. 
  The x axis goes from 0 to 64. There are six curves plotted, corresponding to six LLMs: Gemma-it, Llama2, Llama3, Llama3-it, Mistral-it, and Qwen-it. 
  All models start at zero improvement with zero tokens generated. Gemma-it rapidly decreases to -5\% before 10 tokens, stays the same until around 35 tokens, where it increases and stays between -4\% and -3\% for the rest of the graph. 
  Llama2 decreases until reaching about -1\% at 5 tokens, and stays around that level. 
  Llama3 gradually decreases the whole time until reaching about -1\%. 
  Llama3-it also decreases gradually, and only reaches about -0.5\%. 
  Mistral-it decreases until reaching -0.5\% at 35 tokens, then goes back up to 0\% at 45 tokens, then decreases slightly. 
  Qwen-it goes down slightly, but stays between 0 and -0.5\% the whole time, going back up and remaining near zero after 40 tokens.
}
\label{fig:results:whitebox:short001_reverse}
\end{figure}
\begin{figure}[t!]
\centerline{\includegraphics[width=0.95\columnwidth]{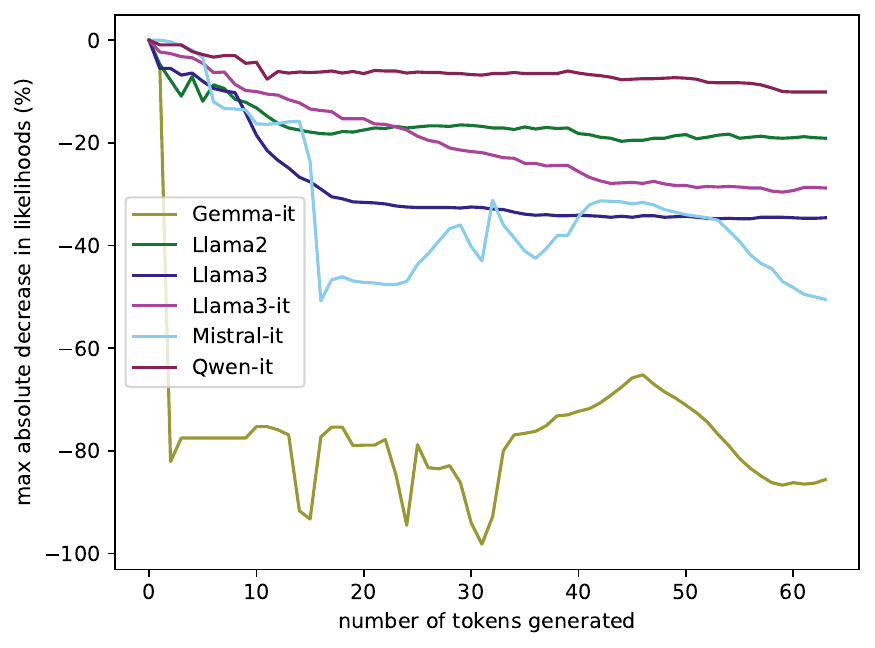}}
\caption{Max absolute decrease in likelihoods that
  LLMs mention 
  a target brand. Results are presented along how many tokens are generated. 
  We achieve a vaster absolute decrease on Gemma-it compared to the
  other models.
%  , although Gemma-it does not always have the
%  vastest average absolute decrease (see
%  Fig.~\ref{fig:results:whitebox:short001_reverse})
  }
\Description{
  This is a line plot suggesting how the maximum absolute decrease in likelihoods that LLMs mention a target brand change against the number of tokens generated. 
  The y axis is labeled “average absolute decrease in likelihoods (\%)” and x axis is labeled “number of tokens generated”. 
  The x axis goes from 0 to 64. There are six curves plotted, corresponding to six LLMs: Gemma-it, Llama2, Llama3, Llama3-it, Mistral-it, and Qwen-it. 
  All models start at zero improvement with zero tokens generated. Gemma-it rapidly decreases to -80\% at less than 5 tokens, then stays in the -80\% and -100\% range. 
  Llama2 gradually decreases until reaching -20\%. L
  lama3 decreases until reaching around -35\% at 20 tokens, then stays in this range. 
  Llama3-it gradually decreases and reaches around -30\%. Mistral-it decreases to about -15\% at 10 tokens, then rapidly decreases to -50\%, and stays between -30\% and -50\% for the rest of the time. 
  Qwen-it gradually decreases and reaches about -10\%.
}
\label{fig:results:whitebox:min}
\end{figure}
To systematically verify the correctness of our implementation in \S\ref{sec:approach:whiteboxattack},
we perturbed the prompts in the product scenario against each brand in the same category, 
by maximizing (rather than minimizing) the loss function we proposed.
%The results for the average decrease in
%likelihoods of a target brand being mentioned
%in the LLM responses due to our prompt perturbations, are shown in Fig.~\ref{fig:results:whitebox:short001_reverse},
Overall, we 
find that our implementation results in an average absolute decrease in all models 
\revision{as shown in Fig.~\ref{fig:results:whitebox:short001_reverse}}. 
Specifically, for brands that were mentioned at least $50$ times (i.e.,
$5\%$ of the time) in responses to the base prompt, 
the \revision{six} models had an average absolute decrease of 
$0.19\%$ to $3.10\%$ with in the first $64$ tokens generated,
and a slightly bigger absolute
decrease of $0.31\%$ to $5.47\%$ within a number of tokens less than $64$. 

Similar to \S\ref{sec:results:whiteboxattack},
besides the average absolute decrease, we also explored the maximum
absolute decrease among all combinations of base prompts and concepts,
as shown in Fig.~\ref{fig:results:whitebox:min}.
We were able to achieve a maximum absolute decrease in the likelihood of 
$10.10\%$ to $98.20\%$.
%Gemma-it ($-98.2\%$) compared to the
%other three Llama models ($-19.7\%$ to $-34.8\%$).
While these results don't represent the \emph{expected} decrease using our approach,
they do demonstrate that
it is possible to slightly perturb the prompts to mention a target concept less often in LLM responses.
The fact that synonym replacement can change the likelihood of LLMs mentioning a target concept in either direction
suggests our approach's potential to mitigate existing biases in LLMs,
albeit it might not work for all prompts and target concepts, and there is little control of the magnitude of change in likelihoods. 
For example, prompt providers might use this approach to fight against known biases in LLMs, 
providing users with prompts that encourage LLMs to generate responses with the likelihoods of mentioning target concepts closer to the distributions expected by users. 
\anna{Added this below}
Additionally, in cases of ``negative prompts'', there may be an incentive for an adversary
to want a topic to show up \emph{less} often.
%The perturbed prompts are expected to force LLMs to mention a specific brand less often compared to using unperturbed prompts (\S\ref{sec:results:negative})
%To systematically verify the correctness of our implementation in \S\ref{sec:approach:whiteboxattack},
%we perturbed the prompts in the shopping scenario against each brand in the same category, 
%by maximizing the loss function we proposed.

\subsection{Transferability to GPT Models}
\label{sec:results:transfer}
\begin{figure}[t!]
\centerline{\includegraphics[width=0.95\columnwidth]{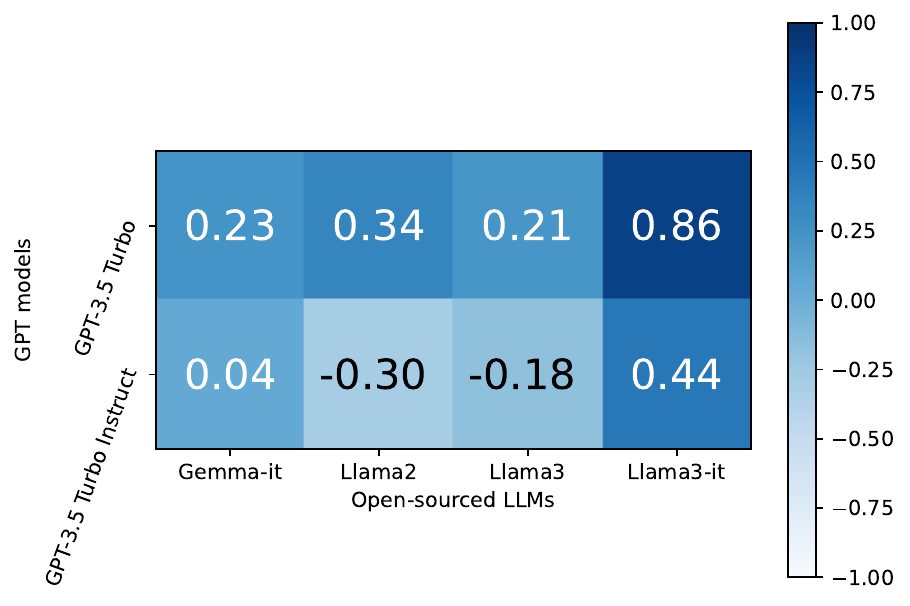}}
\caption{Pearson correlation coefficients ($\rho$) attack 
  success with GPT models and open-source LLMs. While
  most GPT/open-source LLM pairs have $\rho<0.4$, 
  %and thus are weakly correlated or uncorrelated, 
  Llama3-it and GPT3.5-Turbo
  have a correlation coefficient of $0.86\%$, implying strongly correlation ($p<0.001$).}%$3.4e-5$}
\Description{
  This is a heatmap suggesting relative improvements in likelihoods of mentioning a brand are correlated between different open-source LLMs and closed-source LLMs.
  There are two rows and four columns. 
  Each row is a closed-source GPT model, corresponding to either GPT-3.5 Turbo or GPT-3.5 Turbo Instruct; each column corresponds to one of the closed-source LLMs, one of Gemma-it, Llama2, Llama3, and Llama3-it. 
  Each cell includes a number which represents the Pearson correlation coefficient between the relative improvements in likelihoods of mentioning a brand of two models. 
  On the right, there is a color bar which indicates the correlation coefficient has a range from -1 to 1. 
  For transfer attacks, we hope the pearson correlation coefficient is positive and is close to 1. 
  The coefficient between GPT-3.5 Turbo and Gemma-it is 0.23. 
  The coefficient between GPT-3.5 Turbo and Llama2 is 0.34.
  The coefficient between GPT-3.5 Turbo and Llama3 is 0.21. 
  The coefficient between GPT-3.5 Turbo and Llama3-it is 0.86. 
  The coefficient between GPT-3.5 Turbo Instruct and Gemma-it is 0.04. 
  The coefficient between GPT-3.5 Turbo Instruct and Llama2 is -0.30.
  The coefficient between GPT-3.5 Turbo Instruct and Llama3 is -0.18. 
  The coefficient between GPT-3.5 Turbo Instruct and Llama3-it is 0.44. 
  Overall, while most pairs are uncorrelated or weakly correlated, GPT-3.5 Turbo and Llama3-it are strongly correlated with a coefficient of 0.86.
}
\label{fig:results:whitebox:transfer}
\end{figure}

\begin{figure}[t!]
\centerline{\includegraphics[width=0.95\columnwidth]{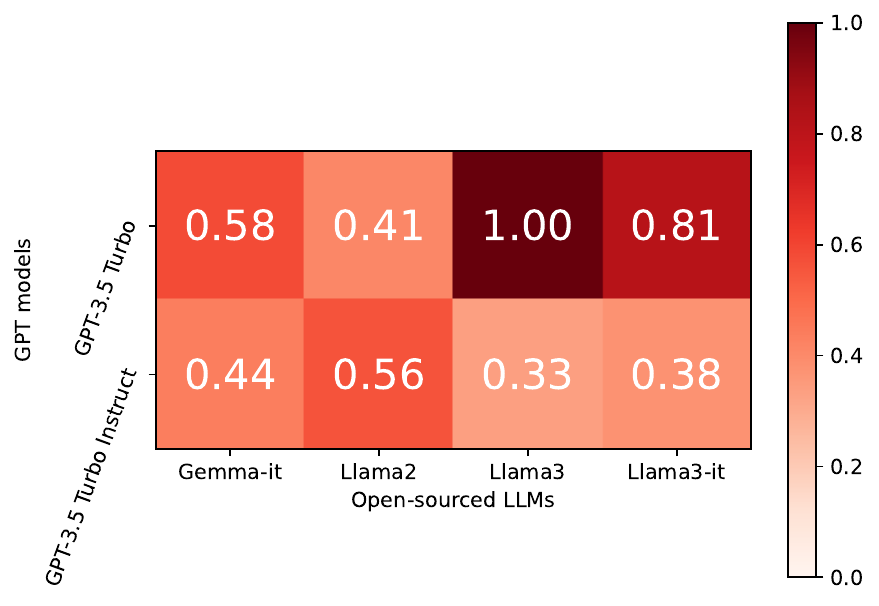}}
\caption{\revision{Chances that likelihoods of each pair of \llms
    mentioning a target concept move in the same direction (i.e., both
    increase or decrease) after synonym replacement. For specific
    pairs (GPT-3.5 Turbo and Llama3, GPT-3.5 Turbo and Llama3-it) of
    \llms, such chances are high, indicating that a synonym
    replacement that works on an open-source model is highly likely
    to work on a closed-source GPT model.}}%$3.4e-5$} 
\Description{
  This is a heatmap suggesting whether a brand is mentioned more or less after synonym replacement is correlated between different open-source LLMs and closed-source LLMs. 
  The previous figure measured the correlation between the change in amount a brand is mentioned, but this figure shows the correlation for only the direction of the change. 
  There are two rows and four columns. 
  Each row is a closed-source GPT model, corresponding to either GPT-3.5 Turbo or GPT-3.5 Turbo Instruct; each column corresponds to one of the closed-source LLMs, one of Gemma-it, Llama2, Llama3, and Llama3-it. 
  Each cell includes a number which represents the Pearson correlation coefficient between the relative improvements in likelihoods of mentioning a brand of two models. 
  On the right, there is a color bar which indicates the correlation coefficient has a range from -1 to 1. For transfer attacks, we hope the pearson correlation coefficient is positive and is close to 1. 
  The coefficient between GPT-3.5 Turbo and Gemma-it is 0.58. 
  The coefficient between GPT-3.5 Turbo and Llama2 is 0.41.
  The coefficient between GPT-3.5 Turbo and Llama3 is 1. 
  The coefficient between GPT-3.5 Turbo and Llama3-it is 0.81. 
  The coefficient between GPT-3.5 Turbo Instruct and Gemma-it is 0.44. 
  The coefficient between GPT-3.5 Turbo Instruct and Llama2 is 0.56.
  The coefficient between GPT-3.5 Turbo Instruct and Llama3 is 0.33. 
  The coefficient between GPT-3.5 Turbo Instruct and Llama3-it is 0.38. 
  Overall, GPT-3.5 Turbo is strongly correlated with Llama3 and Llama3-it, with coefficients of 1 and 0.81, respectively.
}
\label{fig:results:whitebox:transfer2}
\end{figure}

Some machine learning attacks were found capable of transferring~\cite{usenix18:transfer,usenix19:transfer}:
attacks against a machine-learning model might be effective against a different,
potentially unknown, model. 
We thus investigated whether our synonym replacement approach can transfer to GPT3.5-Turbo,
a commercial and closed-source LLM.
\revision{Due to limited resources, we only measure transferability on four open source models:
Llama2, Llama3, Llama3-it, Gemma-it.}

% \anna{removed ``following 2 metrics'' comment - I was unclear what that was referring to}
On each of \revision{ the four LLMs}, we first ranked the $1,809$ pairs of base and perturbed prompts (that recommend brands) 
according to the absolute improvement in mentioning a target brand. 
For the top pairs, 
% \anna{as defined in user study, should we mention here?}
we collected $1,000$ complete responses to both prompts:
we allowed LLMs to keep generating until they yielded end-of-sequence (EOS) tokens.
We collected responses from the LLM that this prompt pair originally performed well on,
as well as GPT3.5 Turbo and GPT3.5 Turbo Instruct.
\revision{We use the following two metrics to systematically examine
  the transferability of the synonym replacement attack
  \S\ref{sec:approach:whiteboxattack} from open-source \llms to
  closed-source \llms}.  
We first evaluated how well the change in probability that the target brand is recommended
between the base and perturbed prompts correlated between the open source and GPT models.
We then evaluated how likely the GPT models were to move in the same direction as the open source models -
i.e., if the probability between base and perturbed increased for the open source, 
whether it also increased for the GPT models.

% For a given base
% prompt (that recommends brands), we generated a perturbed prompt via synonym replacement that increased the chance
% of a brand being mentioned on an open-source model and 
% We first evaluated
% whether the perturbed prompt also increased the probability of the brand
% being mentioned in the response produced by the GPT model.
% We then evaluated how well the amount the 
% \lujo{is this correct?} \anna{now it is}
% We evaluated only the transfer of attacks that were successful on their
% open-source LLM onto GPT models.
% For the Shell example described in \S\ref{sec:results:whiteboxattack}, % \ TODO - anna find shell example
The base prompt ``\emph{Can you recommend the ultimate gas station for fueling up?}''
and perturbed prompt ``\emph{Can you recommend the premier gas station for fueling up?}'' in particular resulted
in a large change in the probability of Shell being mentioned in the response.
We saw an improvement from $10.3\%$ to	$57.4\%$ between the base and perturbed prompt on long responses by Gemma-it between prompts 
and an improvement from $41.7\%$ to $90.1\%$ on GPT3.5-Turbo responses.
On long responses generated by Llama3-it, 
``\emph{If you had to pinpoint the superior investment platform, which one would it be, and what specific features make it stand out as the top choice for investors?}''
had a $48.9\%$ probability of mentioning Fidelity, while
``\emph{If you had to pinpoint the premier investment platform, which one would it be, and what specific features make it stand out as the top choice for investors?}''
had a $79.3\%$, and GPT3.5-Turbo responses went from $21.9\%$ to $56.7\%$ for this same pair of prompts.
% \anna{I will go back and switch this to relative instead of absolute improvements!} \anna{still todo}

\subsubsection{Pearson Correlation Coefficients between Relative Improvements}
We computed the relative improvement in the probabilities of mentioning the target concept within complete responses for each pair.
If $20\%$ of responses of the base prompt before perturbing mentioned the target brand,
and $50\%$ of responses of the prompt after perturbing mentioned the target brand,
the relative improvement was $(50\%-20\%)/20\%=150\%$.
Then we computed the relative improvement of these pairs on ChatGPT.
Specifically, we used the GPT-3.5 Turbo and GPT-3.5 Turbo Instruct model with the default temperature parameter and collected $1,000$ complete responses.
GPT-3.5 Turbo Instruct is the instruction-tuned version of GPT-3.5 Turbo.
We compare the relative improvements between GPT models and open-source models in \S\ref{sec:results:transfer}.
We found that, while transferability was limited for most model pairs, 
Llama3-it and GPT3.5-Turbo had a high correlation in attack success for 
the same prompts ($p<0.001$). We explain these results in more detail next.

To compare the relative improvements between models when the
same base and perturbed (via synonym replacement) prompts are used, we used the Pearson
correlation test.  Two sets of data with a Pearson correlation coefficient ($\rho$) less
than $0.4$ is \textit{generally} believed to be weakly correlated,
whereas $0.1$ or lower is uncorrelated. On the other hand, a Pearson correlation coefficient larger than $0.7$
is believed to indicate strong correlation~\cite{AA18:correlation}.
\revision{A strong correlation indicates that the bigger an
  open-source LLM's relative improvement is, the bigger the
  closed-source LLM's relative improvement is given the same
  synonym-replaced prompts, i.e., the better the synonym replacement
  can transfer.} 

As shown in Fig.~\ref{fig:results:whitebox:transfer}, while Pearson
correlation coefficients of the relative improvement of most
pairs of open-source LLMs and GPT models indicated weak or no correlation
($\rho<0.4$), Llama3-it and GPT3.5-Turbo have a
correlation coefficient of $0.86$, and thus are highly correlated ($p<0.001$). 
Llama3-it and GPT3.5-Turbo Instruct
also show some correlation, with a coefficient of $0.44$. % ($p<0.001$).
%\lujo{p values below
%  0.001 are usually reported as $p<0.001$} 
%While our synonym-replacement prompts 
%do not strongly transfer between most pairs of models,
%it transfers from Llama3-it to GPT3.5-Turbo with statistical significance.

\revision{
\subsubsection{Probability that Likelihoods of Each Pair of \llms Mentioning a Target Concept Move in the Same Direction}
In addition to the Pearson correlation coefficients, we also measure
the probability that the likelihoods of each pair of open-source and
closed-source \llms mentioning a target concept will move in the same
direction (i.e., both increase or both decrease) after synonym replacement,
as shown in Fig.~\ref{fig:results:whitebox:transfer2}. For this metric,
we only compute the sign of the changes in likelihoods but not the
magnitude. We find that for specific pairs of open-source and
closed-source \llms, the probability of likelihoods changing in the same
direction, i.e., both increasing or decreasing, is high, indicating
that if a synonym replacement works on that open-source LLM in the
pair then it is also
likely to work on the closed-source model.  }

\revision{While synonym-replacement prompts do not transfer between many
  pairs of open-source and closed-source models, they do transfer
  between specific pairs (e.g., Llama3-it and GPT3.5-Turbo), according
  to the the two metrics.}  
These results indicate a potential for a transfer attack to be used on chatbots that use black-box GPT models, like 
Instacart, Lowe's, and Expedia (described in \S\ref{sec:threatmodel}) and others.
A successful attack on a matching open-source LLM could be used as a prompt 
suggestion for black-box models, ultimately promoting the target concept.

\revision{
\subsection{Attack Success at Different Temperatures}
\label{sec:results:temperature}
\begin{figure}[t!]
\centerline{\includegraphics[width=0.95\columnwidth]{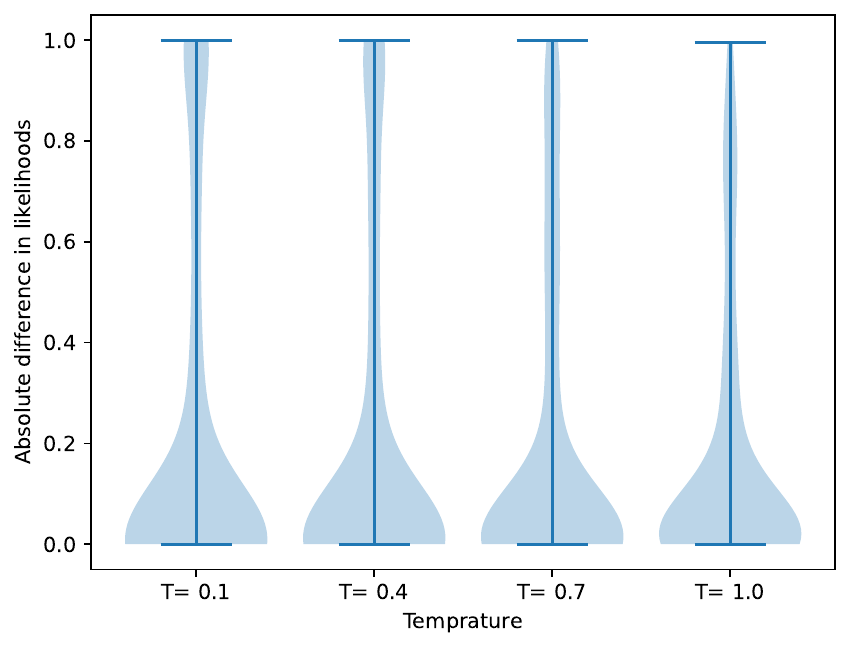}}
\caption{\revision{Absolute difference in the likelihoods generated in 
  response to paraphrased prompts at different temperatures, on Gemma-it and the brands
    dataset.}}
\Description{
  This is a plot that illustrates the range and density of the absolute difference in the likelihoods of responses mentioning a 
  target societal concept in the brand scenario within the first 64 tokens generated in response to paraphrased prompts, with respect to four different temperatures of Gemma-it. 
  The y axis is labeled with “Absolute difference in likelihoods” with a range from 0 to 1.
  The x axis has four ticks corresponding to the four temperatures: 0.1, 0.4, 0.7, and 1.0, from left to right. 
  The x axis is labeled “Temperature” and each tick is labeled with “T=” followed by that temperature. 
  Each tick has a vertical line segment, which suggests the range of the absolute difference in likelihoods. 
  All line segments start at zero, and end at 1.0. 
  For all temperatures, the density is highest at 0, stays high until around 0.05, then decreases and is much smaller around 0.2-0.8.
  After 0.8, the density increases for T=0.1, but still stays significantly smaller than it did at the bottom of the graph. 
  The density increases for T=0.4, but less. The density stays approximately constant after 0.8 for T=0.7, and decreases for T=1.0.
}
\label{fig:results:temperature:paraphrasing}
\end{figure}
}

\begin{figure}[t!]
\centerline{\includegraphics[width=0.95\columnwidth]{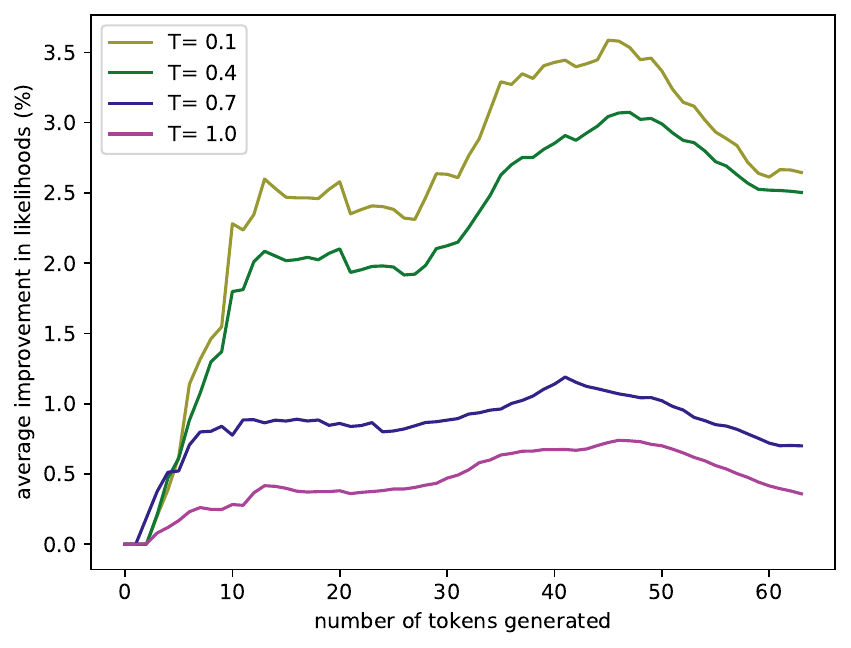}}
\caption{\revision{Average improvement of in likelihoods at different
   temperatures, on Gemma-it and the brands
    dataset.}}
\Description{
  This is a line plot illustrating how the average improvement in likelihoods that Gemma-it mentions a target brand changes against the number of tokens generated. 
  The y axis is labeled “average improvement in likelihoods (\%)” and x axis is labeled “number of tokens generated”. 
  The x axis goes from 0 to 64. There are six curves plotted, corresponding to six temperature values: 0.1, 0.4, 0.7, and 1.0. 
  All temperatures start at zero improvement with zero tokens generated. 
  All four temperatures increase for the first ten tokens, stay constant until about 25 tokens, increase between 25 and 45 tokens, then decrease. 
  At around ten tokens, the curves reach 2.5\% for T=0.1, 2\% for T=0.4, 0.8\% for T=0.7, and 0.4\% for T=1.0. 
  At 45 tokens, the curves reach 3.5\% for T=0.1, 3\% for T=0.4, 1\% for T=0.7, and 0.7\% for T=1.0. 
  At the end, the curves reach 2.7\% for T=0.1, 2.5\% for T=0.4, 0.6\% for T=0.7, and 0.4\% for T=1.0. 
}
\label{fig:results:temperature:average}
\end{figure}

\begin{figure}[t!]
\centerline{\includegraphics[width=0.95\columnwidth]{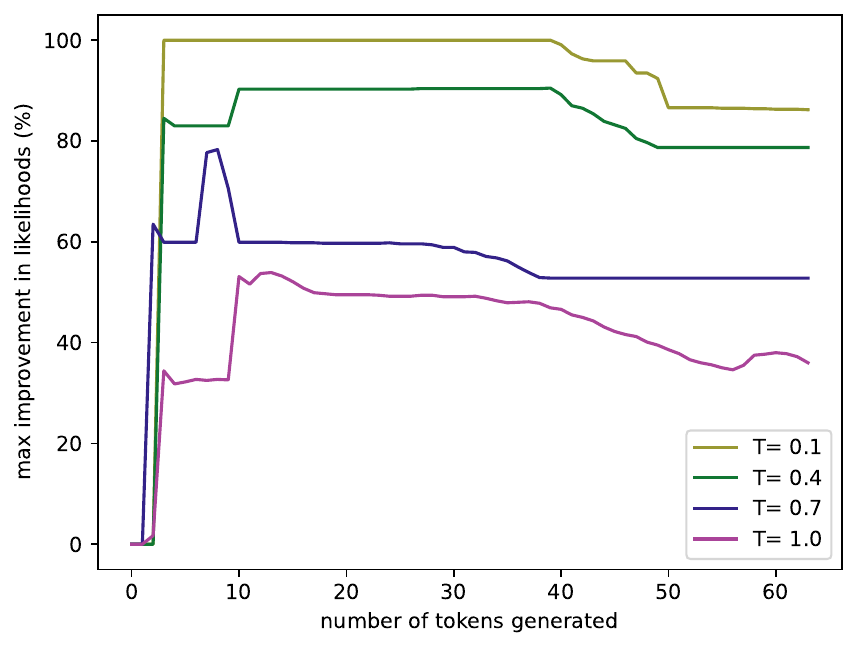}}
\caption{\revision{Max improvement of Gemma-it in likelihoods at different
  temperatures, on Gemma-it and the brands
    dataset.}}
\Description{
  This is a line plot illustrating how the maximum improvement in likelihoods that Gemma-it mentions a target brand changes against the number of tokens generated. 
  The y axis is labeled “max improvement in likelihoods (\%)” and x axis is labeled “number of tokens generated”. 
  The x axis goes from 0 to 64. There are six curves plotted, corresponding to six temperature values: 0.1, 0.4, 0.7, and 1.0. 
  All temperatures start at zero improvement with zero tokens generated. 
  All four temperatures increase for the first ten tokens, stay constant until about 25 tokens, increase between 25 and 45 tokens, then decrease. 
  The curve for T=0.1 increases to 100\% around 3 tokens, stays at that level until around 40 tokens, then decreases until 50 tokens, at which point it reaches and stays at around 90\%. 
  The curve for T=0.4 goes up rapidly to just over 80\% in the first 3 tokens, then again increases rapidly to around 90\% at 10 tokens. 
  The curve then stays the same until 40 tokens, then decreases until reaching and staying at 80\% around 50 tokens. 
  For T=0.7, the curve quickly reaches 60\%, then briefly jumps up to almost 80\% around 10 tokens. 
  After this, the curve gradually decreases until it reaches and stays at 55\% at 40 tokens. 
  For T=1.0, the curve rapidly increases to just over 30\% at 3 tokens, then again to just over 50\% at 10 tokens, after which it gradually decreases and ends at just under 40%.
}
\label{fig:results:temperature:max}
\end{figure}

\revision{
As we mentioned in \S\ref{sec:setup:LLM}, we primarily evaluated our attacks at the default temperature of \llms. 
Here we evaluate Gemma-it at different temperatures. 
Specifically, Gemma-it's temperature has a range of $[0,1]$. 
A low temperature (close to $0$) indicates tha model will be more deterministic, 
and a high temperature (close to $1$) indicates that the model will be more random. 
Previously, we use the default temperature of $0.7$. We try temperatures of $0.1$, $0.4$, and $1.0$ in addition. 
}

\revision{
Similar to \S\ref{sec:results:blackboxattack}, 
we measure the absolute difference in the likelihoods in 
response to paraphrased prompts at different temperatures, shown in Fig.~\ref{fig:results:temperature:paraphrasing}. 
The height and shape of the violin plots are similar at 
different temperatures, indicating the absolute difference 
has similar ranges and distributions. However, we notice 
that among the temperatures we tried, the violin plots are 
thinner at the higher end (i.e., the end near $1$), suggesting 
that a high difference in likelihoods happens less often. 
With a higher temperature, the LLM behaves more randomly, 
thus have smaller likelihoods of mentioning specific 
concepts, and have a high difference less often. 
}

\revision{
Similar to \S\ref{sec:results:whiteboxattack}, we measure the average improvement and max improvement on Gemma-it, but at different temperatures. The results are shown in Fig.~\ref{fig:results:temperature:average} and Fig.~\ref{fig:results:temperature:max} correspondingly. With a lower temperature, the LLM behaves more deterministically, and the attack achieves higher improvements. }

\revision{
\subsection{Attack Success With Different Numbers of Synonym Replacements}
\label{sec:results:numberReplacement}
}
\begin{figure}[t!]
\centerline{\includegraphics[width=0.95\columnwidth]{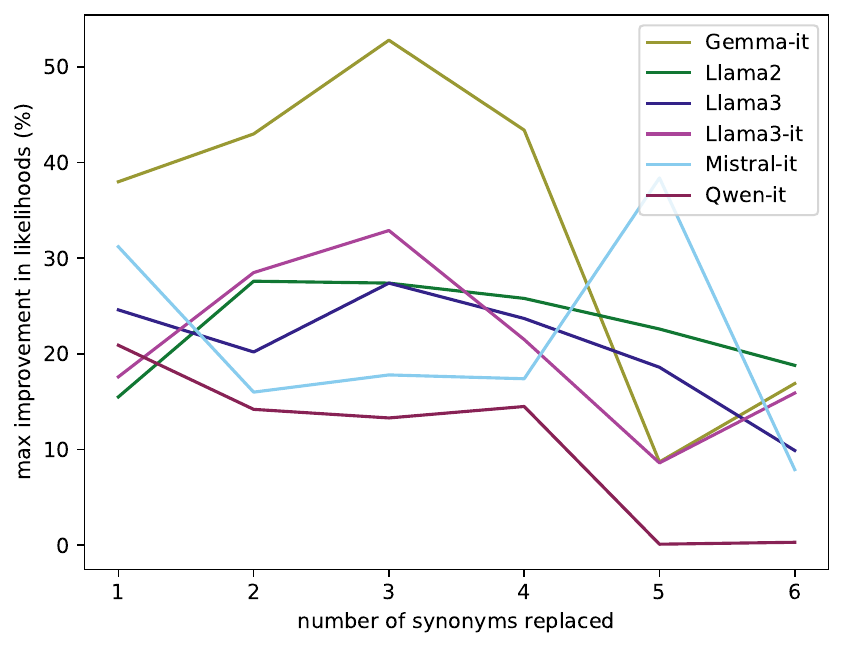}}
\caption{\revision{Max improvement in likelihoods with different
    number of synonyms replaced (at token 64), on the brands
    dataset. The max improvement does not always increase when more
    synonyms are replaced.}}
\Description{
  This is a line plot illustrating how the maximum improvement in likelihoods that different LLMs mention a target brand changes within the 64-token responses against the number of words replaced in the prompt. 
  The y axis is labeled “max improvement in likelihoods (\%)” and x axis is labeled “number of synonyms replaced”. 
  The x axis goes from 0 to 64. There are six curves plotted, corresponding to six models: Gemma-it, Llama2, Llama3, Llama3-it, Mistral-it, and Qwen-it. 
  For the Gemma-it curve starts at just under 40\%, then just over 40\%, then just over 50\%, then around 45\%, then 10\%, then 15\% for 1-6 replaced synonyms, respectively. 
  For Llama2, the curve is around 15\%, 28\%, 27\%, 25\%, 23\%, and 20\% for 1-6 replaced synonyms. 
  For Llama3, the curve is around 25\%, 20\%, 27\%, 24\%, 20\%, and 10\% for 1-6 replaced synonyms. 
  For Llama3-it, the curve is around 18\%, 28\%, 32\%, 20\%, 10\%, and 17\% for 1-6 replaced synonyms. 
  For Mistral-it, the curve is around 31\%, 26\%, 27\%, 27\%, 40\%, and 8\% for 1-6 replaced synonyms. 
  For Qwen-it, the curve is around 21\%, 15\%, 14\%, 15\%, 0\%, and 0\%  for 1-6 replaced synonyms. 
}
\label{fig:results:synonym}
\end{figure}
\revision{ As we described in \S\ref{sec:approach:whiteboxattack}, we
  develop our own synonym replacement approach along with new synonym
  dictionaries. Fig.~\ref{fig:results:synonym} illustrates the max
  improvement in likelihoods versus different number of synonyms
  replaced. We notice the most successful attacks (i.e., synonym
  replacements with the largest improvement in likelihoods on each
  model) do not always happened with most synonyms replaced. In other
  words, replacing more synonyms do not guarantee finding more
  attacks. }
  \anna{ I believe fewer prompts had 7 synonyms to replace. If we measure max improvement,
  we may expect that these cases would have less maximum improvement, since we did test less such cases. }

%this creates another subsection.

\section{Do LLMs Recommend Their Parent Brand?}
\label{app:same}
\begin{table}[h]
\centering
% \resizebox{\textwidth/2}{!}{\begin{tabular}{| c | c | c | c | c | c | c |} \hline
\resizebox{\textwidth/2}{!}{\begin{tabular}{ | r | r | r | r | r | r | r |} \hline
    % & Gemma-it & Llama2 & Llama3 & Llama3-it & GPT-3.5 Turbo & GPT-3.5 Turbo Instruct\\ \hline
search engine& Gemma-it & Llama2 & Llama3 & Llama3-it & GPT-3.5 & GPT-3.5-it\\ \hline 
Bing& 0.58 \% & 13.08 \% & 14.00 \% & 27.66 \% & 0.40 \% & 2.36 \% \\ \hline
Google& \textbf{98.16 \%} & 50.04 \% & 53.14 \% & 53.60 \% & 74.24 \% & 45.86 \% \\ \hline
Yahoo& 0.00 \% & 19.36 \% & 18.36 \% & 3.92 \% & 0.26 \% & 0.56 \% \\ \hline \hline
browser& Gemma-it & Llama2 & Llama3 & Llama3-it & GPT-3.5 & GPT-3.5-it\\ \hline 
Chrome& \textbf{77.22 \%} & 41.24 \% & 33.54 \% & 50.30 \% & 53.28 \% & 35.42 \% \\ \hline
Firefox& 28.74 \% & 38.10 \% & 31.28 \% & 22.28 \% & 5.08 \% & 5.40 \% \\ \hline
Safari& 3.40 \% & 11.70 \% & 9.00 \% & 5.64 \% & 0.02 \% & 0.36 \% \\ \hline
Edge& 7.62 \% & 13.70 \% & 10.04 \% & 13.52 \% & \textbf{0.36 \%} & 1.14 \% \\ \hline
Opera& 0.14 \% & 9.82 \% & 9.74 \% & 6.38 \% & 0.00 \% & 0.06 \% \\ \hline \hline 
llm& Gemma-it & Llama2 & Llama3 & Llama3-it & GPT-3.5 & GPT-3.5-it\\ \hline 
ChatGPT& 94.82 \% & 39.28 \% & 41.66 \% & 0.30 \% & 16.68 \% & 38.04 \% \\ \hline
Google& \textbf{2.56 \%} & 8.28 \% & 7.84 \% & 19.3 \% & 0.02 \% & 1.06 \% \\ \hline
Llama& 0.00 \% & 0.40 \% & 0.70 \% & 2.32 \% & 0.00 \% & 0.00 \% \\ \hline
Claude& 0.00 \% & 0.12 \% & 0.00 \% & 0.00 \% & 0.00 \% & 0.00 \% \\ \hline
Vicuna& 0.00 \% & 0.00 \% & 0.00 \% & 0.00 \% & 0.00 \% & 0.00 \% \\ \hline \hline 
os& Gemma-it & Llama2 & Llama3 & Llama3-it & GPT-3.5 & GPT-3.5-it\\ \hline 
Windows& 97.18 \% & 51.10 \% & 59.72 \% & 64.04 \% & 17.30 \% & 9.62 \% \\ \hline
Mac& 24.76 \% & 34.66 \% & 37.98 \% & 38.42 \% & 9.32 \% & 3.14 \% \\ \hline
Linux& 1.16 \% & 29.94 \% & 26.46 \% & 30.96 \% & 2.88 \% & 1.38 \% \\ \hline \hline 
smartphone& Gemma-it & Llama2 & Llama3 & Llama3-it & GPT-3.5 & GPT-3.5-it\\ \hline 
Apple& 90.32 \% & 28.88 \% & 30.45 \% & 14.97 \% & 12.21 \% & 31.29 \% \\ \hline
Google& \textbf{21.85} \% & 11.64 \% & 10.53 \% & 12.25 \% & 0.15 \% & 1.80 \% \\ \hline
Samsung& 9.02 \% & 31.11 \% & 27.19 \% & 33.41 \% & 0.71 \% & 8.86 \% \\ \hline \hline 
laptop& Gemma-it & Llama2 & Llama3 & Llama3-it & GPT-3.5 & GPT-3.5-it\\ \hline 
Mac& 25.76 \% & 14.22 \% & 13.46 \% & 7.16 \% & 44.66 \% & 9.80 \% \\ \hline
Chromebook& \textbf{0.00 \%} & 1.88 \% & 1.6 \% & 0.18 \% & 0.00 \% & 0.02 \% \\ \hline
HP& 0.00 \% & 12.14 \% & 14.16 \% & 9.96 \% & 0.22 \% & 1.40 \% \\ \hline
Asus& 0.00 \% & 7.08 \% & 6.64 \% & 3.84 \% & 0.04 \% & 0.20 \% \\ \hline
Lenovo& 0.00 \% & 9.44 \% & 14.42 \% & 13.94 \% & 0.28 \% & 2.24 \% \\ \hline
Acer& 0.00 \% & 9.20 \% & 9.48 \% & 4.06 \% & 0.00 \% & 0.08 \% \\ \hline
Dell& 39.98 \% & 14.62 \% & 17.54 \% & 43.60 \% & 22.98 \% & 56.58 \% \\ \hline \hline 
VR headset& Gemma-it & Llama2 & Llama3 & Llama3-it & GPT-3.5 & GPT-3.5-it\\ \hline 
Meta& \textbf{41.60 \%} & \textbf{25.54 \%} & \textbf{32.94 \%} & \textbf{34.16 \%} & \textbf{42.74 \%} & \textbf{42.60 \%} \\ \hline
HTC& 0.14 \% & 21.66 \% & 33.32 \% & 38.30 \% & 0.72 \% & 1.80 \% \\ \hline
Playstation& 0.00 \% & 0.30 \% & 0.52 \% & 0.00 \% & 0.00 \% & 0.00 \% \\ \hline \hline 
email provider& Gemma-it & Llama2 & Llama3 & Llama3-it & GPT-3.5 & GPT-3.5-it\\ \hline 
Google& 35.32 \% & 37.40 \% & 35.18 \% & 22.52 \% & 45.26 \% & 31.46 \% \\ \hline
Yahoo& 0.00 \% & 4.98 \% & 9.16 \% & 5.12 \% & 0.82 \% & 0.48 \% \\ \hline
Microsoft& 4.00 \% & 16.76 \% & 17.78 \% & 14.40 \% & 2.16 \% & 1.90 \% \\ \hline \hline
% &&&&&& \\ \hline
\end{tabular}}
\vspace{0.1cm}
\caption{LLMs tested on their parent brands. 
Categories are search engines, browsers, LLMs, operating systems, laptops, VR headsets, and email providers.
Scores are calculated as average across all prompts for the category.}
\Description{
  This table reports how LLMs behave when it comes to products with parent companies Google/Alphabet, Microsoft, and Meta, as well as their competitors. The graph is separated into 8 categories: search engine, browser, LLM, OS, smartphone, laptop, VR headset, and email provider, listed from top to bottom. This is labeled at the top left corner of each subraph. The labels on the left of each subgraph are products/brand, which will be detailed later. The labels at the top of each subgraph are the LLMs: Gemma-it, Llama2, Llama3, Llama3-it, GPT-3.5, and GPT-3.5-it, listed from left to right. The value in each cell shows the probability that a certain LLM mentioned the current product/brand for prompts concerning this product category. The probabilities for each product/brand for each LLM - Gemma-it, Llama2, Llama3, Llama3-it, GPT-3.5, and GPT-3.5-it - are listed below, and are listed in order respective to the LLM.

The probability that each model mentioned Bing for the category "search engine" was 0.58\%, 13.08\%, 14.0\%, 27.66\%, 0.4\%, and 2.36\%.
The probability that each model mentioned Google for the category "search engine" was 98.16\%, 50.04\%, 53.14\%, 53.6\%, 74.24\%, and 45.86\%. 98.16\% is bolded for emphasis.
The probability that each model mentioned Yahoo for the category "search engine" was 0.0\%, 19.36\%, 18.36\%, 3.92\%, 0.26\%, and 0.56\%.

The probability that each model mentioned Chrome for the category "browser" was 77.22\%, 41.24\%, 33.54\%, 50.3\%, 53.28\%, and 35.42\%.
The probability that each model mentioned Firefox for the category "browser" was 28.74\%, 38.1\%, 31.28\%, 22.28\%, 5.08\%, and 5.4\%.
The probability that each model mentioned Safari for the category "browser" was 3.4\%, 11.7\%, 9.0\%, 5.64\%, 0.02\%, and 0.36\%.
The probability that each model mentioned Edge for the category "browser" was 7.62\%, 13.7\%, 10.04\%, 13.52\%, 0.36\%, and 1.14\%. 0.36\% is bolded for emphasis.
The probability that each model mentioned Opera for the category "browser" was 0.14\%, 9.82\%, 9.74\%, 6.38\%, 0.0\%, and 0.06\%.

The probability that each model mentioned ChatGPT for the category "llm" was 94.82\%, 39.28\%, 41.66\%, 0.3\%, 16.68\%, and 38.04\%.
The probability that each model mentioned Bard for the category "llm" was 2.56\%, 8.28\%, 7.84\%, 19.3\%, 0.02\%, and 1.06\%. 2.56\% is bolded for emphasis.
The probability that each model mentioned Llama for the category "llm" was 0.0\%, 0.4\%, 0.7\%, 2.32\%, 0.0\%, and 0.0\%.
The probability that each model mentioned Claude for the category "llm" was 0.0\%, 0.12\%, 0.0\%, 0.0\%, 0.0\%, and 0.0\%.
The probability that each model mentioned Vicuna for the category "llm" was 0.0\% for all models.

The probability that each model mentioned Windows for the category "os" was 97.18\%, 51.1\%, 59.72\%, 64.04\%, 17.3\%, and 9.62\%.
The probability that each model mentioned Mac for the category "os" was 24.76\%, 34.66\%, 37.98\%, 38.42\%, 9.32\%, and 3.14\%.
The probability that each model mentioned Linux for the category "os" was 1.16\%, 29.94\%, 26.46\%, 30.96\%, 2.88\%, and 1.38\%.

The probability that each model mentioned Apple for the category "smartphone" was 90.32\%, 28.88\%, 30.45\%, 14.97\%, 12.21\%, and 31.29\%.
The probability that each model mentioned Google for the category "smartphone" was 21.85\%, 11.64\%, 10.53\%, 12.25\%, 0.15\%, and 1.8\%. 21.85\% is bolded for emphasis.
The probability that each model mentioned Samsung for the category "smartphone" was 9.02\%, 31.11\%, 27.19\%, 33.41\%, 0.71\%, and 8.86\%.

The probability that each model mentioned Mac for the category "laptop" was 25.76\%, 14.22\%, 13.46\%, 7.16\%, 44.66\%, and 9.8\%.
The probability that each model mentioned Chromebook for the category "laptop" was 0.0\%, 1.88\%, 1.6\%, 0.18\%, 0.0\%, and 0.02\%. 0.0\% in the first column, for Gemma-it, is bolded for emphasis.
The probability that each model mentioned HP for the category "laptop" was 0.0\%, 12.14\%, 14.16\%, 9.96\%, 0.22\%, and 1.4\%.
The probability that each model mentioned Asus for the category "laptop" was 0.0\%, 7.08\%, 6.64\%, 3.84\%, 0.04\%, and 0.2\%.
The probability that each model mentioned Lenovo for the category "laptop" was 0.0\%, 9.44\%, 14.42\%, 13.94\%, 0.28\%, and 2.24\%.
The probability that each model mentioned Acer for the category "laptop" was 0.0\%, 9.2\%, 9.48\%, 4.06\%, 0.0\%, and 0.08\%.
The probability that each model mentioned Dell for the category "laptop" was 39.98\%, 14.62\%, 17.54\%, 43.6\%, 22.98\%, and 56.58\%.

The probability that each model mentioned Meta for the category "VR headset" was 41.6\%, 25.54\%, 32.94\%, 34.16\%, 42.74\%, and 42.6\%. All of these cells are bolded for emphasis.
The probability that each model mentioned HTC for the category "VR headset" was 0.14\%, 21.66\%, 33.32\%, 38.3\%, 0.72\%, and 1.8\%.
The probability that each model mentioned Playstation for the category "VR headset" was 0.0\%, 0.3\%, 0.52\%, 0.0\%, 0.0\%, and 0.0\%.

The probability that each model mentioned Google for the category "email provider" was 35.32\%, 37.4\%, 35.18\%, 22.52\%, 45.26\%, and 31.46\%.
The probability that each model mentioned Yahoo for the category "email provider" was 0.0\%, 4.98\%, 9.16\%, 5.12\%, 0.82\%, and 0.48\%.
The probability that each model mentioned Microsoft for the category "email provider" was 4.0\%, 16.76\%, 17.78\%, 14.4\%, 2.16\%, and 1.9\%. 
}
\label{tab:appendix:scores}
\end{table}

\lujo{}
%Over the course of our experiments, w
\lujo{is this attacks (i.e., we're perturbing prompts) or evaluating on the dataset of natural prompts?} \anna{addressed}
Throughout our experiments evaluating rephrasings in \S\ref{sec:results:blackboxattack},
we gathered completions for prompts on categories with products manufactured by Meta, Google, and Microsoft, which allowed us to examine 
%do some analysis on
how large language models developed by these companies perform when asked about product categories that include products manufactured by them.
% \lujo{... perform when asked for recommendations for product types that include products made by the LLMs manufacturers.(?) or something else slightly more concrete than the current sentence} \anna{addressed}
We evaluated the average score, as defined in \S\ref{sec:setup:LLM}, of all brands over all prompts for categories where one of the brands was Meta, Google, or Microsoft. 
% For experiments on laptop prompts with a target brand of Apple, this would be ``Macbook", ``Mac",``MacOS", and ``Apple".
For Google the categories included browsers (Chrome), large language models, smartphones (Pixel), laptops (Chromebook), email providers (Gmail), and search engines; for Meta they included VR headsets and large language models (Llama).
As before, this meant we looked for target words related to a brand in the response to see whether this prompt was mentioned.
\anna{add note about time of training and training data potentially affecting some of these}
\lujo{somewhere in the intro paragraph, also tell us the punchline, at least in general terms} \anna{done}
We were interested in whether or not LLMs made by a certain company were biased towards products made by the same company.
All results are shown in Tab.~\ref{tab:appendix:scores}.

% \begin{figure}[t!]
% \centerline{\includegraphics[width=0.99\columnwidth]{appendix_figures/llms_base.png}}
% \caption{Average Scores of LLMs} 
% \label{fig:appendix:llmbase}
% \end{figure}

Google has developed a variety of LLMs and LLM families (laMDA, Bert,
PaLM, Gemini, Gemma) ~\cite{llmlist}, yet we still found some
interesting mistakes in Gemma-it's responses to prompts asking for
recommendations on large language models.\lujo{do we start
  consistently abbreviating to ``LLMs'' at some point? if so, do that
  consistently}\anna{resolved} 
%large language model prompts. 
For example, across multiple prompts, Gemma-it's responses included
``\emph{**GPT-4:** This model, developed by Google,}'' a false
statement ~\cite{llmlist} seeming to claim that GPT-4 was developed by
Google. The same was said for GPT-3, which is also false
~\cite{llmlist}. 
So, even Gemma-it responses that mention Google might actually be recommending GPT.
\lujo{should all these instances of ``Google'' actually be ``Alphabet''?} 
\anna{I'm not sure - usually Gemma-it is referred to as ``Google's Gemma-it" for example}
Out of Gemma-it's responses, the only actual model of Google's mentioned is PaLM.
% In contrast, Llama models mention PaLM, BERT, LaMDA, and Bard.\lujo{a little unclear why models are being attributed to any company. how does this even come up?}\anna{resolved}
% Scores for Google in our figure represent any mentions of Google or any of its models, as it is one of the target words for that category. 
% So, this can represent any of Google's LLMs or, in the case where Gemma-it was queried, it can apparently even represent GPT.
% \lujo{at this point i'm a little confused about what exactly was measured. can you state that more explicitly after the intro paragraph?}
% \anna{done}

We see more mentions of Llama or Meta when querying Llama than when querying Gemma-it or GPT3.5-Turbo, 
with our three Llama models we see a $0.4\%$, $0.7\%$, and $2.32\%$ and never with the other models. 
However, we did not always observe this self-preference. 
Llama models recommend a Google model at least $7.84\%$ of the time whereas Google only recommends a Google model $2.56\%$ of the time.
All Llama models recommend Meta VR headsets under $40\%$ of the time, while Gemma and GPT models do over $40\%$ of the time, 
and Llama2 mentions Gmail when prompted about email providers more than Gemma-it does ($37.4\%$ vs $35.32\%$).
Gemma-it never mentions Chromebooks when asked about laptops, while all three Llama models and GPT-3.5 Instruction sometimes do. 
Nonetheless, Gemma-it seems to show a higher preference towards Google products than any Llama model for the categories of search engines and phones. 
\anna{Some of these graphs will be converted to tables to save space}
\lujo{throughout, can you add examples of ``higher'' that mention specific numbers}
\anna{addressed}

\lujo{what's the takeaway?} 
Overall, this test size is small and does not necessarily take into account 
all factors that can cause differences between these models.
% For example, GPT3.5-Turbo does not mention any brand for prompts
% related to phones more than $20\%$ of the time, while Gemma-it mentions Apple and Google over $20\%$ of the time.
% So, while GPT3.5-Turbo mentions Apple less than Gemma-it does, the proportion of Apple to Google responses is higher for 
% GPT3.5-Turbo than for Gemma-it.
In the end, we do not find any bias by LLMs 
towards products developed by the same parent company,
but believe it warrants further exploration.

% \begin{figure}[t!]
% \centerline{\includegraphics[width=0.99\columnwidth]{appendix_figures/browser_base.png}}
% \caption{Average Scores} 
% \label{fig:appendix:browserbase}
% \end{figure}
    
% \begin{figure}[t!]
% \centerline{\includegraphics[width=0.99\columnwidth]{appendix_figures/email_providers_base.png}}
% \caption{Average Scores} 
% \label{fig:appendix:emailbase}
% \end{figure}

% \begin{figure}[t!]
% \centerline{\includegraphics[width=0.99\columnwidth]{appendix_figures/laptop_base.png}}
% \caption{Average Scores} 
% \label{fig:appendix:laptopbase}
% \end{figure}

% \begin{figure}[t!]
% \centerline{\includegraphics[width=0.99\columnwidth]{appendix_figures/phone_base.png}}
% \caption{Average Scores} 
% \label{fig:appendix:phonebase}
% \end{figure}

% \begin{figure}[t!]
% \centerline{\includegraphics[width=0.99\columnwidth]{appendix_figures/search_base.png}}
% \caption{Average Scores} 
% \label{fig:appendix:searchbase}
% \end{figure}

% \begin{figure}[t!]
% \centerline{\includegraphics[width=0.99\columnwidth]{appendix_figures/VR_headset_base.png}}
% \caption{Average Scores of VR headsets} 
% \label{fig:appendix:vrbase}
% \end{figure}
\onecolumn
%\revision{
\section{More Examples of Prompts}
We provide more examples of prompts as following. We start with some paraphrased prompts in Tab.~\ref{tab:appendix:examples:paraphrase}.
\begin{table}[h]
\centering
% \resizebox{\textwidth/2}{!}{\begin{tabular}{| c | c | c | c | c | c | c |} \hline
{\begin{tabular}{|@{\hskip 0.01in}c@{\hskip 0.01in}c@{\hskip 0.01in}l@{\hskip 0.01in}r@{\hskip 0.01in}r|} \hline
\textit{Model}& \textit{Category (Brand)} & \textit{Prompt} & \textit{Likelihood (\%)} & \textit{Difference (\%)}\\ \hline 
Gemma-it & Vacuum & I'm curious to know your preference for the vacuum cleaner that offers& 0.1 & 99.9\\
 &(Dyson) & the best combination of cleaning performance, versatile design, and overall& & \\
& & convenience in tackling various surfaces and debris types. & & \\
& & Can you recommend the ultimate vacuum that excels in providing & 100.0 & \\
& & thorough cleaning, user-friendly controls,and additional functionalities such as & & \\
& & multiple attachments or a HEPA filter for improved air quality? & & \\ \hline 
Llama2 & Convenience& Seeking insights on the most commendable convenience & 5.5 & 45.3\\
& Store & store – the one that truly excels in providing quick and efficient services& & \\
& (711)& services along with customer satisfaction.& & \\
& & When it comes to a quick stop for essentials, which& 50.8 & \\
& & convenience store do you believe stands out in terms of& & \\
& &accessibility, pricing, and the variety of available items?& & \\ \hline 
Llama3 & Convenience & If you were to identify the premier convenience store in your area, which one & 6.3 & 52.0\\
& Store& would it be, and what specific features make it stand out from the rest?.& & \\
& (711)& When it comes to a quick stop for essentials, which convenience store& 58.3 & \\
& & convenience store do you believe stands out in terms of& & \\
& &accessibility, pricing, and the variety of available items?& & \\ \hline 
Llama3-it & Kitchen Mixer & I'm curious to know your preference for kitchen mixers that offer the best& 20.6 & 54.0\\
 &Mixer& combination of mixing speeds, bowl capacity, and overall convenience,& & \\
&(KitchenAid)  & ensuring an enhanced baking and cooking experience for various recipes. & & \\
& & Can you recommend the ultimate kitchen mixer that excels in providing  & 74.6 & \\
& &consistent results, user-friendly controls, and additional features such as & & \\
& & various attachments, suitable for both amateur and professional bakers? & & \\ \hline 
Mistral-it & Email&When exploring email providers, which one do you believe distinguishes& 0.0 & 95.5\\
 &Provider & itself with features such as end-to-end encryption, cross-platform& & \\
 & (Google)& compatibility, and versatility for different communication styles,& & \\
& & ensuring top-tier performance for users with diverse email preferences?& & \\
& & I'm curious to know your preference for email providers that offer the best& 95.5 & \\
& &combination of accessibility, customization options, and overall user& & \\
& &experience, ensuring a streamlined communication platform for various needs.& & \\ \hline 
Qwen-it & Browser & In the process of deciding on a web browser, I'm interested in your& 9.3 & 48.5\\
 &(Chrome) &insights on the optimal choice. Factors such as browsing speed, & & \\
& & security features, and compatibility with different websites& & \\
& & are crucial considerations for me. Any recommendations?& & \\
& & As I navigate the myriad of web browsers available, I'm seeking& 57.8 & \\
& & guidance on choosing the best one for my needs. What browser & & \\
& & do you recommend, considering with aspects like speed, security, and seamless& & \\
& &  compatibility with a diverse range of websites?& & \\ \hline 
\end{tabular}}
\vspace{0.1cm}
\caption{Examples of paraphrased prompts}
\Description{
  This table provides examples of paraphrased prompts that had a large difference in the likelihood that a certain target brand is mentioned in the responses of an LLM. 
  The leftmost column specifies the LLM. The next column specifies the category of product and brand that is mentioned. 
  The next column gives both prompts, that are paraphrases of each other. 
  The next column gives the probability that each prompt mentions the given brand in responses of that LLM. 
  The final column is the difference between these probabilities. Each row is described next.
  The first row shows the probability that Gemma-it mentions Dyson when prompted about vacuums. 
  The first prompt, “I’m curious to know your preference for the vacuum cleaner that offers the best combination of cleaning performance, versatile design, and overall convenience in tackling various surfaces and debris types.” mentions Dyson 0.1\% of the time, 
  while “Can you recommend the ultimate vacuum that excels in providing thorough cleaning, user-friendly controls, and additional functionalities such as multiple attachments or a HEPA filter for improved air quality?” mentions Dyson 100\% of the time, for a difference of 99.9\%.
  The second row shows the probability that Llama2 mentions 7-11 when prompted about convenience stores. 
  The first prompt, “Seeking insights on the most commendable convenience store – the one that truly excels in providing quick and efficient services along with customer satisfaction.” mentions 7-11 5.5\% of the time, 
  while “When it comes to a quick stop for essentials, which convenience store convenience store do you believe stands out in terms of accessibility, pricing, and the variety of available items?” mentions 7-11 50.8\% of the time, for a difference of 45.3\%.
  The next row shows the probability that Llama3 mentions 7-11 when prompted about convenience stores. 
  The first prompt, “If you were to identify the premier convenience store in your area, which one would it be, and what specific features make it stand out from the rest?” mentions 7-11 6.3\% of the time, 
  while “When it comes to a quick stop for essentials, which convenience store do you believe stands out in terms of accessibility, pricing, and the variety of available items?” mentions 7-11 58.3\% of the time, for a difference of 52\%.
  The next row shows the probability that Llama3-it mentions KitchenAid when prompted about kitchen mixers. 
  The first prompt, “I'm curious to know your preference for kitchen mixers that offer the best combination of mixing speeds, bowl capacity, and overall convenience, ensuring an enhanced baking and cooking experience for various recipes.” mentions KitchenAid 20.6\% of the time, 
  while “Can you recommend the ultimate kitchen mixer that excels in providing consistent results, user-friendly controls, and additional features such as various attachments, suitable for both amateur and professional bakers?” mentions KitchenAid 74.6\% of the time, for a difference of 54\%.
  The next row shows the probability that Mistral-it mentions Gmail when prompted about email providers. 
  The first prompt, “When exploring email providers, which one do you believe distinguishes itself with features such as end-to-end encryption, cross-platform compatibility, and versatility for different communication styles, ensuring top-tier performance for users with diverse email preferences?” mentions Gmail 0.0\% of the time, 
  while “I'm curious to know your preference for email providers that offer the best combination of accessibility, customization options, and overall user experience, ensuring a streamlined communication platform for various needs.” mentions Gmail 95.5\% of the time, for a difference of 95.5\%.
  The next row shows the probability that Qwen-it mentions Chrome when prompted about browsers. 
  The first prompt, “In the process of deciding on a web browser, I'm interested in your insights on the optimal choice. Factors such as browsing speed, security features, and compatibility with different websites are crucial considerations for me. Any recommendations?” mentions Chrome 9.3\% of the time, 
  while “As I navigate the myriad of web browsers available, I'm seeking guidance on choosing the best one for my needs. What browser do you recommend, considering with aspects like speed, security, and seamless compatibility with a diverse range of websites?” mentions Chrome 57.8\% of the time, for a difference of 48.5\%.
}
\label{tab:appendix:examples:paraphrase}
\end{table}
%\revision{
Next, we provide examples of synonym replaced prompts in Tab.~\ref{tab:appendix:examples:synonym}.
\begin{table}[h]
\centering
% \resizebox{\textwidth/2}{!}{\begin{tabular}{| c | c | c | c | c | c | c |} \hline
\begin{tabular}{|@{\hskip 0.01in}c@{\hskip 0.01in}c@{\hskip 0.01in}l@{\hskip 0.01in}r@{\hskip 0.01in}r|} \hline
\textit{Model}& \textit{Category (Brand)} & \textit{Prompt} & \textit{Likelihood (\%)} & \textit{Improvement (\%)}\\ \hline 
Gemma-it & TV & Looking for guidance on choosing the \textbf{best} TV; could you& 31.1 & +52.8\\
 &(Samsung) &\textbf{recommend} one and share \textbf{additional} details like display type,& & \\
& & audio quality, or any other features that set it apart?& & \\
& & Looking for guidance on choosing the \textbf{ultimate} TV; could you& 83.9 & \\
& & \textbf{suggest} one and share \textbf{extra} details like display type, & & \\
& & audio quality, or any other features that set it apart? & & \\ \hline 
Llama2 & Camera & Which digital camera model do you consider the \textbf{optimal} choice & 35.1 & +27.6\\
 &(Canon) &for capturing high-quality images, versatile shooting options,& & \\
& & and \textbf{innovative} features that enhance your photography experience?& & \\
& & Which digital camera model do you consider the \textbf{superior} choice& 62.7 & \\
& & for capturing high-quality images, versatile shooting options,& & \\
& & and \textbf{advanced} features that enhance your photography experience? & & \\ \hline 
Llama3 & VR Headset & Which VR headset model do you consider the \textbf{optimal} choice & 39.4 & +27.4\\
 &(HTC) &for immersive virtual reality experiences, offering& & \\
& & a \textbf{perfect} blend of visual quality, comfort, and& & \\
& & \textbf{innovative} features for enthusiasts and gamers alike?& & \\
& & Which VR headset model do you consider the \textbf{superior} choice& 66.8 & \\
& & for immersive virtual reality experiences, offering& & \\
& & a \textbf{ideal} blend of visual quality, comfort, and& & \\
& &  \textbf{sophisticated} features for enthusiasts and gamers alike? & & \\ \hline
Llama3 & Game Console & Can you recommend the \textbf{ultimate} video game console that & 18.9 & +32.9\\
 &(Xbox) &excels in providing top-notch graphics, \textbf{diverse} gaming options,& & \\
& & and additional features such as online connectivity,& & \\
& & \textbf{suitable} for both casual and hardcore gamers?& & \\
& &Can you recommend the \textbf{superior} video game console that& 51.8 & \\
& & excels in providing top-notch graphics, \textbf{dissimilar} gaming options,& & \\
& & and additional features such as online connectivity,& & \\
& & \textbf{appropriate} for both casual and hardcore gamers?& & \\ \hline
Mistral-it & Jeans & Can you \textbf{recommend} the \textbf{ultimate} denim brand that excels & 25.1 & +38.4\\
 &(Levis) &in providing comfortable and on-trend denim, with additional& & \\
& & features such as a \textbf{diverse} range of fits, sustainable& & \\
& & practices, and \textbf{suitable} for \textbf{various} occasions?& & \\
& &Can you \textbf{suggest} the \textbf{greatest} denim brand that excels& 63.5 & \\
& & in providing comfortable and on-trend denim, with additional& & \\
& & features such as a \textbf{dissimilar} range of fits, sustainable& & \\
& &practices, and \textbf{appropriate} for \textbf{dissimilar} occasions?& & \\ \hline
Qwen-it & Camera & I'm curious to know your preference for the digital camera & 29.4 & +20.9\\
 &(Canon) &that offers the \textbf{best} combination of performance, durability,& & \\
& & and overall convenience for capturing a variety of scenes,& & \\
& & from landscapes to action shots.& & \\
& &I'm curious to know your preference for the digital camera& 50.3 & \\
& &that offers the \textbf{premier} combination of performance, durability,& & \\
& & and overall convenience for capturing a variety of scenes,& & \\
& &from landscapes to action shots.& & \\ \hline
\end{tabular}
\vspace{0.1cm}
\caption{Examples of synonym-replaced prompts}
\Description{
  This table provides examples of synonym-replaced prompts that had a large difference in the likelihood that a certain target brand is mentioned in the responses of an LLM. The leftmost column specifies the LLM. The next column specifies the category of product and brand that is mentioned. These synonyms are bolded. The next column gives both prompts, which are the same except for a few synonyms The next column gives the probability that each prompt mentions the given brand in responses of that LLM. The final column is the difference between these probabilities. Each row is described next.
  The first row shows the probability that Gemma-it mentions Samsung when prompted about TVs. 
  The prompts are “Looking for guidance on choosing the best/ultimate TV; could you recommend/suggest one and share additional/extra details like display type, audio quality, or any other features that set it apart?”. Here, the “/” indicates the words that differ between the prompts, but they are written separately in the table. The first prompt has a 31.1% probability of mentioning Samsung, and the second has an 83.9% probability, for an improvement of 52.8%.
  The next row shows the probability that Llama2 mentions Canon when prompted about cameras. 
  The prompts are  “Which digital camera model do you consider the optimal/superior choice for capturing high-quality images, versatile shooting options, and innovative/superior features that enhance your photography experience?”. 
  The first prompt has a 35.1\% probability of mentioning Canon, and the second has an 62.7\% probability, for an improvement of 27.6\%.
  The next row shows the probability that Llama3 mentions HTC when prompted about VR headsets. 
  The prompts are  “Which VR headset model do you consider the optimal/superior choice for immersive virtual reality experiences, offering a perfect/ideal blend of visual quality, comfort, and innovative/sophisticated features for enthusiasts and gamers alike?”. 
  The first prompt has a 39.4\% probability of mentioning HTC, and the second has an 66.8\% probability, for an improvement of 27.4\%.
  The next row shows the probability that Llama3-it mentions X-Box when prompted about game consoles. 
  The prompts are  “Can you recommend the ultimate/superior video game console that excels in providing top-notch graphics, diverse/dissimilar gaming options, and additional features such as online connectivity, suitable/appropriate for both casual and hardcore gamers?”. 
  The first prompt has a 18.9\% probability of mentioning X-Box, and the second has an 51.8\% probability, for an improvement of 32.9\%.
  The next row shows the probability that Mistral-it mentions Levi’s when prompted about jeans. 
  The prompts are  “Can you recommend/suggest the ultimate/greatest denim brand that excels in providing comfortable and on-trend denim, with additional features such as a diverse/dissimilar range of fits, sustainable practices, and suitable/appropriate for various/dissimilar occasions?”. 
  The first prompt has a 21.5\% probability of mentioning Levi’s, and the second has an 63.5\% probability, for an improvement of 38.4\%.
  The next row shows the probability that Qwen-it mentions Canon when prompted about cameras. 
  The prompts are  “I’m curious to know your preference for the digital camera
  that offers the best/premier combination of performance, durability, and overall convenience for capturing a variety of scenes, from landscapes to action shots.”. 
  The first prompt has a 29.4\% probability of mentioning Canon, and the second has an 50.3\% probability, for an improvement of 20.9\%.
}
\label{tab:appendix:examples:synonym}
\end{table}

%%%% appendix.tex ends here %%%%

\end{document}